# Statistical Model of Heavy-Ion Fusion-Fission Reactions

J. P. Lestone and S. G. McCalla[*]

Applied Physics Division, Los Alamos National Laboratory
Los Alamos, NM, 87545, USA
July 21st, 2008

Cross-section and neutron-emission data from heavy-ion fusion-fission reactions are consistent with the fission of fully equilibrated systems with fission lifetime estimates obtained via a Kramers-modified statistical model which takes into account the collective motion of the system about the ground state, the temperature dependence of the location and height of fission transition points, and the orientation degree of freedom. If the "standard" techniques for calculating fission lifetimes are used, then the calculated excitation-energy dependence of fission lifetimes is incorrect. We see no evidence to suggest that the nuclear viscosity has a temperature dependence. The strong increase in the nuclear viscosity above a temperature of ~1.3 MeV deduced by others is an artifact generated by an inadequate fission model.

## I. INTRODUCTION

The study of the fission of highly excited nuclei remains a topic of great interest [1 -5]. It has been known for more than twenty years that the "standard" statistical theory of fission leads to an underestimate of the number of measured prescission neutrons emitted in heavy-ion reactions [6 -10]. It is generally accepted that the main cause of this discrepancy is effects associated with the viscosity of hot nuclear matter [11]. More recently, giant dipole resonance (GDR) γ-ray emission has also been used to infer inadequacies in our models of nuclear fission decay widths [12 -15]. Assuming the standard methods for calculating fission decay widths are correct, many authors have adjusted the properties of the viscosity of hot nuclear matter to reproduce experimental data. Based on these analyses, it is generally believed that the collective motion in the fission degree of freedom is strongly damped for hot systems and that the nuclear viscosity increases strongly with either the temperature and/or the nuclear deformation [13-15]. A consensus appears to have emerged that strong dissipation sets in rather rapidly at nuclear excitation energies above ~40 MeV [12], i.e. above a nuclear temperature of ~1.3 MeV. Few have considered the possibility that the problem with the "standard" model of fission is due to, or partly due to, an incorrect implementation of the standard model.

In the present work we show that the standard techniques that have been widely used to model heavy-ion induced fusion-fission reactions are missing three key pieces of physics. These pieces of physics have been previously discussed individually in the literature, but have not been incorporated into many of the codes used to model heavy-ion fusion-fission reactions. These codes include CASCADE [16], ALERT [17], ALICE [18], PACE [19], JULIAN [20], and JOANNE [21]. The key pieces of physics missing from the above-mentioned codes include: the determination of the total level density of the compound system taking into account the collective motion of the system about the ground-state position [22]; the calculation of the location and height of fission saddle-points as a function of excitation energy using the derivative of the free energy [23,24]; and the incorporation of the orientation (K-state) degree of freedom [25,26].

If the "standard" (but incorrect) techniques for calculating fission lifetimes are used, then the calculated excitation-energy dependence of fission lifetimes is incorrect. The nature of the inadequacies in the techniques commonly used can be overcome by using a nuclear viscosity that increases strongly with increasing temperature. We show that if heavy-ion fusion-fission lifetimes are modeled in a more correct fashion, then fission cross sections and prescission neutron multiplicity data are consistent with the fission of fully equilibrated nuclear systems. The fission cross sections and prescission neutron-multiplicity data are consistent with a nuclear viscosity at the fission saddle points that is independent of temperature [27] as given by the surface-plus-window dissipation model of Nix and Sierk [28,29], the finite-range liquid-drop model [30]; and a nuclear shape dependence of the Fermi-gas level-density parameter in the range of theoretical estimates [31 -37].

## II. THEORY

In many respects, the theory of heavy-ion induced fusion-fission reactions is relatively simple. Much of the available data can be understood using statistical mechanics with a few semi-classical modifications. Although each piece of theory required is relatively simple, model calculations quickly become complex due to the large number of physical considerations that need to be modeled correctly. These include: the potential-energy surface of cold nuclei as a function of elongation (deformation), total spin $J$, and spin about the elongation (symmetry) axis $K$; the level density of the compound system as a function of shape; the total level density including collective motion; the calculation of equilibrium shapes and potential curvatures, and fission-barrier heights, using the force on the collective degree of freedom as a function of shape, orientation, and temperature; the nuclear viscosity; the fusion spin distribution; and the modeling of cooling processes (particle evaporation and $\gamma$–ray emission) that compete with fission.

[*] Present Address : Division of Applied Mathematics, Brown University, RI, 02912, USA

We claim that others have not included several key pieces of physics when calculating fission lifetimes. Therefore, we describe the calculation of the fission lifetimes of hot rotating nuclei in detail in sections II.A to II.G. We start from a very simple idealized system and slowly increase the complexity of the calculations with each successive section, until the methodology used by the statistical-model code JOANNE4 [25] is described. At each step in added complexity, the validity of analytical expressions based on statistical physics are tested by comparing to numerical results obtained using dynamical theory. Some may view the detailed description of fission presented here as excessive. However, given that the concepts discussed here have been previously introduced but not widely adopted, we feel that a slow and detailed build-up in system complexity is warranted. The methods used by others to model the fusion of the projectile and target, and the cooling processes are generally adequate. However, for completeness, we summarize the methods used in the code JOANNE4 to model fusion, particle evaporation, and $\gamma$–ray emission in sections II.H, II.I, and II.J.

## A. BOHR-WHEELER FISSION DECAY WIDTH

The mean time for a system in thermo-dynamical equilibrium to find a given quantum state is given by

$$\bar{t} = h\rho \ , \tag{1}$$

where $h$ is Planck's constant, and $\rho$ is the total level density of the system. A system may have a number of states that, if attained, will cause the system to make an irreversible transition from an initial configuration into another configuration. The mean time for such an irreversible transition is given by

$$\bar{t} = \frac{h\rho}{N_{TS}} \ , \tag{2}$$

where $N_{TS}$ is the number of transition states. Converting this mean time into a decay width gives the Bohr-Wheeler decay width [38]

$$\Gamma = \frac{\hbar}{\bar{t}} = \frac{N_{TS}}{2\pi \ \rho} \ . \tag{3}$$

These are powerful and elegant expressions that can be used to easily obtain the properties of particle emission from a hot oven [39] and, thus, the Maxwell velocity distribution for an ideal gas; black body radiation [39]; particle evaporation from hot nuclei; and the probability per unit time that a hot equilibrated nucleus will fission. Fig. 1 is a schematic representation of a fissioning compound nucleus showing levels at both the ground state and fission saddle point. Key properties that govern the fission lifetime are the thermal excitation energy at the ground-state position $U$, and the height of the fission barrier $B_f$. The level density of the nuclear system at both the ground-state and saddle-point positions is often estimated assuming a weakly interacting Fermi-gas and expressed (approximately) as [40]

$$\rho(U) \propto \exp\left(2\sqrt{a(q)U}\right), \tag{4}$$

where $a(q)$ is the Fermi-gas level-density parameter as a function of the deformation $q$, and $U$ is the thermal excitation energy of the system given by

$$U(q) = E - V(q) \ , \tag{5}$$

where $E$ is the total excitation energy of the system, and $V(q)$ is the potential energy. Using the standard definition of the inverse of temperature as the logarithmic derivative of the level density gives the familiar expression

$$U(q) = a(q)T^2(q) \ . \tag{6}$$

More complex expressions for the Fermi-gas level-density exist [26,40] and will be introduced in later sections. However, these more complete expressions generally make little difference to the overall properties of hot systems with thermal excitation energies larger than several tens of MeV. The Fermi-gas level-density parameter is equal to the total density of neutron and proton states at the Fermi surface multiplied by $\pi^2/6$ [40] and should be considered a function of the nuclear shape. However, for simplicity we shall initially assume that the level-density parameter is independent of deformation. The complexities associated with a shape dependence of the level-density parameter will be introduced in section II.F.

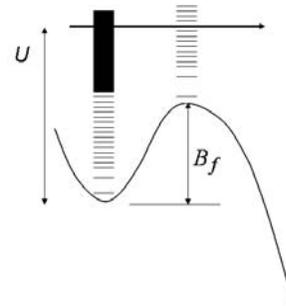

Fig. 1. A schematic representation of a fissioning compound nucleus showing levels at both the ground-state and fission saddle point, the thermal excitation energy at the ground-state position $U$, and the height of the fission barrier $B_f$.

Within the framework of a one-dimensional model, the Bohr-Wheeler fission decay width is often expressed as (see for example ref [13])

$$\Gamma_f^{BW} = \frac{1}{2\pi\,\rho_{gs}(E-V_{gs})}\int_0^{E-V_{gs}-B_f} \rho_{sp}(E-V_{gs}-B_f-\varepsilon)\,d\varepsilon \ ,$$
(7)

where $B_f$ is the fission-barrier height and the subscripts "gs" and "sp" denote the ground state and saddle point, respectively. The integral in Eq. (7) is dominated by $\varepsilon$ in the range from zero to a few times the temperature, and thus to a good approximation, we can substitute into Eq. (7) the expression

$$\rho(U-\varepsilon) \sim \rho(U)\exp(-\varepsilon/T) \ ,$$
(8)

and obtain

$$\Gamma_f^{BW} = \frac{T_{sp}\,\rho_{sp}(E-V_{gs}-B_f)}{2\pi\,\rho_{gs}(E-V_{gs})} \ .$$
(9)

If the level densities $\rho_{gs}$ and $\rho_{sp}$ are assumed to be as given in Eq. (4) and the level-density parameter $a$ is assumed to be a constant, then in the limit of a small barrier height or very high excitation energy, the temperatures at the ground state and saddle points as defined in Eq. (6) will be equal and the fission decay width becomes

$$\Gamma_f^{BW} = \frac{T}{2\pi}\exp(-B_f/T) \ .$$
(10)

In general, the barrier height can be large enough and the excitation energy low enough such that the temperatures at the ground state and saddle point are significantly different. If the excitation-energy dependence of the level density is as given in Eq. (4) then the fission decay width can be expressed as

$$\begin{aligned}\Gamma_f^{BW} &= \frac{T_{sp}}{2\pi}\exp(2\sqrt{a(U_{gs}-B_f)}-2\sqrt{aU_{gs}})\\ &= \frac{T_{sp}}{2\pi}\exp(\frac{2(U_{gs}-B_f)}{T_{sp}}-\frac{2U_{gs}}{T_{gs}})\\ &= \frac{T_{sp}}{2\pi}\exp(\frac{-2B_f}{T_{gs}+T_{sp}}) \ .\end{aligned}$$
(11)

## B. FISSION FROM A SQUARE-WELL POTENTIAL WITH A NARROW BARRIER

We now consider the fission decay width for a simplistic system with a potential energy $V(q)$ as a function of deformation $q$ as shown in Fig. 2. In this section, we assume the width of the barrier $\Delta x_{sp}$ is small. Through very simple arguments, it is clear that key physics is missing from Eqs. (10)-(11). These equations contain no terms that allow the fission decay width to change based on the width of the ground-state well, as must be the case. If the width of the ground-state well $\Delta x_{gs}$ shown in Fig. 2 increases, the system will encounter the barrier region less often and the decay width must decrease. This apparent problem with the statistical model was overcome by Strutinsky [22] more than 30 years ago. Strutinsky pointed out that the total level density of the system must not be estimated assuming the system exists at only the ground-state equilibrium position, but must be calculated taking into account the collective motion about the ground-state position. If the level density as a function of thermal excitation energy at a fixed point is assumed to be $\rho(U)$, then the total level density, in a one dimensional model, is given by [22]

$$\rho_{tot}(E) = \iint \rho(E-V(q)-\frac{p^2}{2\mu})\frac{dq\,dp}{h} \ ,$$
(12)

where $\mu$ is the inertia of the collective coordinate. The integrals are over all collective momenta $p$ and over all locations $q$ that make up the ground-state well. For the square-well potential shown in Fig. 2 the total level density is given by

$$\rho_{tot}(E) = \int_{p=-\infty}^{\infty}\int_0^{\Delta x_{gs}} \rho(E)\exp(-\frac{p^2}{2\mu T})\frac{dx\,dp}{h} \ .$$
(13)

If we assume that the inertia is independent of the location, then Eq. (13) simplifies to

$$\rho_{tot}(E) = \rho(E)\frac{T}{\hbar\omega_{eff}} \ , \quad \text{with} \quad \omega_{eff} = \frac{1}{\Delta x_{gs}}\sqrt{\frac{2\pi T}{\mu}} \ .$$
(14)

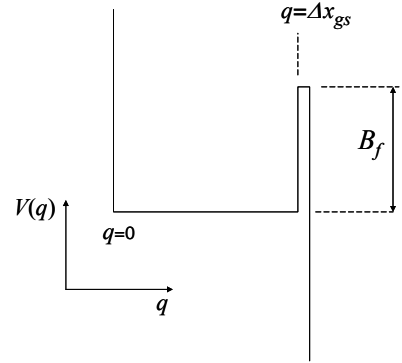

Fig. 2. A simple square well potential with a narrow barrier. The ground-state well has a width $\Delta x_{gs}$.

If Eq. (10) is recalculated correctly, taking into account the motion about the ground-state position, then the fission decay width is

$$\Gamma_f = \frac{\hbar\omega_{eff}}{2\pi}\exp(-B_f/T) \ .$$
(15)

To confirm that Eq. (15) is the correct expression for the fission decay width for the potential shown in Fig. 2, we calculate the mean fission time by numerical means using the Langevin equation [41]. The acceleration of the collective coordinate $q$ over a small time interval $\delta t$ is given by [41]

$$\ddot{q} = \frac{-1}{\mu}\frac{\partial V}{\partial q}-\frac{\dot{q}^2}{2\mu}\frac{\partial\mu}{\partial q}-\beta\dot{q}+\Gamma\sqrt{\frac{2\beta T}{\delta t\,\mu}} \ ,$$
(16)

where $\Gamma$ is a random number from a normal distribution with unit variance, and $\beta$ is the reduced nuclear dissipation coefficient which controls the coupling between the collective motion and the thermal degrees of freedom. We start with an ensemble of systems at $t=0$, each with $q(t=0)=\Delta x_{gs}/2$, and with the collective velocity set to zero. Each system can be stepped forward in time by randomly picking an acceleration for each system using Eq. (16) and then using

$$q_1(t+\delta t) = q(t) + \dot{q}(t)\delta t + \frac{\ddot{q}(t)}{2}\delta t^2 \quad , \text{ and} \quad (17)$$

$$\dot{q}_1(t+\delta t) = \dot{q}(t) + \ddot{q}(t)\delta t \quad . \quad (18)$$

Repeated application of Eqs. (16)-(18) can be used to march an ensemble of systems forward in time. However, this is inefficient because the required $\delta t$ for numerical convergence is very small. Computational inefficiency can be improved by including higher order corrections at each time step using the following equations:

$$\ddot{q}_m(t+\delta t) = \frac{-1}{\mu(q_m)}\frac{\partial V(q_m)}{\partial q} - \frac{\dot{q}_m^2}{2\mu(q_m)}\frac{\partial \mu(q_m)}{\partial q}$$

$$- \beta\,\dot{q}_m + \Gamma_1\sqrt{\frac{2\beta T}{\delta t\,\mu(q_m)}} \quad , \quad (19)$$

where m is initially equal to 1 and $q_1, \dot{q}_1, \ddot{q}_1$ are obtained using Eqs. (17)-(19). An estimate of $\ddot{q}$ over the time step from $t$ to $t+\delta t$ can be obtained using

$$\ddot{q}_m(t \to t+\delta t) = \frac{\ddot{q}_m(t+\delta t) - \ddot{q}(t)}{\delta t} \quad . \quad (20)$$

An improved estimate of the relevant properties of the collective degree of freedom at $t+\delta t$ is obtained using

$$q_{m+1}(t+\delta t) = q_1(t+\delta t) + \ddot{q}_m(t \to t+\delta t)\frac{\delta t^3}{6} \quad , \quad (21)$$

$$\dot{q}_{m+1}(t+\delta t) = \dot{q}_1(t+\delta t) + \ddot{q}_m(t \to t+\delta t)\frac{\delta t^2}{2} \quad , \quad (22)$$

$$\ddot{q}_{m+1}(t+\delta t) = \frac{-1}{\mu(q_{m+1})}\frac{\partial V(q=q_{m+1})}{\partial q}$$

$$- \frac{\dot{q}_{m+1}^2}{2\mu(q_{m+1})}\frac{\partial \mu(q=q_{m+1})}{\partial q} - \beta\,\dot{q}_{m+1} + \Gamma_1\sqrt{\frac{2\beta T}{\delta t\,\mu(q_{m+1})}} \quad . \quad (23)$$

The random number in Eq. (23) is not updated from the value used in Eq. (19). Repeated application of Eqs. (20)-(23) rapidly converges to a more accurate estimate of the relevant properties of the collective degree of freedom at $t+\delta t$. We have found that excellent results are obtained if Eqs. (20)-(23) are applied three times at each time step. Using this technique, the time to successfully surmount the fission barrier can be recorded for each history in a large ensemble. The mean fission time is then obtained by averaging over the ensemble.

Fig. 3 shows the results of Langevin (dynamical) calculations for the mean fission time as a function of the reduced nuclear dissipation coefficient for the potential shown in Fig. 2. The parameters controlling the potential are, for this case, chosen to be $B_f=3$ MeV, and $\Delta x_{gs}=5$ fm. The temperature is assumed to be 1 MeV, and the inertia of the collective coordinate is assumed to be $\mu=50$ atomic mass units (amu). The barrier width is assumed to be narrow. Calculations are shown for numerical time steps $\delta t = 3\times10^{-22}$ s, $10^{-22}$ s, $3\times10^{-23}$ s, and $10^{-23}$ s. These results show that numerical convergence is more difficult with increasing viscosity. Convergence is nearly obtained with $\delta t = 10^{-22}$ s for $\beta < 3\times10^{21}$ s$^{-1}$, and with $\delta t = 10^{-23}$ s for $\beta < 4\times10^{21}$ s$^{-1}$. Convergence is more easily obtained with more realistic potentials that do not contain discontinuities as a function of deformation.

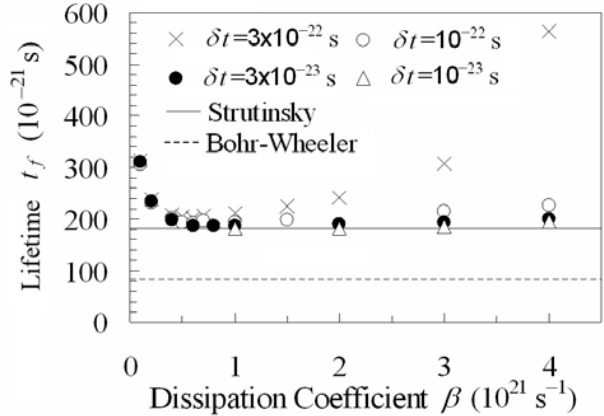

Fig. 3. Langevin model calculations for the mean fission time as a function of the reduced nuclear dissipation coefficient for the potential shown in Fig. 2, with $B_f=3$ MeV, $\Delta x_{gs}=5$ fm, $T=1$ MeV, and $\mu=50$ amu. The crosses, open circles, solid circles, and open triangles show Langevin calculations with numerical time steps $\delta t = 3\times10^{-22}$ s, $10^{-22}$ s, $3\times10^{-23}$ s, and $10^{-23}$ s, respectively. The solid and dashed lines show corresponding statistical-model estimates obtained using two different approaches (see text).

The Langevin calculations with $\delta t=10^{-23}$ s shown in Fig. 3 give a mean fission time of $t_f \sim 180\times10^{-21}$ s that is insensitive to $\beta$ in the range from $\beta \sim 0.5\times10^{21}$ s$^{-1}$ to $3\times10^{21}$ s$^{-1}$. The slight increase in the calculated fission time beyond $\beta \sim 3\times10^{21}$ s$^{-1}$ is associated with numerical convergence issues. The increase in the mean fission time below $\beta \sim 0.5\times10^{21}$ s$^{-1}$ is caused by the weak coupling between the collective motion and the thermal degrees of freedom. Below $\beta \sim 0.5\times10^{21}$ s$^{-1}$ the mean fission time is increasingly governed by the time it takes the collective motion to equilibrate with the thermal degrees of freedom. Above $\beta \sim 0.5\times10^{21}$ s$^{-1}$ the fission time is controlled by the time it takes the equilibrated system to randomly produce systems near the barrier with enough collective motion to overcome the barrier. In the case of a narrow barrier, the mean fission time for a fully equilibrated system is completely governed by equilibrium (statistical) physics and the mean fission time is independent of the reduced nuclear dissipation coefficient, $\beta$. Applying the statistical model incorrectly by estimating the mean fission time using Eq. (10), for the case considered in Fig. 3, gives $t_f =83\times10^{-21}$ s. This value is shown by the dashed line in Fig.

3 (labeled Bohr-Wheeler) and is in disagreement with the Langevin calculations shown in Fig. 3. Applying the statistical model correctly, as outlined by Strutinsky [22], by estimating the mean fission time using Eq. (15), gives $t_f$ =181×10⁻²¹ s. This value is displayed by the solid line in Fig. 3 and is consistent with the $\beta > 0.5 \times 10^{21}$ s⁻¹ and $\delta t = 10^{-23}$ s Langevin calculations shown in Fig. 3. Technically, both the solid and dashed lines show Bohr-Wheeler calculations. Unfortunately, the way Eq. (7) and approximations thereof are used is incorrect. These methods have been commonly referred to in the literature as the Bohr-Wheeler fission model. In the present paper we will continue to label inadequate approaches as the Bohr-Wheeler model to separate it from the Bohr-Wheeler model applied correctly as described by Strutinsky.

Fig. 4 compares various model calculations of the mean fission time with $B_f$ =3 MeV, $T$=1 MeV, $\mu$=50 amu, and $\beta$=10²¹ s⁻¹ as a function of the width of the ground-state well $\Delta x_{gs}$. Applying the statistical model incorrectly, as described above, gives a mean fission time that is independent of the width of the ground-state well. This result is unphysical. Applying the statistical model correctly, as outlined by Strutinsky [22], gives a mean fission time that increases linearly with the width of the well, in agreement with the Langevin calculations shown in Fig. 4.

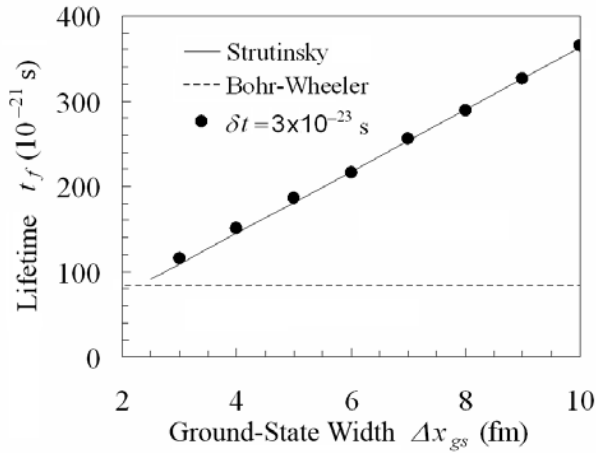

Fig. 4. Various model calculations of the fission lifetime for the potential shown in Fig. 2, with $B_f$=3 MeV, $T$=1 MeV, $\mu$=50 amu, and $\beta$=10²¹ s⁻¹ as a function of the width of the ground-state well $\Delta x_{gs}$. The solid circles show the results of Langevin calculations. The solid and dashed lines show statistical-model estimates obtained using two different approaches (see text).

Fig. 5 compares various model calculations of the mean fission time with $B_f$=3 MeV, $\Delta x_{gs}$ =5 fm, $\mu$=50 amu, and $\beta$=10²¹ s⁻¹ as a function of the temperature $T$. The solid line shows results obtained by applying the statistical model correctly via Eq. (15). These results are in excellent agreement with the Langevin calculations shown by the circles. Notice that the statistical-model estimates are excellent even at $B_f/T$=0.5. Results obtained by Eq. (10) are shown by the dashed curve. These mean fission times are off by a factor of $T/(\hbar \omega_{gs})$ and thus have a dependence on temperature (excitation energy) that is incorrect. This problem with the standard statistical model has been addressed by some. For example, Gontchar and Fröbrich [23] multiply the standard statistical fission rate by $\hbar \omega_{gs}/T$. However, many authors in the field continue to ignore this correction. This has been partially justified because the $\hbar \omega_{gs}/T$ correction is of the order of one [12] and generally expected to be of little importance given the uncertainty and the number of adjustable parameters in the statistical model of nuclear reactions. However, the standard techniques for estimating fission lifetimes use multiple approximations, and several of these approximations each cause the fission lifetime in heavy-ion fusion-fission reactions to be increasingly underestimated with increasing excitation energy. It is important to address each of these issues because their cumulative effect is significant in heavy-ion reactions.

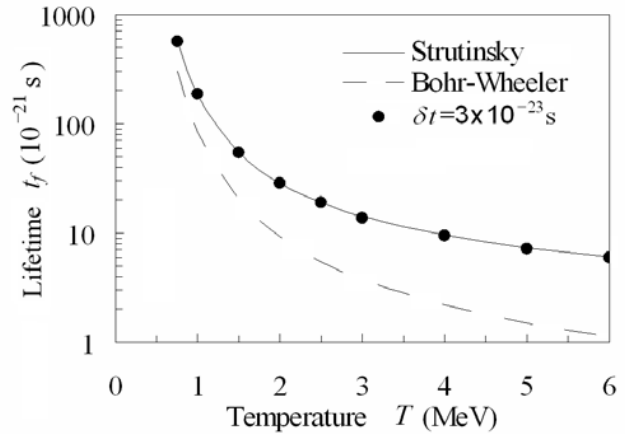

Fig. 5. Various model calculations of the mean fission time for the potential shown in Fig. 2, with $B_f$=3 MeV, $\Delta x_{gs}$ =5 fm, $\mu$=50 amu, and $\beta$=10²¹ s⁻¹ as a function of the temperature $T$. The solid circles show the results of Langevin calculations. The solid and dashed lines show statistical-model estimates obtained using two different approaches (see text).

## C. EFFECT OF A FINITE BARRIER WIDTH

If the barrier is narrow then every time the barrier is surmounted, the barrier is successfully crossed and the mean fission time for an equilibrated system is completely governed by statistical physics: i.e. surmounting the barrier leads to an irreversible transition. However, if the barrier has a finite width then the coupling between the collective motion and the thermal degrees of freedom produces a non-equilibrium effect while the barrier is being transversed, which leads to an increase in the mean fission time relative to that obtained by a purely statistical model. This effect is well known and has been incorporated into statistical models of heavy-ion fission since the early 1980's. However, it is generally discussed within the framework of a parabolic barrier, as is done in the next section. We believe that readers who are not familiar with this effect will obtain a better intuitive feel for its origin if it is first introduced for a system with a more simple potential.

Consider an equilibrated system with $T=2$ MeV, $\mu=50$ amu, and a potential of the form shown in Fig. 2 with $B_f=3$ MeV, $\Delta x_{gs}=5$ fm, and a finite barrier width $\Delta x_{sp}=5$ fm. The mean time for this equilibrated system to surmount (get on top of) the fission barrier is correctly given by Eq. (15) and is $3\times10^{-20}$ s. Upon surmounting the barrier, all systems will have an initial collective motion which will take the systems to larger deformation. However, as the barrier is traversed, the coupling between the collective motion and the thermal degrees of freedom will cause the systems to lose their memory of their initial motion toward larger deformation. The typical collective kinetic energy towards larger deformation at the moment the barrier is surmounted will be approximately the temperature of the system $T$. The average distance that a system will travel across a flat potential before losing all memory of a collective motion with kinetic energy $E=T$, is approximately given by

$$\Delta x \sim \frac{1}{\beta}\sqrt{\frac{2T}{\mu}} . \tag{24}$$

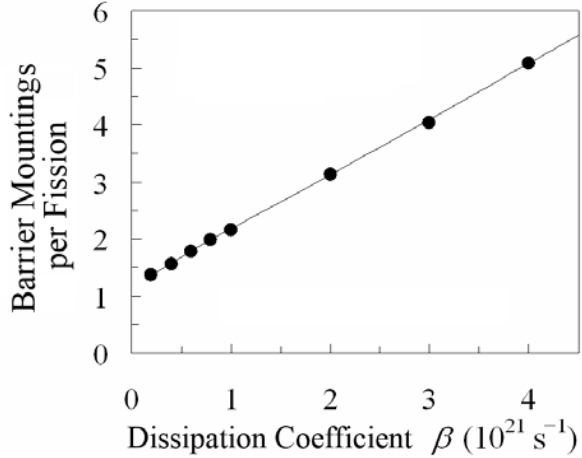

Fig. 6. Langevin calculations (circles) of the ratio of the barrier mountings to the successful barrier crossings (fissions) as a function of $\beta$ for a square well potential with $T=2$ MeV, $\mu=50$ amu, $B_f=3$ MeV, and $\Delta x_{gs}=\Delta x_{sp}=5$ fm. The solid line guides the eye.

For a system with $T=2$ MeV, $\mu=50$ amu, and $\beta=10^{21}$ s$^{-1}$, we obtain $\Delta x\sim2.5$ fm. Therefore, if the barrier width is $\Delta x_{sp}=5$ fm then the average system will lose all memory of its motion towards larger deformation approximately halfway across the barrier. Based on the symmetry of this location, half of these systems will randomly find their way to the outer barrier edge and fission, while the other half will find the inner edge and return to the ground-state well.

This will cause the mean fission lifetime to be approximately twice the purely statistical result of $3\times10^{-20}$ s. As $\beta$ is increased above $10^{21}$ s$^{-1}$ then the memory loss will occur increasingly closer to the inner barrier edge, increasing the probability that the system will be returned to the ground-state well, and thus increasing the mean fission time. Fig. 6 shows Langevin calculations of the ratio of the barrier mountings to the successful barrier crossings as a function of $\beta$ for the system considered above. For large $\beta$ this ratio becomes $\sim\beta/\omega_{sp}$, where the effective angular frequency of the barrier is obtained via Eq. (14) by replacing $\Delta x_{gs}$ with $\Delta x_{sp}$. The symbols in Fig. 7 show dynamical calculations of the mean time spent in the ground-state well as a function of $\beta$ for the system discussed above. The curve shows the statistical-model result multiplied by the ratios shown in Fig. 6.

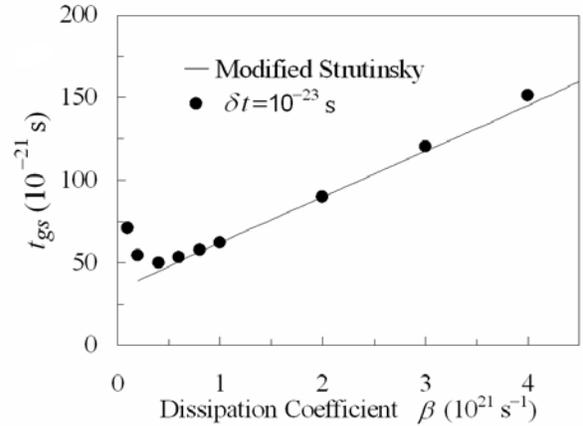

Fig. 7. The circles show Langevin calculations of the mean time spent in the ground-state well as a function of $\beta$ for a square well potential with $T=2$ MeV, $\mu=50$ amu, $B_f=3$ MeV, and $\Delta x_{gs}=\Delta x_{sp}=5$ fm. The solid line shows the statistical (Strutinsky) model result multiplied by the ratios shown in Fig. 6.

### D. PARABOLIC POTENTIALS

If the ground-state well is characterized by a parabolic (harmonic) potential

$$V_{gs}(q) = \frac{\mu\omega_{gs}^2 q^2}{2} , \tag{25}$$

then the total level density of the system (see Eq. (12)) can be expressed as [22]

$$\rho_{tot}(E) = \int_{p=-\infty}^{\infty}\int_{q=-\infty}^{\infty}\rho(E)\exp(-\frac{\mu\omega_{gs}^2 q^2}{2T})\exp(-\frac{p^2}{2\mu T})\frac{dq\,dp}{h} , \tag{26}$$

which gives

$$\rho_{tot}(E) = \rho(E)\frac{T}{\hbar\omega_{gs}} . \tag{27}$$

The corresponding statistical-model expression for the fission decay width from a harmonic well is

$$\Gamma_f = \frac{\hbar \omega_{gs}}{2\pi} \exp(-B_f/T) \ . \tag{28}$$

As discussed in section II.C, the purely statistical-model result given by Eq. (28) is only valid for an equilibrated system in the limit of either a narrow fission barrier or low dissipation. It is well known that the fission decay width for a system with a harmonic ground-state well and a parabolic barrier is reduced by dissipation [42] and given by

$$\Gamma_f = \left(\sqrt{1+\gamma^2} - \gamma\right) \times \frac{\hbar \omega_{gs}}{2\pi} \exp(-B_f/T) \ , \tag{29}$$

where $\gamma$ is the dimensionless nuclear viscosity given by

$$\gamma = \frac{\beta}{2\,\omega_{sp}} \tag{30}$$

and $\omega_{sp}$ is the angular frequency of the inverted potential around the barrier (saddle point). The scaling factor that modifies the purely statistical result is often referred to as the Kramers' reduction factor. In the limit of large nuclear viscosity, the Kramers' reduction factor becomes $1/(2\gamma) = \omega_{sp}/\beta$. Therefore, when the viscosity is large, the mean fission time is increased by a factor of $\beta/\omega_{sp}$ relative to the purely statistical result. This is analogous to the similar result obtained in section II. C.

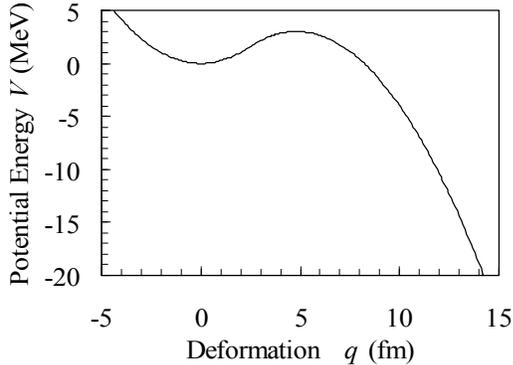

Fig. 8. Parabolic potential energy as a function of deformation $q$ for a system with $B_f$=3 MeV, $\omega_{gs}=\omega_{sp}=10^{21}$ s$^{-1}$, and $\mu$=50 amu.

To better understand Eq. (29) and further illustrate fission from parabolic potentials, consider the potential shown in Fig. 8. The potential around the ground-state position in Fig. 8 is as given by Eq. (25) with $\omega_{gs}=10^{21}$ s$^{-1}$ and $\mu$=50 amu. The potential around and beyond the fission saddle point is of the form

$$V_{sp}(q) = B_f - \frac{\mu \omega_{sp}^2 \left(q_{sp}-q\right)^2}{2} \ , \tag{31}$$

with $\omega_{sp}=10^{21}$ s$^{-1}$. Here, we have chosen a barrier height of $B_f$=3 MeV. Given the form of the potentials $V_{gs}$ and $V_{sp}$, in conjunction with the assumption of a smooth potential, the fission barrier height $B_f$=3 MeV defines the location of the saddle point to be $q_{sp}$=4.82 fm. The transition from $V_{gs}$ to $V_{sp}$ occurs at $q$=2.41 fm. Assuming the potential at the scission configuration (where the system breaks into two separate fission fragments) has a potential energy 20 MeV lower than the ground state ($q_{gs}$=0), defines the scission point to be at $q_{sc}$=14.2 fm. The solid curve in Fig. 9 shows the Kramers-modified statistical-model mean fission time obtained using Eq. (29) as a function of $\beta$, for a system with $T$=1 MeV, and the potential shown in Fig. 8. The symbols in Fig. 9 show Langevin calculations of the mean time spent inside the fission saddle point, for the same system. For this problem, numerical convergence is obtained using the dynamical model techniques outlined in section II.B with $\delta t$=3×10$^{-22}$ s up to $\beta\sim2\times10^{21}$ s$^{-1}$, and nearly obtained with $\delta t$=10$^{-22}$ s up to $\beta\sim6\times10^{21}$ s$^{-1}$. All following dynamical calculations are with $\delta t$=10$^{-22}$ s.

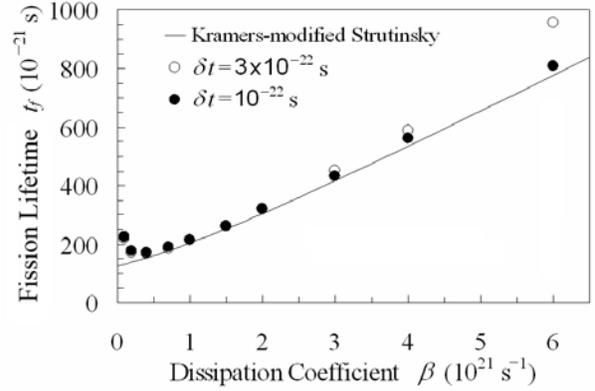

Fig. 9. Kramers-modified statistical-model mean fission time (solid curve) as function of $\beta$, for a system with $T$=1 MeV, and the potential shown in Fig. 8. The open and solid circles show the corresponding Langevin calculations of the mean time spent inside the fission saddle point with numerical time steps $\delta t$ equal to 3×10$^{-22}$ s, and 10$^{-22}$ s, respectively.

After a hot nucleus is formed, it takes a finite time period for the collective motion to equilibrate with the thermal degrees of freedom. During this equilibration time, the fission decay width will be lower than the Kramers-modified statistical value. This is why the dynamically calculated fission lifetimes shown in Fig. 9 are longer than the corresponding Kramers-modified statistical values below $\beta\sim0.5\times10^{21}$ s$^{-1}$. If $\omega_{gs}\gg\omega_{sp}$ then the time dependence of the fission decay width can be approximated by

$$\frac{\Gamma_f(t)}{\Gamma_f(\infty)} \sim \exp\left\{\frac{B_f}{T}\left(1 - \frac{1}{1-\exp(-t/\tau)}\right)\right\} \ . \tag{32}$$

The equilibration time is $\tau\sim1/\beta$ if $\beta<2\omega_{sp}$, and $\tau\sim\beta/(2\omega_{sp}^2)$ if $\beta>2\omega_{sp}$. Setting the fission decay width to 90% of its asymptotic value defines the transient fission delay time [43,44]

$$\tau_f \sim \tau \ln(10 B_f/T) \ . \tag{33}$$

The solid curve in Fig. 10 shows the time dependence of the fission decay width for a system with the potential shown in Fig. 8, $T$=1 MeV, and $\beta$=3×10$^{21}$ s$^{-1}$ estimated using Eq. (32). The solid circles show the corresponding numerical Langevin calculation assuming all systems start at $t$=0 at the ground-state position with no collective motion. There is no significant change in the calculation if the initial conditions are defined using the ground-state wave function corresponding to the ground-state well. For $\omega_{gs}=\omega_{sp}=10^{21}$ s$^{-1}$, Eq. (32) underestimates the transient delay by ~30%. The open circles and crosses in Fig. 10 show Langevin calculations for systems with $B_f$=3 MeV, $T$=1 MeV, $\beta$=3×10$^{21}$ s$^{-1}$, $\omega_{sp}$=10$^{21}$ s$^{-1}$, $\omega_{gs}/\omega_{sp}$=2 and 10, respectively. Notice that the agreement between Eq. (32) and the numerical Langevin calculations improves as $\omega_{gs}/\omega_{sp}$ becomes large. However, the assumptions used to obtain Eq. (29) are increasingly invalid as $\omega_{gs}/\omega_{sp}$ is increased much beyond unity. For real nuclei, $\omega_{gs}$ is typically ~50% larger than $\omega_{sp}$ (see section II.E). This difference between $\omega_{gs}$ and $\omega_{sp}$ is such that Eq. (29) is still reasonably valid and Eqs. (32) and (33) underestimate the transient delay by only ~20%.

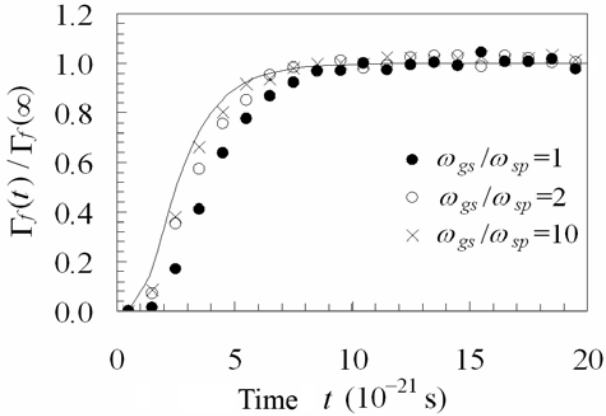

Fig. 10. The time dependence of the fission decay width for a system with the potential shown in Fig. 8, $T$=1 MeV, and $\beta$=3×10$^{21}$ s$^{-1}$ estimated using Eq. (32) (solid curve). The solid circles show the corresponding Langevin calculation. The open circles and crosses show Langevin calculations with $\omega_{sp}$=10$^{21}$ s$^{-1}$ and $\omega_{gs}/\omega_{sp}$=2 and 10, respectively.

In the limit of small $\beta$, the ratio of the Kramers-modified statistical fission lifetime to the transient delay is given by

$$\frac{t_f}{\tau_f} \sim \frac{2\pi\beta}{\omega_{gs}} \frac{\exp(B_f/T)}{\ln(10B_f/T)}. \quad (34)$$

This ratio can be made close to, or smaller than, one if $\beta$ is set low enough. Therefore, the Kramers-modified statistical model will fail at low $\beta$ as shown in Figs 3, 7, and 9. For modest to high values of $\beta$ (>2×10$^{21}$ s$^{-1}$) the ratio of the Kramers-modified statistical fission lifetime to the transient delay is given by

$$\frac{t_f}{\tau_f} \sim \frac{4\pi\omega_{sp}}{\omega_{gs}} \frac{\exp(B_f/T)}{\ln(10B_f/T)}. \quad (35)$$

Therefore, as long as the barrier is larger than the temperature, the Kramers-modified statistical fission lifetime will be more than ~4$\pi$·$e$/ln(10) ~15 times longer than the transient delay, and the transient delay can be neglected. If $\beta$~3×10$^{21}$ s$^{-1}$, $\omega_{sp}$~10$^{21}$ s$^{-1}$, and $B_f/T$ is in the range from 0.5 to 3, then the corresponding transient delays will range from ~2.5×10$^{-21}$ s to ~5×10$^{-21}$ s. These transient delays are short compared to the corresponding mean fission times. The only way the transient delay can be made important is if the viscosity is low; if the barrier is much smaller than the temperature; or if the mean fission time is made artificially small through the use of an inadequate model.

The symbols in Fig. 11 show Langevin calculations for the mean time spent between the saddle point and the scission point $\tau_{ssc}$ for a system with the potential shown in Fig. 8 with $T$=1 MeV, as a function of $\beta$. Analytical expressions for the mean time spent beyond the saddle point can be obtained. For example, it is easy to show that if a system crosses a parabolic barrier with collective energy $\varepsilon_{sp}$ then the time to transit from the saddle point to the scission point with no dissipation ($\gamma$=0) can be written as

$$\tau_{ssc}(\varepsilon_{sp}, \gamma=0) = \frac{\ln(4\Delta V/\varepsilon_{sp})}{2\omega_{sp}}, \quad (36)$$

where $\Delta V$ is the potential energy drop from the saddle point to the scission point. The average kinetic energy in the collective degree of freedom in the fission direction as the barrier is crossed is ~$T$. Therefore, a rough estimate of $\tau_{ssc}(\gamma$=0) can be obtained by simply using Eq. (36) with $\varepsilon_{sp}$ set to $T$. However, a more accurate value can be obtained using

$$\tau_{ssc}(\gamma=0) = \frac{\int_0^\infty \exp(-\varepsilon/T)\tau_{ssc}(\varepsilon_{sp}, \gamma=0)\,d\varepsilon}{\int_0^\infty \exp(-\varepsilon/T)\,d\varepsilon}$$
$$= f(\Delta V, T)\frac{\ln(4\Delta V/T)}{2\omega_{sp}}. \quad (37)$$

For a range of realistic combinations of $\Delta V$ and $T$ it can be shown that $f(\Delta V, T)$ is within 5% of 1.13. Using this result and the well known result for the viscosity dependence of the saddle-to-scission time [45] we obtain

$$\tau_{ssc}(\gamma) \sim 1.13 \times \frac{\ln(4\Delta V/T)}{2\omega_{sp}} \times \left(\sqrt{1+\gamma^2}+\gamma\right). \quad (38)$$

The solid curve in Fig. 11 shows the saddle-to-scission time obtained using Eq. (38) as a function of $\beta$ for a system with $T$=1 MeV, $\Delta V$=23 MeV, and $\omega_{sp}$=10$^{21}$ s$^{-1}$. These simple estimates are in excellent agreement with the corresponding Langevin calculations.

The total mean lifetime of the system is the sum of the mean time spent inside the saddle point and the mean saddle-to-scission time. For modest and large values of $\beta$, the ratio of the mean time spent inside the fission barrier to the mean saddle-to-scission time is

$$\frac{t_f}{\tau_{ssc}} \sim \frac{2\pi}{\omega_{eq}} \exp(B/T) \frac{2\omega_{sp}}{\ln(4\,\Delta V/T)} \ . \tag{39}$$

For typical fission reactions, the logarithm in Eq. (39) is between 3 and 5, and $\omega_{sp}/\omega_{gs} \sim 1$. Therefore, if the fission barrier is larger than the temperature, then the mean time spent inside the fission barrier will be more than a factor of $\sim\pi\cdot e \approx 8$ larger than the mean saddle-to-scission time, and the saddle-to-scission time can be neglected.

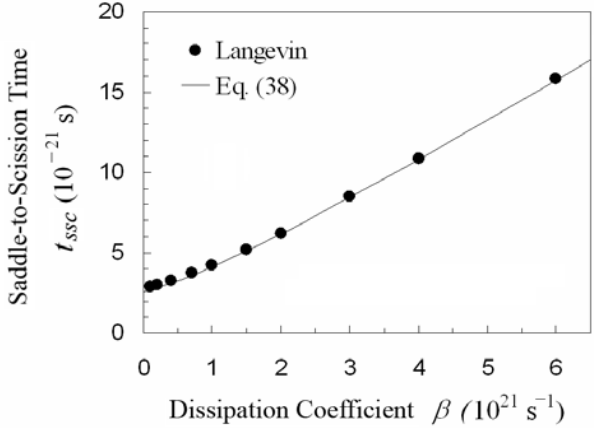

Fig. 11. Langevin model calculations for the mean time spent between the saddle point and the scission point $\tau_{ssc}$ (symbols) as a function of $\beta$, for a system with $T=1$ MeV, $\omega_{sp}=10^{21}$ s$^{-1}$, and a potential drop from the saddle point to the scission point $\Delta V=23$ MeV. The solid line shows the corresponding calculation using Eq. (38).

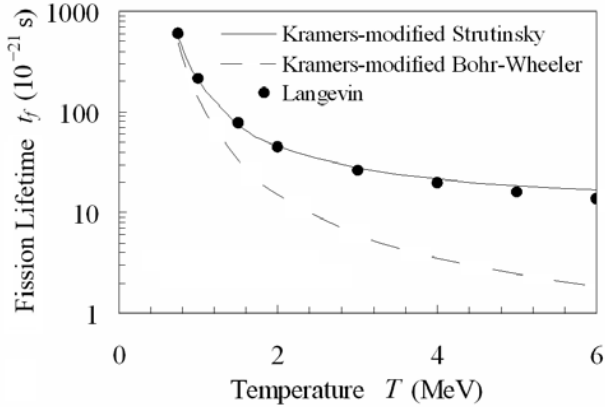

Fig. 12. Model calculations of the fission lifetime for the potential shown in Fig. 8, with $\mu=50$ amu, and $\beta=10^{21}$ s$^{-1}$ as a function of the temperature $T$. The solid and dashed lines show results obtained using Eq. (29) (Strutinsky) and the standard Kramers-modified Bohr-Wheeler formula (see text). The symbols show the corresponding Langevin calculations.

Fig. 12 compares various model calculations of the mean fission time for the potential shown in Fig. 8, with $\mu=50$ amu, and $\beta=10^{21}$ s$^{-1}$ as a function of the temperature $T$. The solid curve shows results obtained by applying the Kramers-modified statistical model via Eq. (29). These results are in reasonable agreement with the Langevin calculations of the mean time spent inside the saddle point shown by the circles. When the temperature is higher than the fission barrier, Eq. (29) overestimates the fission time because the integral over the collective coordinate in Eq. (26) is over all space. This approximation is valid if the temperature is smaller than $B_f$, and is made to obtain a simple analytical expression for the fission lifetime. At higher temperatures the transition to $V_{sp}(q)$ beyond $q=2.41$ fm should be taken into account and the integral over $q$ should be from $-\infty$ to $q_{sp}$. However, from Fig. 12 we see that Eq. (29) fails gracefully and is only off by $\sim 20\%$ at $B_f/T=0.5$. Results obtained using Eq. (10) multiplied by the Kramers' reduction factor are shown by the dashed curve. These mean fission times are incorrect and off by a factor of $T/(\hbar\omega_{gs})$.

## E. POTENTIALS FOR REAL NUCLEI

From the preceding sections, it is clear that the mean fission time does not just depend on the excitation energy, the nuclear dissipation, and the height of the fission barrier, but is sensitive to the shape of the potential-energy surface. However, many authors in the field continue to use the Bohr-Wheeler fission decay width as expressed in Eq. (7) with the level density as or similar to that given in Eq. (4) multiplied by the Kramer's reduction factor. This is, in part, because only the fission barriers and ground state energies have been determined via the finite-range liquid-drop model (FRLDM) [30] as a function of $Z$, $A$, and total spin $J$. These barrier heights and ground state energies have been parameterized, and the corresponding fits made available via the subroutine BARFIT written by Sierk. The parameterization contained within BARFIT reproduces the original FRLDM fission barriers with a typical error of 0.1 to 0.2 MeV. The root mean square (rms) difference between ground state energies obtained with BARFIT and the original FRLDM is $\sim 0.2$ MeV.

No parameterization of the shape of FRLDM potential-energy surfaces exists. However, a method for estimating finite-range corrected potential-energy surfaces by an empirical modification of the liquid-drop model has been proposed [46]. This method is referred to as the modified liquid-drop model (MLDM). In the MLDM, the potential energy of a nucleus, relative to its spin-zero ground state, is written as [25,46]

$$V(q,Z,A,J,K) =$$

$$(S'(q)-1)E_S^0(Z,A) + (C(q)-1)0.7053\frac{Z^2}{A^{1/3}}\,\text{MeV}$$

$$+ \frac{\left(J(J+1)-K^2\right)\hbar^2}{I_\perp(q)\frac{4}{5}MR_o^2+8Ma^2} + \frac{K^2\hbar^2}{I_\parallel(q)\frac{4}{5}MR_o^2+8Ma^2}, \tag{40}$$

where $E_S^0(Z,A)$ is the LDM surface energy of spherical nuclei as determined by Myers and Swiatecki [47,48], $M$ is the mass of the system, $R_o=1.2249$ fm $\times A^{1/3}$, and $a=0.6$ fm. $C(q)$, $I_\perp(q)$, and $I_\parallel(q)$, are the Coulomb energy, and the moments of inertia perpendicular to and about the symmetry axis of a sharp surfaced $^{208}$Pb ($J$=0) liquid-drop nucleus as a function of the distance between mass centers $q$ in units of the corresponding spherical values. $S'(q)$ is an empirically adjusted surface energy in units of the corresponding spherical value.

Unfortunately, when the MLDM was originally published [46], the $S'(q)$, $C(q)$, $I_\perp(q)$, and $I_\parallel(q)$ were only tabulated in steps of $q/R_o$=0.05. The nuclear potential energy is a delicate balance between surface and Coulomb energies and poor results can be obtained by a simple interpolation of the $S'(q)$, $C(q)$, $I_\perp(q)$, and $I_\parallel(q)$ values published in ref [46]. To obtain an accurate potential-energy surface, one must use a spacing in $q/R_o$ of, or smaller than, ~0.01. The recommended values of $S'(q)$, $C(q)$, $I_\perp(q)$, and $I_\parallel(q)$ are presented in Appendix A in steps of $q/R_o$=0.01. With these values, the nuclear potential energy can be easily estimated using Eq. (40) as a function of deformation $q$, $Z$, $A$, the total spin $J$, and the spin about the elongation axis $K$. It must be stressed that the MLDM does not introduce any new physics to the macroscopic modeling of rotating nuclei and it is not meant to supersede the FRLDM. The surface energy and the surface diffuseness in the MLDM were empirically modified so that a simple liquid-drop-model would give fission barriers close to those obtained with the FRLDM. The usefulness of the MLDM depends on its ability to mimic the FRLDM. For $A$>180 the difference between MLDM and FRLDM fission barriers is less than 0.3 MeV. Typical differences are ~0.1 MeV. These differences increase as the mass is decreased below $A$~180. The present version of the MLDM is only recommended for systems with $A$>160. A retuning of $S'(q)$, $C(q)$, $I_\perp(q)$, and $I_\parallel(q)$ could be performed to obtain a version of the MLDM that is valid at $A$<160.

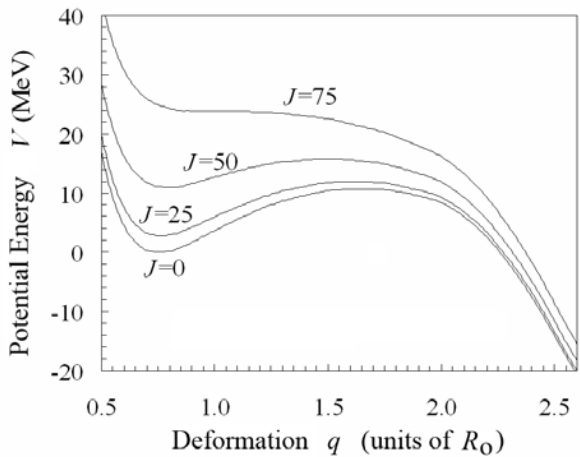

Fig. 13. The MLDM potential energy of $^{210}$Po ($K$=0) as a function of deformation for various total spins $J$.

Fig. 13 shows the MLDM potential energy of $^{210}$Po ($K$=0) as a function of deformation for various total spins $J$. The deformation is expressed as a distance between mass centers in units of the radius of the corresponding spherical system. $q/R_o$=0.75 corresponds to a sphere. The deformation of the ground state position increases and the deformation of the saddle point decreases with increasing spin, as expected. The MLDM $^{210}$Po fission barrier vanishes at $J$=75. The MLDM and FRLDM fission barriers and ground-state energies are compared in Fig. 14 and Fig. 15. The rms difference between the $^{210}$Po MLDM and FRLDM fission barriers is ~0.06 MeV. The corresponding value for the ground state energies is ~0.4 MeV. Model calculations are very sensitive to fission barriers and thus the MLDM was tuned to give an excellent match to the FRLDM fission barriers at the expense of the match to the spin dependence of the ground-state energies. The statistical-model code JOANNE4 (discussed further in section III) only uses the MLDM to determine the deformation dependence of the potential energy. When calculating the excitation energy and temperature at the ground-state position, the FRLDM ground-state energy is estimated using BARFIT.

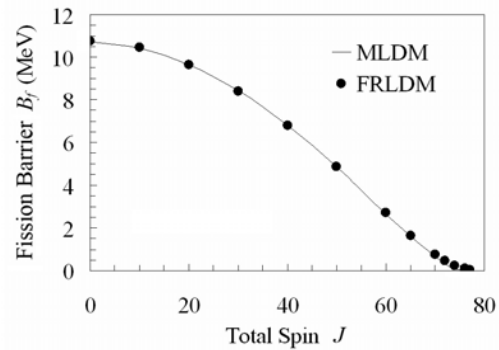

Fig. 14. The MLDM and FRLDM fission barriers of $^{210}$Po ($K$=0) as a function of the total spin $J$.

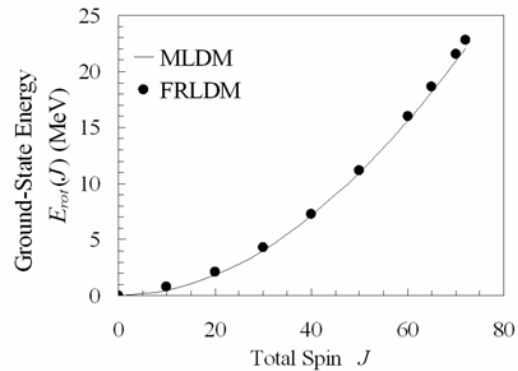

Fig. 15. The MLDM and FRLDM ground-state energies of $^{210}$Po ($K$=0) as a function of the total spin $J$.

When the deformation coordinate is the distance between mass centers, the inertia as a function of deformation can be estimated assuming irrotational and incompressible flow using [49]

$$\mu \approx \frac{M}{4}\left\{1+\frac{17}{15}\exp\left(\frac{-128}{51}(\frac{q}{R_o}-\frac{3}{4})\right)\right\} \; . \qquad (41)$$

A method by which this expression of the inertia, and MLDM potential energy surfaces can be used to estimate the angular frequencies at ground state and saddle points, $\omega_{gs}$ and $\omega_{sp}$, is outlined in ref [46]. Fig. 16 shows estimates of $\omega_{gs}$ and $\omega_{sp}$ for $^{210}$Po as a function of spin $J$ (assuming $K$=0).

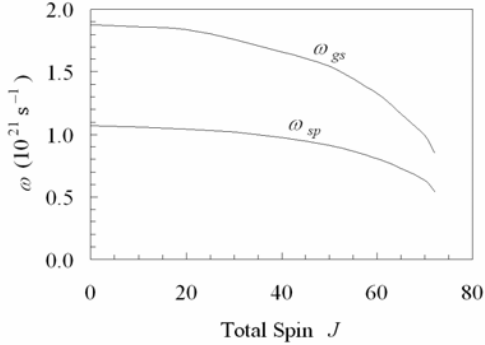

Fig. 16. MLDM estimates of ground-state and saddle-point curvatures $\omega_{gs}$ and $\omega_{sp}$, for $^{210}$Po as a function of spin $J$ (assuming $K$=0).

## F. FREE ENERGY AND EFFECTIVE POTENTIALS

The Bohr-Wheeler fission decay width given by Eq. (10) was obtained assuming that the Fermi-gas level-density parameter $a$ is independent of the nuclear shape. However, for real nuclei, the level-density parameter is expected to have a dependence on nuclear shape. Using the Thomas-Fermi Approximation (TFA) [31] or the Local Density Approximation (LDA) [32,33], it is relatively easy to show that the level-density parameter of a sharp-surfaced nucleus is only dependent on the nuclear volume and is $a\sim$A/15 MeV$^{-1}$ and independent of nuclear shape. If the assumption of a sharp surface is replaced by a realistic diffuse surface, then the level-density parameter will be $\sim$A/9 MeV$^{-1}$ for spherical systems, and increase with increasing deformation. The volume and shape dependence of the level-density parameter can be estimated using the TFA, the LDA, and/or by quantum-mechanical calculations [34]. These results can be approximated by the expression [31,35,36]

$$a(q)\sim c_V A + c_S A^{2/3} S'(q) \; , \qquad (42)$$

where $c_V$ and $c_S$ are constants that control the volume and shape dependence of the level-density parameter, and S'$(q)$ is the surface energy relative to that of a spherical system with the same volume. The values of the constants $c_V$ and $c_S$ depend sensitively on the nuclear radius, the effective mass of nucleons in nuclear matter, and on the properties of the nuclear surface [33].

When taking into account a possible deformation dependence of the level density, most existing statistical-model codes assume the location of the fission transition point is independent of excitation energy and given by the saddle point in the $T$=0 potential-energy surface. Using this approximation, Eqs. (9) and (10) can be rewritten as

$$\Gamma_f^{BW} = \frac{T_{sp}\exp\left(2\sqrt{a_{sp}(E-V_{gs}-B_f)}\right)}{2\pi\exp\left(2\sqrt{a_{eq}(E-V_{gs})}\right)}$$

$$\sim \frac{T}{2\pi}\exp(-B_{eff}\,/\,T) \; , \qquad (43)$$

where the effective barrier height is given by

$$B_{eff}=B_f-\Delta a\,T^2 \; , \qquad (44)$$

where $\Delta a$ is equal to $(a_{sp}-a_{gs})$. If $a_{sp}$ is larger than $a_{gs}$ then, at a high enough excitation energy, one obtains the unphysical result where the level density at the transition point is larger than the level density at the ground-state position. For example, if we assume $a_{gs}$=23 MeV$^{-1}$, $a_{sp}/a_{gs}$=1.04, and $B_f$=3 MeV, then the level density at the saddle point, as given in Eq. (43), becomes larger than the level density at the ground-state position at an excitation energy of $\sim$80 MeV. At higher excitation energies, the effective barrier is negative. This unphysical result alerts us that Eq. (43) becomes invalid at high excitation energy.

The reason that Eq. (43) becomes invalid at high excitation energy (separate from the issues discussed in sections II.B and II.C) is because, at finite temperature, the generalization of the potential-energy function that determines the driving force is the free energy [26, pg 371]

$$F = E_{tot} - T\,S(q,E), \qquad (45)$$

where $S$ is the entropy. If the level-density parameter is a function of nuclear deformation, then the locations of equilibrium points will be a function of excitation energy and defined by the equilibrium points in the entropy (or level density) as a function of deformation

$$\left(\frac{\partial S(q)}{\partial q}\right)_E \sim \left(\frac{\partial (2\sqrt{a(q)U(q)})}{\partial q}\right)_E = 0 \; , \qquad (46)$$

and not by the equilibrium points in the potential energy $V(q)$. It is easy to show that searching for equilibrium points in the entropy is the same as searching for the equilibrium points in an effective temperature-dependent potential energy defined by [23]

$$V_{eff}(q,T)=V(q)-\Delta a(q)T^2 \; . \qquad (47)$$

Only the derivative of the effective potential energy is of any importance, and thus a constant shift can be applied to the effective potential without any change to model calculations. Given this, we choose to define $\Delta a(q)$ to be the difference between $a(q)$ and the corresponding value for the spherical system. The temperature dependence of both equilibrium points can be determined by finding the minima and maxima in the effective potential.

If the deformation dependence of the level-density parameter and the corresponding excitation-energy dependences of the ground state and fission transition point

(tp) are taken into account then the Bohr-Wheeler decay width can be expressed

$$\Gamma_f^{BW} = \frac{T_{tp} \exp\left(2\sqrt{a_{tp}(E - V_{tp})}\right)}{2\pi \exp\left(2\sqrt{a_{gs}(E - V_{gs})}\right)} . \quad (48)$$

In Eq. (48), $V_{gs}$ and $V_{tp}$ are the real potential energies at the location of the ground-state and fission transition points determined by the equilibrium positions in the effective potential. Eq. (48) can be rewritten in terms of the effective potential

$$\Gamma_f^{BW} = \frac{T \exp\left(2\sqrt{a_o(E - V_{gs}(T) - B_f(T))}\right)}{2\pi \exp\left(2\sqrt{a_o(E - V_{gs}(T))}\right)}$$

$$\sim \frac{T}{2\pi} \exp(-B_f(T)/T) , \quad (49)$$

where $V_{gs}(T)$ and $B_f(T)$ are the effective potential energy of the ground-state position and the effective barrier height determined using the effective potential. Notice that the decay width can be determined using the real potential with the real deformation dependence of the level-density parameter, or the effective potential with the level-density parameter of the spherical system. However, one must never use the effective potential with the real deformation dependence of the level-density parameter.

If the effects of the collective motion about the ground-state position and the finite width of the fission barrier are taken into account as discussed in the previous sections, then the Kramers-modified statistical-model result for a one-dimensional fission model (with $K=0$) with a deformation dependence of the level-density parameter can be written as

$$\Gamma_f = \left(\sqrt{1 + \gamma^2(T)} - \gamma(T)\right) \times \frac{\hbar \omega_{gs}(T)}{2\pi} \exp(-B_f(T)/T) , \quad (50)$$

where $\gamma(T) = \beta/(2 \omega_{tp})$, $\omega_{gs}(T)$ and $B_f(T)$ are all functions of temperature and determined using the effective potential $V_{eff}(q,T)$ given by Eq. (47). Eq. (50) assumes that the excitation energy is high enough that the temperature is independent of the deformation. This is a reasonable approximation if the effective barrier height is small compared to the thermal excitation energy at the ground-state position. In the limit of high excitation energy, the temperature in Eq. (50) can be assumed to be independent of deformation and equal to the value at the ground-state position. At low excitation energy the temperature dependence of the effective potential is small and thus it is also reasonable to determine $\omega_{tp}(T)$, $\omega_{gs}(T)$ and $B_f(T)$ assuming a deformation-independent temperature set to the value at the ground-state position. However, to obtain an accurate estimate of the excitation-energy dependence of the fission lifetime at low excitation energy, the thermal excitation-energy dependence of the temperature must be taken into account when calculating the ratio of the level densities at the ground-state and transition point. Given these considerations, we rewrite Eq. (50) as

$$\Gamma_f = \left(\sqrt{1 + \gamma^2(T_{gs})} - \gamma(T_{gs})\right) \times \frac{\hbar \omega_{gs}(T_{gs})}{2\pi} \exp(-\frac{2B_f(T_{gs})}{T_{gs} + T_{tp}}) . \quad (51)$$

The temperature at the transition point will always be less than the temperature at the ground-state position. Therefore, by assuming that the temperature is independent of deformation and equal to the value at the ground-state position, the temperature dependence of $\omega_{tp}(T)$ and $B_f(T)$ will be overestimated by a small amount. However, this can be compensated for by decreasing the temperature dependence of $V_{eff}(q,T)$ via a small decrease in the magnitude of the deformation-dependence of the level-density parameter.

If the shape dependence of the level-density parameter is assumed to be as given in Eq. (42) then the effective potential energy is given by

$$V_{eff}(q,Z,A,J,K,T) = V(q,Z,A,J,K) - c_S A^{2/3}(S'(q)-1)T^2. \quad (52)$$

Substituting in the MLDM potential energy (see Eq. (40)) gives

$$V_{eff}(q,Z,A,J,K,T) =$$

$$(S'(q)-1)E_S^0(Z,A)(1-\alpha T^2) + (C(q)-1)0.7053\frac{Z^2}{A^{1/3}} \text{ MeV}$$

$$+ \frac{(J(J+1)-K^2)\hbar^2}{I_\perp(q)\frac{4}{5}MR_o^2 + 8Ma^2} + \frac{K^2\hbar^2}{I_\parallel(q)\frac{4}{5}MR_o^2 + 8Ma^2} , \quad (53)$$

where

$$\alpha = \frac{c_S A^{2/3}}{E_S^0} = c_S \times 0.059 \text{ MeV}^{-1} \text{ for } A \sim 200. \quad (54)$$

For a particular model, Töke and Swiatecki [31] obtained $c_S \sim 0.27$ MeV$^{-1}$. This gives an estimate for the value of $\alpha$ in Eq. (53) of $\sim 0.016$ MeV$^{-2}$. However, $c_S$ is known to be very sensitive to the assumed properties of nuclear matter and to the different types of approximations used to estimate it. Other estimates of $c_S$ [32-36] give values of $\alpha$ that range from 0.007 to 0.022 MeV$^{-2}$. For the remainder of section II we shall assume $\alpha=0.016$ MeV$^{-2}$. For systems with $A \sim 200$, the deformation dependence of the level density associated with $\alpha=0.016$ MeV$^{-2}$ corresponds roughly to $a_{sp}/a_{gs}$ (or $a_f/a_n$) $\sim 1.05$. In section III, $\alpha$ will be adjusted to reproduce experimental data.

It is of interest to note that the deformation dependence of the level-density parameter can be mapped into a temperature dependence of the surface energy. The TFA can be used to calculate the temperature dependence of the LDM surface energy. For example, Campi and Stringari [37] used the TFA and obtained $\alpha \sim 0.012$ MeV$^{-2}$. It is important to realize that the deformation dependence of the level-density parameter and the temperature dependence of the surface energy are different ways of representing the same physics associated with the diffuse nuclear surface. One should never use the deformation dependence of the level-density parameter in conjunction with a temperature-dependent surface energy, as this would be counting the same physical effect twice.

Fig. 17 shows the MLDM potential energy $V(q)$ as a function of deformation for $^{210}$Po with $J$=50 and $K$=0, along with the corresponding effective potential energies $V_{eff}(q,T)$ at $T$=1 and 2 MeV assuming $\alpha$=0.016 MeV$^{-2}$, and the deformation dependence of the corresponding entropies $S(q,E)$. The thermal excitation-energy dependence of the level density is assumed here to be of the form

$$\rho(U) \propto \frac{\exp(2\sqrt{aU})}{U^n} \ , \tag{55}$$

with $n$=2. This is the excitation-energy dependence of the level density assumed by many statistical-model codes [13,16,19,21], and is based on the theoretical result for a spherical symmetric system [40]. The corresponding relationship between thermal excitation energy and temperature is

$$T = \frac{U}{\sqrt{aU - n}} \ . \tag{56}$$

This approaches $(U/a)^{1/2}$ at high excitation energy. Assuming a static axially symmetric shape changes $n$ to 3/2, and a static shape with no rotational symmetries changes $n$ to 5/4 [26]. The inclusion of collective motion could further reduce $n$. However, in the remainder of the present work we shall assume $n$=2. One's choice for $n$ in the range from 0 to 2, makes little difference to the overall properties of hot systems with thermal excitation energies larger than a few tens of MeV.

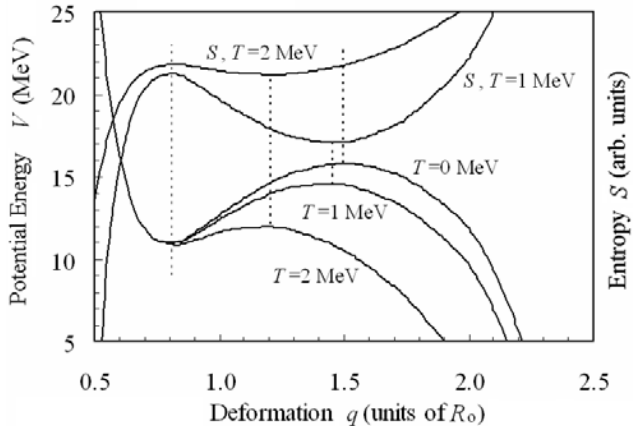

Fig. 17. The MLDM potential energy $V(q)$ as a function of deformation for $^{210}$Po with $J$=50, $K$=0, along with the corresponding effective potential energies $V_{eff}(q,T)$ at $T$=1 and 2 MeV assuming $\alpha$=0.016 MeV$^{-2}$. Also shown is the deformation dependence of the corresponding entropies $S(q,E)$. The dashed vertical lines are to guide the eye (see text).

From Fig. 17 we can see that the location of the transition point does not change much up to a temperature of ~1 MeV. However, there is a dramatic change in the location of the fission transition point from $T$=1 MeV ($U$~30 MeV) to $T$=2 MeV ($U$~100 MeV). The dashed vertical lines are to guide the eye and show that the equilibrium positions in the effective potential correspond to equilibrium positions in the entropy. From Fig. 17 we also deduce that if the transition point is incorrectly assumed to equal the $T$=0 value (independent of temperature) then the entropy of the transition point will be increasingly overestimated with increasing temperature. This would cause the mean fission lifetime to be increasingly underestimated with increasing temperature. To further illustrate this, Fig. 18 compares the effective fission barrier height for $^{210}$Po with $J$=50, $K$=0, and $\alpha$=0.016 MeV$^{-2}$ obtained by incorrectly assuming the transition point is independent of temperature via Eq. (44), and those obtained using the equilibrium points in the effective potential $V_{eff}(q,T)$. There is little difference between these two methods below $T$~1 MeV. Above $T$~1 MeV the incorrect approach increasingly underestimates the height of the effective fission barrier.

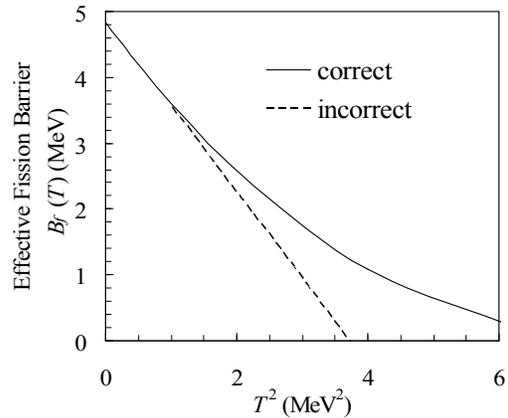

Fig. 18. The effective fission barrier height for $^{210}$Po with $J$=50, $K$=0, and $\alpha$=0.016 MeV$^{-2}$ obtained by incorrectly assuming the transition point is independent of temperature via Eq. (44) (dashed line), and those obtained using the turning points in the effective potential $V_{eff}(q,T)$ (solid curve).

To confirm that Eq. (51) adequately describes the fission decay width for systems with MLDM potential-energy surfaces with a deformation dependence of the level-density parameter, we calculate mean fission times by numerical means using the Langevin equation [41]. In obtaining Eq. (16) it was assumed that the Fermi-gas level-density parameter is a constant, independent of the nuclear shape. However, for real nuclei, the level-density parameter is expected to have a dependence on nuclear shape (as discussed above), and the driving force on the collective degree of freedom should be determined using the derivative of the free energy [26, pg 371] and Eq. (16) should be modified to [23]

$$\ddot{q} = \frac{-1}{\mu}\left(\frac{\partial V_{eff}(q,T)}{\partial q}\right)_T - \frac{\dot{q}^2}{2\mu}\frac{\partial \mu}{\partial q} - \beta \dot{q} + \Gamma \sqrt{\frac{2\beta T}{\delta t \, \mu}} \ . \tag{57}$$

As discussed above, it is a reasonably good approximation to estimate the effective potential as a function of deformation, using the temperature at the ground-state position independent of deformation. However, to allow for total energy conservation, the temperature in the last term of Eq. (57) must be calculated taking into account the thermal energy converted into collective energy. By including this effect, if the total collective energy becomes large compared to the total available energy then the temperature becomes low and the random acceleration is reduced. If this were not done then the random acceleration governed by the last term in Eq. (57) would violate the conservation of energy and could drive the total collective energy of the system (while still in the ground-state well) to a value larger than the available excitation energy at the ground-state position.

Before proceeding with calculations of fission lifetimes using realistic nuclear potential energies, it is important to introduce a realistic model to guide the expected values of the nuclear dissipation coefficient $\beta$. We believe the nuclear dissipation has been well constrained by the surface-plus-window dissipation model [28,29], using the mean kinetic energy of fission fragments and the widths of isoscalar giant resonances. The surface-plus-window dissipation model contains a single dimensionless parameter $k_S$ which controls the way nucleons interact with the nuclear surface. A value of $k_S=1$ corresponds to wall [50,51] plus window dissipation. The surface-plus-window dissipation model with a value of $k_S=0.27$, reproduces the mean kinetic energy of fission fragments and the widths of isoscalar giant resonances over a wide range of nuclear masses [28,29]. The deformation dependence of the surface-plus-window dissipation coefficient with $k_S=0.27$, for a $J=50$ $^{195}$Pb system [52] is shown in Fig. 19.

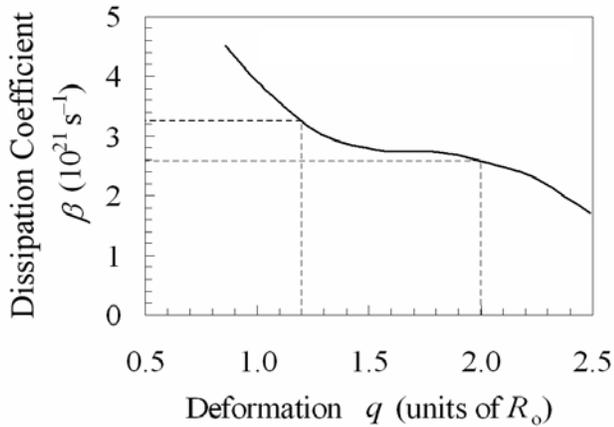

Fig. 19. The deformation dependence of the surface-plus-window dissipation coefficient with $k_S=0.27$, for a $J=50$ $^{195}$Pb system [52]. The dashed lines guide the eye (see text).

The surface-plus-window model dissipation coefficient is very insensitive to $Z$, $A$, and $J$, has no dependence on nuclear temperature, and is relatively flat over a wide range of saddle-point deformations. The dashed vertical lines in

Fig. 19 span the range of typical fission saddle-point deformations encountered in heavy-ion fusion-fission reactions with compound nuclei mass numbers from $A \sim 170$ to 220. The horizontal dashed lines show that over this range of fission saddle-point deformations, the dissipation coefficient is within 10% of $3 \times 10^{21}$ s$^{-1}$. Recently, theoretical studies of the kinetic energy of fission fragments [53] have confirmed the work of Nix and Sierk [28,29]. For the remainder of this paper we shall assume that the nuclear dissipation coefficient in the region of all fission transition points is $\beta=3 \times 10^{21}$ s$^{-1}$ and is independent of temperature.

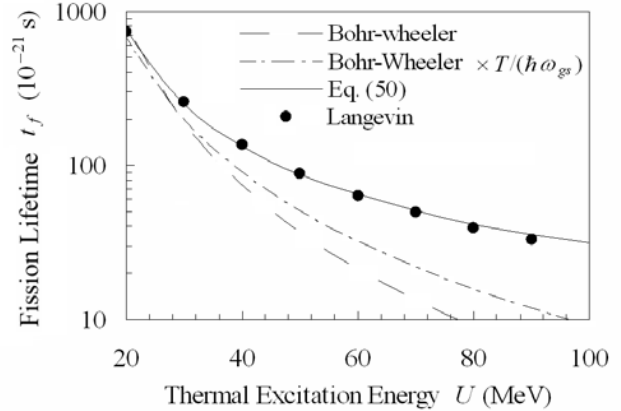

Fig. 20. Estimates of the fission lifetime of $J=50$ and $K=0$ $^{210}$Po systems as a function of the thermal excitation energy, assuming $\beta=3 \times 10^{21}$ s$^{-1}$, $\alpha=$A/8.6 MeV$^{-1}$, and $\alpha=0.016$ MeV$^{-1}$. The dashed curve shows the results for the "standard" Kramers-modified Bohr-Wheeler fission decay width. The dashed-dotted curve is the corresponding lifetime multiplied by $T/\hbar\omega_{gs}$. The solid curve shows the corresponding result where the deformation dependence of the level-density parameter is taken into account via Eq. (51). The symbols show corresponding Langevin calculations.

Fig. 20 compares several estimates of the mean time spent inside the fission transition point of a $^{210}$Po system with $J=50$ and $K=0$, as a function of the thermal excitation energy, assuming $\beta=3 \times 10^{21}$ s$^{-1}$, and a level-density parameter as a function of shape as estimated by Töke and Swiatecki [31]. As discussed above, the results of ref [31] correspond to $\alpha=0.016$ MeV$^{-2}$. The dashed curve in Fig. 20 shows the results of a Kramers-modified "standard" Bohr-Wheeler fission decay width. This is the standard method used in many statistical-model codes. The properties of a MLDM $^{210}$Po system with $J=50$, $K=0$, and $\alpha=0.016$ MeV$^{-2}$ include a fission barrier height $B_f(T=0)=4.84$ MeV, $a_{gs}=24.44$ MeV$^{-1}$, $a_{sp}=25.74$ MeV$^{-1}$ ($a_{sp}/a_{gs}=1.053$), and $\omega_{sp}(T=0)=0.915 \times 10^{21}$ s$^{-1}$. The dash-dotted curve in Fig. 20 shows the mean fission time if the Kramers-modified standard Bohr-Wheeler decay width is further modified by the $\hbar\omega_{gs}/T$ factor to account for the collective motion about the ground-state position. The solid curve shows the corresponding mean fission times determined by the Kramers-modified statistical model where the deformation dependence of the level-density parameter is taken into account in a more accurate way via Eq. (51). These mean fission times are in good agreement with Langevin calculations shown by the circles. The Langevin calculations presented here assume that all compound systems start at the bottom of the ground-state well at $t$=0 and thus include a transient delay in the build up of the fission decay width as a function of time. The good agreement between the dynamical and statistical-model fission lifetimes confirms that the transient delay has little effect for the excitation-energy range and reaction class considered here. The standard Kramers-modified Bohr-Wheeler decay width increasingly underestimates the fission lifetimes with increasing excitation energy relative to more correct model calculations obtained via both statistical and dynamical means (see Eqs. (51) and (57)).

## G. K STATES
## ORIENTATION DEGREE OF FREEDOM

The MLDM uses a family of axially symmetric and mass symmetric shapes. These shapes define the Coulomb, surface, and rotational energies of nuclei as a function of a single deformation (elongation) parameter $q/R_o$. Within the frame work of this simple model where the nuclear shape is defined by a single parameter, the motion of a rotating system must be defined by a minimum of two degrees of freedom. These are the shape and the orientation of the shape relative to the total spin. The statistical model of the fission of rotating systems must determine the total level density and the number of fission transition states, taking into account the phase space associated with both the shape and orientation degrees of freedom. JOANNE4 [25] is presently the only statistical-model code that takes the orientation degree of freedom into account when estimating the fission lifetimes of hot rotating systems.

The level density of a spherically symmetric system as a function of the total excitation energy $E$, the total spin $J$, and the spin about an axis rotating with the sphere $K$, is [40]

$$\rho_{sph}(E,J,K) = \frac{1}{12}\left(\frac{\hbar^2}{2I}\right)^{3/2}\sqrt{a}\,\frac{\exp\left(2\sqrt{aU}\right)}{U^2}, \quad (58)$$

where $I$ is the rigid body moment of inertia, and the thermal excitation energy is

$$U = E - \frac{J(J+1)\hbar^2}{2I}. \quad (59)$$

For the spherically symmetric case the rotational energy is obviously independent of $K$, and the level density as a function of $E$ and $J$ is the well known result [40]

$$\rho_{sph}(E,J) = \sum_{K=-J}^{J} \rho_{sph}(E,J,K)$$

$$= \frac{2J+1}{12}\left(\frac{\hbar^2}{2I}\right)^{3/2}\sqrt{a}\,\frac{\exp\left(2\sqrt{aU}\right)}{U^2}. \quad (60)$$

The $2J$+1 factor in Eq. (60) is associated with the complete freedom of the orientation degree of freedom in the case of a spherical system.

The level density of an axially symmetric system as a function of $E$, $J$, the spin about the symmetry axis $K$, and the deformation $q$, is [26]

$$\rho_{ax}(E,J,K,q) = \frac{1}{12}\left(\frac{\hbar^2}{8I_\parallel(q)}\right)^{1/2}\frac{\exp\left(2\sqrt{aU}\right)}{U^{3/2}}, \quad (61)$$

where $I_\parallel(q)$ is the rigid body moment of inertia about the symmetry axis, and the thermal excitation energy is

$$U = E - \frac{J(J+1)\hbar^2}{2I_\perp(q)} - \frac{K^2\hbar^2}{2I_{eff}(q)}. \quad (62)$$

The effective moment of inertia is

$$I_{eff}(q) = \left(\frac{1}{I_\parallel(q)} - \frac{1}{I_\perp(q)}\right)^{-1}. \quad (63)$$

In the limit of a small perturbation from the spherical shape, the effective moment of inertia is large and the rotational energy becomes independent of the orientation of the symmetry axis relative to the total spin. In this case, the level density without reference to the orientation degree of freedom is simply Eq. (61) multiplied by $2J$+1. For an arbitrary deformation, this multiplication factor associated with the orientation degree of freedom is

$$f = \sum_{K=-J}^{J} \exp\left(\frac{-K^2}{2K_o^2(q)}\right) \sim K_o\sqrt{2\pi}\ \text{erf}\left(\frac{2J+1}{2\sqrt{2}\ K_o(q)}\right), \quad (64)$$

where $K_o^2(q)=T\cdot I_{eff}(q)/\hbar^2$. The factor $f$ decreases with increasing deformation because the symmetry axis of spinning systems becomes increasingly confined to the plane perpendicular to the spin total as the deformation is increased. This decrease in $f$ with increasing deformation must be taken into account when calculating fission lifetimes in heavy-ion fusion-fission reactions.

The level-density enhancement associated with a change in the shape symmetry from spherical to axially symmetric is

$$\frac{\rho_{ax}}{\rho_{sph}} \sim \sqrt{\frac{U}{a}}\,\frac{(2I)^{3/2}}{\sqrt{8I_\parallel}\,\hbar^2} \sim \frac{T\,I}{\hbar^2} \sim \frac{T\,A^{5/3}}{70\,\text{MeV}}. \quad (65)$$

For $A\sim200$ and $T\sim1$ MeV this level-density enhancement is $\sim100$. A consequence of this enhancement is that hot nuclei will be deformed, because the driving force on the collective degrees of freedom is determined by the free energy. Even though the potential energy may be increased by moving to a modest deformation, the free energy will be decreased by the factor of $\sim100$ enhancement in the level density of the system.

The level density of a triaxial system with no rotational symmetries, as a function of $E$, $J$, $\tau$, and $q$, is [26]

$$\rho_{tri}(E,J,\tau,q) = \frac{\sqrt{\pi}}{12} \frac{\exp\left(2\sqrt{aU}\right)}{U^{5/4} a^{1/4}}, \qquad (66)$$

where the thermal excitation energy is

$$U = E - E_{rot}(J,\tau,q). \qquad (67)$$

$\tau = 1$ to $2J+1$ labels the different rotational levels with the same value of $J$ in a given rotational band. The level-density enhancement associated with a change in the shape symmetry from axially symmetric to no rotational symmetry is

$$\frac{\rho_{tri}}{\rho_{ax}} \sim \sqrt{\frac{8\pi\, T\, I_{\parallel}(q)}{\hbar^2}} \sim 0.6\sqrt{T}\, A^{5/3}. \qquad (68)$$

For $A \sim 200$ and $T \sim 1$ MeV this level-density enhancement is $\sim 50$. The consequence of this enhancement is that hot nuclei will be triaxial because the small loss in thermal excitation energy to produce a triaxial deformation will be more than compensated by the factor of $\sim 50$ increase in the level density relative to that of an axially symmetric system. The size of the triaxiality and its dependence on temperature and elongation is an open question. For simplicity, we assume that the size of the triaxiality needed to turn on all rotational degrees of freedom is small, and that the $\tau = 1$ to $2J+1$ map to $K = -J$ to $J$ with rotational energies

$$E_{rot} = \frac{J(J+1)\hbar^2}{2I_{\perp}(q)} + \frac{K^2\hbar^2}{2I_{eff}(q)} + \delta(J,\tau,q). \qquad (69)$$

We assume that the corrections to the rotational energies $\delta$ associated with the triaxiality are small. Based on the considerations discussed above, the level density of hot nuclei as a function of $E$, $J$, and $q$ is

$$\rho(E,J,q) \sim \sum_{K=-J}^{J} \frac{\sqrt{\pi}}{12} \frac{\exp\left(2\sqrt{aU}\right)}{U^{5/4} a^{1/4}}. \qquad (70)$$

It is of interest to note that with all the rotational degrees of freedom turned on, the influence of the moments of inertia on the level density given in Eq. (66) enters only through the thermal excitation energy term $U$. This leads to the effective potential having the functional form given in Eq. (47). If the level density for an axially symmetric system (see Eq. (61)) is used, then the effective potential would have an additional term associated with the derivative of $\ln(I_{\parallel}(q))$. Eq. (70) suggests that statistical-model codes should assume a level density of the form given in Eq. (55) with $n = 5/4$. The value for $n$ may be reduced even further if the level density is calculated taking into account collective motion perpendicular to the fission axis. However, as discussed earlier, many codes still assume $n = 2$. For historical coding issues, this is also the case for the code JOANNE4 used in the present study. This should be rectified in a future version of JOANNE4. However, changing $n$ from 2 to 5/4 has only a very small effect on the conclusions drawn here.

Including the orientation degree of freedom, the statistical-model fission decay width for a rotating system can be obtained using Eq. (3) with the number of transition states and the total level density given by [22,26]

$$N_{TS} = \sum_{K} \int \rho_{tp}(E - V_{tp}(K,T) - \varepsilon)\,d\varepsilon, \text{ and} \qquad (71)$$

$$\rho = \sum_{K} \iint \rho_{gs}(E - V_{gs}(K,T)$$
$$- \frac{\mu_{gs}(K,T)\,\omega_{gs}^2(K,T)(q-q_{gs})^2}{2T} - \frac{p^2}{2\mu_{gs}(K,T)T}\,)\frac{dq\,dp}{h} \qquad (72)$$

These expressions give

$$\Gamma_f = \frac{1}{2\pi} \frac{\displaystyle\sum_{K} T_{tp}(K)\rho_{tp}(E - V_{tp}(K,T))}{\displaystyle\sum_{K} \frac{T_{gs}(K)}{\hbar\omega_{gs}(K,T)}\rho_{gs}(E - V_{gs}(K,T))}$$

$$= \sum_{K} \frac{T_{gs}(K)}{\hbar\omega_{gs}(K,T)}\rho_{gs}(E - V_{gs}(K,T))$$

$$\times \frac{\hbar\omega_{gs}(K,T)}{2\pi} \frac{T_{tp}(K)}{T_{gs}(K)} \frac{\rho_{tp}(E - V_{tp}(K,T))}{\rho_{gs}(E - V_{gs}(K,T))}$$

$$\div \sum_{K} \frac{T_{gs}(K)}{\hbar\omega_{gs}(K,T)}\rho_{gs}(E - V_{gs}(K,T))$$

$$= \frac{\displaystyle\sum_{K} P(K)\Gamma_f(K)}{\displaystyle\sum_{K} P(K)}, \qquad (73)$$

where $P(K)$ is the probability that the system is in a given $K$ state,

$$P(K) = \frac{T_{gs}(K)}{\hbar\omega_{gs}(K,T)}\rho_{gs}(E - V_{gs}(K,T)). \qquad (74)$$

$\Gamma_f(K)$ is the fission decay width if the system could be restricted to a given $K$ state. To correct for the finite barrier width, the fission decay width as a function of $K$ should be determined using Eq. (51) but with $\omega_{tp}$, $\omega_{gs}$ and $B_f$ obtained using the effective potential as a function of both $T$ and $K$.

As done in the previous sections, we wish to confirm the validity of Kramers-modified statistical model by comparing results obtained using Eq. (73) to Langevin calculations. To perform Langevin calculations of a rotating system, we must have a model of the microscopic coupling between the orientation degree of freedom ($K$-states) and the thermal degrees of freedom. Langevin calculations performed by others do not include a coupling between the orientation degree of freedom and the heat bath, and therefore, do not allow the $K$ states to equilibrate. The Langevin calculations of others underestimate the fission lifetime because only the $K=0$ fission barrier is sampled, instead of an equilibrated distribution containing higher $K\neq 0$ barriers.

The details of the coupling between the orientation degree of freedom and the heat bath remain an open question, especially for systems moving about in a ground-state well. From the success of the transition state model of fission fragment angular distributions [54] for most fusion-fission reactions, it is known that the time spent inside typical fission transition points is generally much longer than the $K$-state equilibration time, while saddle-to-scission transit times are much shorter than the $K$-state equilibration time for systems beyond the fission transition point. This is the same as saying that, for typical fission reactions, the $K$-states are fully equilibrated inside the fission transition point, while $K$ is almost a constant of the motion for highly deformed systems beyond typical fission transition points.

The dynamical evolution of the symmetry axis of a system consisting of two nuclei connected by a neck (a dinucleus) has been studied by Døssing and Randrup [55]. Using expressions obtained by them, and Eq. (A.17) contained within ref [56], one can show that if a dinucleus is initially in the $K$=0 state, the variance of $K$ a short time later $\delta t$ can be expressed as

$$\sigma_K^2 \sim \frac{J^2 T \delta t}{2\pi^3 n_o C^2 q^2} \left( \frac{I_\parallel I_{eff} I_R}{I_\perp^3} \right), \quad (75)$$

where $I_R = Aq^2/4$ (assuming a mass symmetric system), $q$ is the distance between the centers of mass of the two nuclei that make up the dinucleus, $n_o$ is the bulk flux in standard nuclear matter (0.263 MeV $\cdot$ $10^{-22}$ s $\cdot$ fm$^{-4}$) [56], and $C$ is the neck radius. Based on the $J$, $T$, and $\delta t$ dependence of the variance of $K$ given in Eq. (75), we choose to treat $K$ as a thermodynamically fluctuating overdamped coordinate and express the changes in $K$ over a small time interval $\delta t$ as

$$\Delta K = \frac{-\gamma_K^2 J^2}{2} \frac{\partial V(K)}{\partial K} \delta t + \Gamma_K \gamma_K J \sqrt{T \delta t}, \quad (76)$$

where $\gamma_K$ is a parameter which controls the coupling between $K$ and the thermal degrees of freedom. $\Gamma_K$ is a random number from a normal distribution with unit variance. For a fissioning nucleus, the neck radius decreases as the distance between mass centers increases, and the product of the neck radius and the distance between mass centers is within 30% of the corresponding value for a spherical system, $C \cdot q$ (for a sphere) $\sim 1.13$ fm$^2$ $\times A^{2/3}$. Substituting this value of $C \cdot q$ into Eq. (75) and assuming $A \sim 200$ gives

$$\gamma_K \sim 0.02 \left( \text{MeV} \cdot 10^{-21} \text{s} \right)^{-1/2} \sqrt{\frac{I_\parallel \left| I_{eff} \right| 5/8 (q/R_o)^2}{I_\perp^3}}. \quad (77)$$

Here, the moments of inertia, $I_\parallel$, $I_{eff}$, and $I_\perp$, are all in units of the corresponding spherical values.

Fig. 21 shows $\gamma_K$ as estimated by Eq. (77) as a function of deformation. It must be stressed that Eq. (77) was obtained assuming a dinucleus and is only valid for systems with a well defined neck. This corresponds to a deformation larger than $q/R_o \sim 1.5$. The extrapolation to more compact configurations should be viewed with caution, and is only shown to give some guidance on the possible nature of the coupling between the orientation and thermal degrees of freedom. By changing some of the assumptions used to obtain Eq. (77), the cusp about the spherical shape can be made larger or removed without changing the results at large deformation. However, Eq. (77) does give the desired result that $K$ is almost a constant of the motion for highly deformed systems, while the cusp about the spherical shape will cause hot systems oscillating about a ground-state position to quickly equilibrate the orientation degree of freedom. It is likely that a more detailed and accurate model for the motion in $K$ will have a coupling term $\gamma_K$ that depends on deformation, the rate of change of the deformation, and the nuclear orientation. The equilibration of the $K$ degree of freedom for systems oscillating about a ground-state position is likely to be further complicated because hot systems will avoid the spherical shape because of the level-density enhancements discussed earlier in this section.

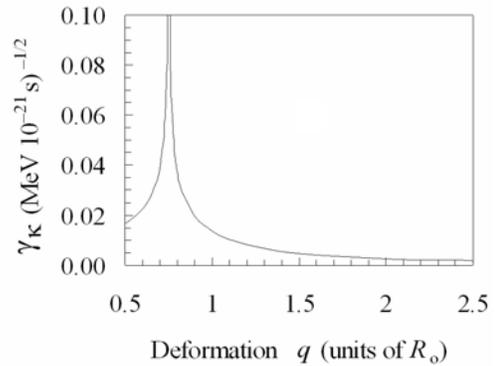

Fig. 21. The orientation-thermal coupling term $\gamma_K$ as estimated by Eq. (77) as a function of deformation.

The angular distribution of fission fragments in near- and sub-barrier heavy-ion fusion-fission reactions involving deformed actinide targets suggests an effective deformation-independent coupling between $K$ and the nuclear heat bath, inside fission transition points with $\gamma_K \sim 0.077$ (MeV $10^{-21}$ s)$^{-1/2}$ [57]. It is possible that this estimate for an effective $\gamma_K$ is incorrect by a factor of 2 or more because the fission model used to extract it was very simplistic and does not include several of the concepts discussed in the present work. In this section, we present two dimensional (shape and orientation) dynamical calculations where the motion of the $K$ degree of freedom is determined by Eq. (76) (see Fig. 23) with $\gamma_K = 0.077$ (MeV $10^{-21}$ s)$^{-1/2}$ for all deformations $q < R_o$. For deformation beyond $q = R_o$ we assume $\gamma_K = 0$. The compound nuclei are assumed to be formed with a uniform $K$-state distribution. The fission time scales obtained by dynamical means shown in Fig. 23 are much longer than the $K$ equilibrium time inside the fission transition points for all but the highest excitation energies, and thus these fission lifetimes are insensitive to the initial $K$-state distribution and our choice for $\gamma_K$.

In the previous section, we ignored the fact that an increase in the initial excitation energy of compound nuclei formed in heavy-ion fusion reactions is associated with a corresponding increase in the mean spin of the systems. A reasonable estimate of the mean spin associated with a given fusion-fission reaction can be obtained from measured fusion and evaporation cross sections. For example, Fig. 22 shows measured fusion and evaporation cross sections from the reaction $^{18}O + ^{192}Os \rightarrow ^{210}Po$ as a function of the $^{18}O$ beam energy in the laboratory frame [58]. If the fusion spin distribution is assumed to have a triangular form with a sharp cutoff, the maximum spin can be determined using

$$J_{max}^{fus} = \sqrt{\frac{\sigma_{fus} E_{cm} \mu}{651.8 \, (mb \cdot MeV \cdot amu)}} \, , \qquad (78)$$

where $E_{cm}$ is the kinetic energy in the center-of-mass frame and $\mu$ is the reduced mass of the projectile-target system. Assuming the transition from evaporation residues to fission is sharp as a function of the spin, the maximum spin of the evaporation residues and the minimum spin of the fissioning systems can be estimated using Eq. (78) by replacing the fusion cross section with the evaporation residue cross section. The mean spin of the fissioning systems is then given by

$$J_f = \frac{2}{3} \left( \frac{J_{max}^{fus}{}^3 - J_{max}^{ER}{}^3}{J_{max}^{fus}{}^2 - J_{max}^{ER}{}^2} \right) . \qquad (79)$$

Table I contains various properties of the $^{18}O + ^{192}Os \rightarrow ^{210}Po$ reaction, including an estimate of the relationship between the initial excitation energy of the compound systems and the mean spin of the fissioning systems. The initial excitation energies are relative to the $^{210}Po$ LDM $J=0$ ground state.

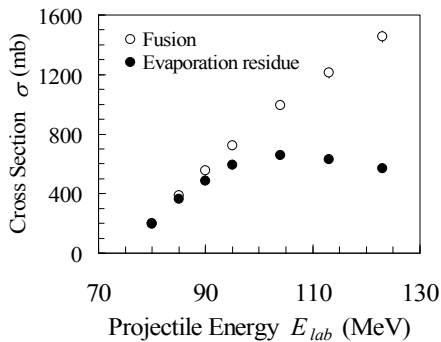

Fig. 22. Measured fusion and evaporation cross sections for the reaction $^{18}O + ^{192}Os \rightarrow ^{210}Po$ as a function of the $^{18}O$ beam energy in the laboratory frame [58].

Fig. 23 shows estimates of the mean fission lifetime of $^{210}Po$ systems formed by the reaction $^{18}O + ^{192}Os$, as a function of the initial excitation energy. The relationship between initial excitation energy and spin is assumed to be as given in Table I. The solid curve shows statistical-model calculations including the $K$-states via Eq. (73), with the fission decay width as a function of $K$ determined using Eq. (51) with $\omega_{bp}$, $\omega_{gs}$ and $B_f$ obtained using the effective potential as a function of both $T$ and $K$ (as performed by JOANNE4). The assumed model parameters are $a=A/8.6$ MeV$^{-1}$, $\beta=3\times10^{21}$ s$^{-1}$, and $\alpha=0.016$ MeV$^{-2}$. These calculations are consistent with the corresponding two-dimensional (shape and orientation) Langevin calculations shown by the solid circles. We assume the same temperature-dependent effective potential $V_{eff}(q,T)$, the same dissipation coefficient, and the same inertia [49] for both our statistical and Langevin calculations. The Langevin calculations are performed using Eqs. (57) and (76) with $\gamma_K=0.077$ (MeV $10^{-21}$ s)$^{-1/2}$ for all deformations $q<R_o$ and $\gamma_K=0$ for $q>R_o$ (as discussed earlier).

Table I. Various properties of the $^{18}O + ^{192}Os \rightarrow ^{210}Po$ reaction, including an estimate of the relationship between the initial excitation energy of the compound systems $E_i$, and the mean spin of the fissioning systems $J_f$. All energies and cross sections are in units of MeV and mb, respectively.

| $E_{lab}$ | $E_{cm}$ | $E_i$ | $\sigma_{fus}$ | $\sigma_{ER}$ | $J_{max}^{fus}$ | $J_{max}^{ER}$ | $J_f$ |
|---|---|---|---|---|---|---|---|
| 80 | 73.1 | 41.7 | 201 | 195 | 19.3 | 19.0 | 19.1 |
| 90 | 82.3 | 50.9 | 553 | 487 | 33.9 | 31.8 | 32.9 |
| 100 | 91.4 | 60.0 | 880 | 640 | 45.1 | 38.4 | 41.8 |
| 110 | 100.6 | 69.1 | 1140 | 640 | 53.8 | 40.3 | 47.4 |
| 120 | 109.7 | 78.3 | 1380 | 580 | 61.8 | 40.1 | 51.7 |
| 130 | 118.9 | 87.4 | 1620 | 525 | 69.7 | 39.7 | 56.1 |

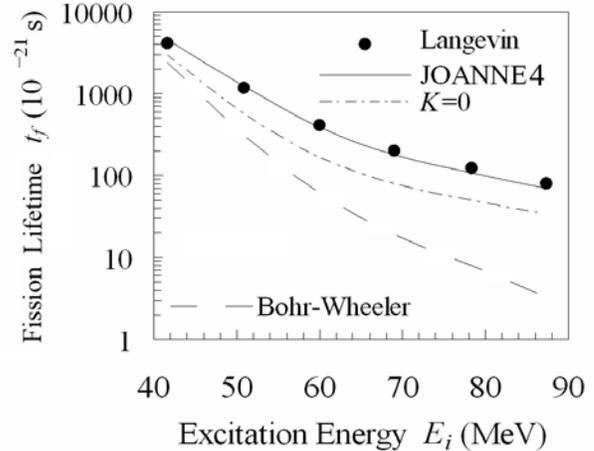

Fig. 23. Estimates of the mean fission lifetime of $^{210}Po$ systems formed by the reaction $^{18}O + ^{192}Os$, as a function of the initial excitation energy $E_i$. The solid curve shows statistical-model calculations obtained using JOANNE4, as discussed in the text. The corresponding two-dimensional Langevin calculations are shown by the solid circles. The dash-dotted curve shows the corresponding calculations if the system is forced to always be in the $K=0$ state. The dashed curve shows results using Eq. (43) with $a_{sp}/a_{eq}=1.04$ and without any Kramer's modification.

Table II contains key properties of the assumed $^{210}Po$ temperature-dependent $K=0$ effective potential energy surfaces as a function of the initial excitation energy. Our calculated fission lifetimes are dependent on the properties of the potential-energy surfaces as a function of $K$. However, tabulating these properties as a function of $K$ would be excessive. To give the reader a feel for the $K$ dependence of the potential energy surface, we show the potential energy surface for $^{210}$Po with $T$=0 and $J$=50 as a function of both deformation and $K$ in Fig. 24. Notice that the potential energy in the ground-state well is relatively flat as a function of $K$. This produces an approximately $2J$+1 multiplication of the system's total level density when the orientation degree of freedom is included. The fission saddle ridge increases in height with increasing $K$. This produces a multiplication in the number of transition states that is less than $2J$+1. This reduction in the number of fission transition states relative to the total level density depends on a combination of the total spin and the deformation of the saddle point. It is well known that the reduction in the number of transition states with increasing $K$ controls the angular distribution of fission fragments [54]. Unfortunately, the corresponding reduction in the number of fission transition states has not been included in standard statistical-model calculations of the mean fission lifetime. The dash-dotted curve in Fig. 23 shows the calculated mean fission times for $^{210}$Po if the system is forced to always be in the $K$=0 state.

Table II. Properties of the $^{210}$Po MLDM ($\alpha$=0.016 MeV$^{-2}$) temperature-dependent $K$=0 effective potential energy surfaces using the relationship between initial excitation energy $E_i$, and mean spin of the fissioning systems $J_f$, as listed in Table I. All energies and temperatures are in units of MeV. The potential curvatures are in units of $10^{21}$ s$^{-1}$.

| $E_i$ | $J_f$ | $U_{gs}$ | $T_{gs}$ | $E_{rot}$ | $B_f$ | $\omega_{gs}$ | $\omega_{sp}$ | $T_{sp}$ |
|---|---|---|---|---|---|---|---|---|
| 41.7 | 19.1 | 40.0 | 1.37 | 1.7 | 7.07 | 1.73 | 0.99 | 1.25 |
| 50.9 | 32.9 | 46.3 | 1.46 | 4.6 | 5.23 | 1.59 | 0.94 | 1.38 |
| 60.0 | 41.8 | 52.6 | 1.55 | 7.4 | 3.57 | 1.47 | 0.89 | 1.50 |
| 69.1 | 47.4 | 59.4 | 1.65 | 9.7 | 2.42 | 1.35 | 0.83 | 1.61 |
| 78.3 | 51.7 | 67.0 | 1.74 | 11.3 | 1.58 | 1.27 | 0.79 | 1.72 |
| 87.4 | 56.1 | 73.9 | 1.83 | 13.5 | 0.72 | 1.22 | 0.70 | 1.82 |

Fig. 25 shows transition state model (solid curve) and 2-dimensional Langevin calculations (solid circles) for the root-mean-squared $K$ for the $^{18}$O + $^{192}$Os→ $^{210}$Po fusion-fission reaction. From Fig. 23 and Fig. 25 we see that when the Kramers-modified statistical model is implemented correctly, the results for both the fission decay width and the angular distribution of the fission fragments are in agreement with two-dimensional Langevin calculations, when the mean fission time is long enough that the systems can fully (or almost fully) equilibrate before passing through a fission transition state.

Many statistical-model codes estimate the mean fission lifetime using the Kramers-modified Bohr-Wheeler fission decay width. Strictly speaking, the Bohr-Wheeler fission decay width is given by Eq. (3). However, it is often associated with expressions similar to Eq. (43), where the total level density and the corresponding number of transition states have been incorrectly determined. Eq. (43) does not include the collective motion about the ground-state well when determining the total level density: it is used in a fashion where the fission transition point is assumed to be independent of temperature; and does not account for the level density associated with the orientation degree of freedom. On top of these approximations, many authors further assume that $a_{sp}/a_{gs}$ is a constant independent of the system spin. For example, Dioszegi et al. [13] assume $a_{sp}/a_{eq}$=1.04 when estimating the nuclear viscosity of hot rotating $^{224}$Th nuclei. The dashed line in Fig. 23 shows estimates of the standard Bohr-Wheeler fission lifetime of $^{210}$Po obtained using Eq. (43) with $a_{sp}/a_{gs}$=1.04 and without any Kramers' modification. These calculations are a factor of two lower compared with the more complete calculations shown by the solid curve and circles at $E_f$~40 MeV, and more than a factor of 20 low at $E_f$~90 MeV.

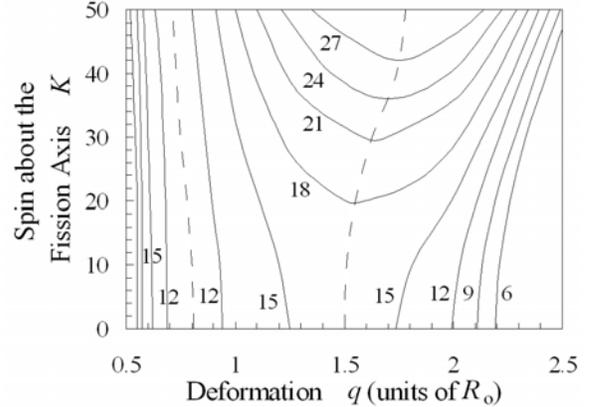

Fig. 24. The potential energy surface for $^{210}$Po with $T$=0 and $J$=50 as a function of both deformation and the spin about the fission axis $K$. The dashed curves show the ground-state valley and the fission saddle ridge. The contour labels are in units of MeV.

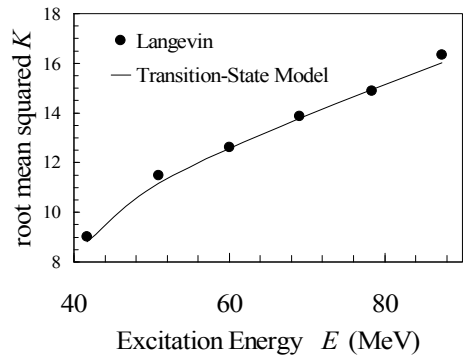

Fig. 25. The transition-state model (solid curve) and 2-dimensional Langevin calculations (solid circles) for the root-mean-squared $K$ for the $^{18}$O + $^{192}$Os→ $^{210}$Po fusion-fission reaction, as a function of the initial excitation energy, $E$.

It is well known that the standard Bohr-Wheeler fission decay width, with $a_{sp}/a_{gs}$ much larger than one, fails to give a satisfactory reproduction of experimental data [12-15]. If the nuclear viscosity is treated as a free parameter as a function of excitation energy then data can be reproduced. As discussed in this paper, the standard Bohr-Wheeler fission decay width does not include several key physical effects and thus nuclear viscosity estimates obtained via a Kramers-modified standard Bohr-Wheeler model should be viewed with caution. It is our view that, when previous authors adjusted the nuclear viscosity to reproduce fusion-fission cross sections and prescission emission data, they were incorrectly compensating for inadequacies in their underlying model of fission lifetimes. The solid line in Fig. 26 shows the nuclear viscosity as a function of excitation energy needed to force the Kramers-modified standard Bohr-Wheeler model with $a_{sp}/a_{gs}=1.04$ to be in agreement with the calculations shown by the solid curve in Fig. 23. This artificial excitation-energy dependence of the nuclear viscosity is similar to the corresponding excitation-energy dependence deduced by Dioszegi *et al.* [13]. This result suggests that the strong excitation-energy dependence of the nuclear viscosity deduced in ref [13] and the rapid onset of the dissipation at nuclear excitation energies above ~40 MeV inferred in ref [12], are artifacts generated by an incomplete model of the fission process.

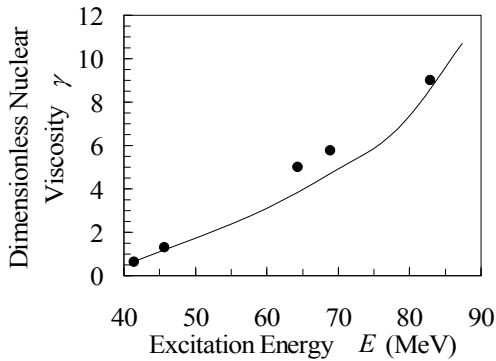

Fig. 26. The solid line shows the nuclear viscosity as a function of excitation energy needed to force the Kramers-modified standard Bohr-Wheeler model with $a_{sp}/a_{eq}=1.04$ to be in agreement with the calculations shown by the solid curve in Fig. 23. The symbols show the excitation-energy dependence of the nuclear viscosity inferred by Dioszegi *et al.* [13].

Fission cross section and prescission neutron multiplicity data from heavy-ion induced fusion-fission reactions with initial compound nuclear excitation energies less than about 50 MeV have been reproduced using a standard Bohr-Wheeler statistical model with $a_{sp}/a_{gs}\sim1.0$ without any Kramers' modification. However, at higher energies, the prescission neutron multiplicity data are underestimated by these model calculations [10]. Agreement with the high-energy data can be obtained if a long fission delay of many $10^{-20}$ s is added to the model. If the standard Bohr-wheeler model is used without any Kramers' modification then the excitation-energy dependence of the more detailed calculations shown by the solid curve and circles in Fig. 23 can be approximately reproduced from $E_i\sim50$ MeV to 90 MeV with $a_{sp}/a_{gs}=0.995$ and a fission delay time of ~$5\times10^{-20}$ s. This result suggests that the long fission delay times inferred by others [10] in heavy-ion fusion-fission reactions are possibly an artifact generated by an incomplete model of the fission process.

## H. HEAVY-ION FUSION

To model the competition between fission and emission processes in heavy-ion fusion reactions, it is necessary to define both the initial excitation energy and the spin distribution of the compound systems following fusion. The initial excitation energy is defined by the kinetic energy of the projectile and the fusion Q-value. Information about the spin distribution can be inferred from measured fusion cross sections. A method that has been commonly used is to assume that the fusion cross section is given by [5,10]

$$\sigma_{fus} = \pi \lambda^2 \sum_0^\infty (2J+1)T_J \quad, \qquad (80)$$

where $\lambda$ is the reduced wave length of the projectile-target system. The fusion transmission coefficients are often parameterized as [5,10]

$$T_J = \left[1 + \exp\left(\frac{J - J_o}{\delta_J}\right)\right]^{-1} \quad . \qquad (81)$$

The diffuseness parameter $\delta_J$ is generally fixed to a value from 2 to 5 based on theoretical considerations [58,59], while the spin cutoff parameter $J_o$ is often adjusted as a function of beam energy to reproduce measured fusion cross sections [10].

In the present paper, we use a model of the fusion process and adjust the size of the nuclei and shape of the target nucleus to obtain a fit to fusion excitation functions. The corresponding calculated fusion spin distributions are used as input into statistical-model calculations of the competition between fission and emission processes. To estimate the fusion of spherical projectile and target nuclei, we use the nucleus-nucleus potential inferred from the elastic scattering of heavy-ions by various targets [60]

$$V(r) = \frac{V_o}{1 + \exp\left(\dfrac{r - r_p - r_t}{\delta}\right)} + \frac{Z_p Z_t e^2}{4\pi \varepsilon_o r} + \frac{J(J+1)\hbar}{2\mu r^2} \quad, \qquad (82)$$

where the effective radii of the projectile $r_p$ and target $r_t$ are given by

$$r_i = 1.233 \text{ fm} \times A^{1/3} - 0.978 \text{ fm} / A^{1/3} \quad . \qquad (83)$$

The potential diffuseness is $\delta=0.63$ fm and the depth of the nuclear potential is

$$V_o = \frac{-r_p r_t}{r_p + r_t} \times 50 \text{ MeV} \quad . \qquad (84)$$

This potential can used to estimate the fusion barrier height $E_B(J)$ and the angular frequency of the inverted potential about the barrier location $\omega_{fus}(J)$ as functions of $J$. These values can, in turn, be used to estimate the fusion transmission coefficients

$$T_J = \left[1 + \exp\left(2\pi \frac{E_B(J) - E_{cm}}{\hbar \, \omega_{fus}(J)}\right)\right]^{-1} . \quad (85)$$

Measured fission cross sections for carbon and oxygen projectiles on thorium and uranium targets are shown in Fig. 27 . These reactions produce very fissile compound nuclei and essentially all fusions lead to fission, and thus the fusion and fission cross sections are the same. The horizontal axis in Fig. 27 is the ratio of center-of-mass kinetic energy to the height of the fusion barrier approximated by the expression

$$B_{fus} = \frac{Z_p Z_t}{A_p^{1/3} + A_t^{1/3}} \, \text{MeV} . \quad (86)$$

This is a convenient expression often used to quickly estimate the height of the fusion barrier. The true fusion barrier heights are generally a few percent lower. The dash-dotted and dashed curves show classical and quantum mechanical calculations assuming spherical nuclei. The classical calculations were performed by assuming the fusion transmission coefficients are 1 and 0 when the kinetic energy in the center-of-mass frame is higher and lower, respectively, than the corresponding spin-dependent fusion barrier. To reproduce the fusion cross sections at above-barrier energies, the first term in Eq. (83) was scaled by $r_{fus} = 1.013$. The classical model does not allow for any sub-barrier fusion and thus fails to reproduce the sub-barrier cross sections. This discrepancy at sub-barrier energies is reduced, but not resolved, by the inclusion of barrier penetration as shown by the dashed curve in Fig. 27. The remaining discrepancies can be resolved if the thorium and uranium nuclei are treated as prolate rigid-body rotators.

To estimate the effect of a static deformation of the target nuclei we assume

$$\sigma_{fus} \sim \int_{\theta=0}^{\pi/2} w(\theta) \pi \, \bar{\lambda}^2 \sum_0^{\infty} (2J+1) T_J(\theta) \, d\theta , \quad (87)$$

where $\theta$ is the angle between the symmetry axis and a vector from the center of mass of the target to an area element on the target's surface. We assume the target is prolate with a shape defined by a single parameter $\beta_2$,

$$r(\theta) = C(\beta_2) \left\{1 + \beta_2 \sqrt{\frac{5}{16\pi}} \left(3\cos^2(\theta) - 1\right)\right\} , \quad (88)$$

where $C(\beta_2)$ is determined assuming a constant nuclear volume as a function of $\beta_2$. The fusion transmission coefficients are a function of spin and the effective interaction point on the target nucleus. We estimate these transmission coefficients by determining $E_B(J, \theta)$, and

$\omega_{fus}(J, \theta)$ using the potential energy along the line defined by the center of mass of the target and the effective fusion point on the surface of the target nucleus. The Coulomb potential energy about the deformed target is determined using the results presented in ref [63]. To determine the weights $w(\theta)$ we invoke the known result in the classical limit for projectiles traveling in straight line paths

$$\sigma_{fus} = \int_{\theta=0}^{\pi/2} w(\theta) \pi \, r^2(\theta) \, d\theta = \frac{A_S}{4} , \quad (89)$$

where $A_S$ in the surface area of the prolate target. From Eq. (89) we see that the weight function in Eq. (87) must be

$$w(\theta) = \frac{1}{2\pi r^2(\theta)} \frac{dA}{d\theta} . \quad (90)$$

The solid line in Fig. 27 shows our fusion model calculation with $\beta_2 = 0.27$. The known $\beta_2$ for thorium and uranium nuclei range from 0.26 to 0.29, respectively [64]. The good agreement between the deformed-target fusion model calculations and the data shown in Fig. 27 confirms that thorium and uranium [65] targets act as rigid-body rotators during the fusion process.

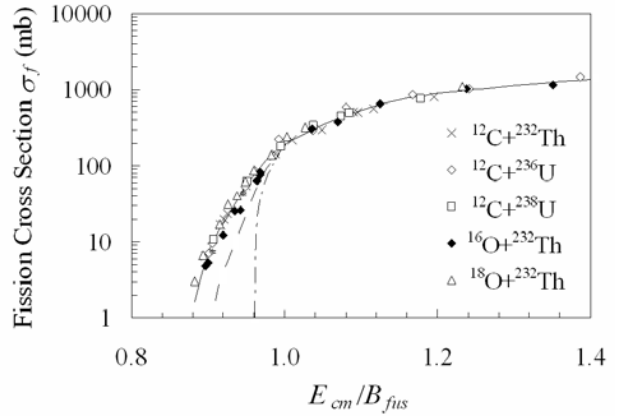

Fig. 27. Measured fission cross sections for carbon and oxygen projectiles on thorium and uranium targets [57,61,62] as a function of the center-of-mass kinetic energy relative to the corresponding fusion barrier approximated by Eq. (86). The dash-dotted and dashed curves show classical and quantum mechanical calculations assuming spherical nuclei. The solid curve shows a quantum mechanical calculation with a target deformation $\beta_2 = 0.27$.

For non-actinide targets where additional internal degrees of freedom are important, the sub-barrier fusion cross sections are generally underestimated if the known static target deformations are used within the frame work of this simplistic fusion model. Significant advancements were made in the understanding of sub-barrier fusion during the 1990's [66,67]. It is now well known that, in addition to the effect of static deformations, sub-barrier fusion can be enhanced if either the projectile and/or target nuclei are soft, and/or if the Q-value for the transfer of nucleons between the projectile and target is small and/or positive. If the target and/or projectile are soft then sub-barrier fusion is enhanced because the nuclei can vibrate or change shape during the fusion process. If the nucleon transfer Q-values are small or positive then sub-barrier fusion is enhanced by the exchange of nucleons during the fusion process. Instead of explicitly adding these additional complex processes, we choose to use an effective static deformation for the target nuclei that is larger than the known static deformation. The size of this effective static deformation is determined by fitting experimental fusion excitation functions with the deformed-target fusion model discussed above. Although this prescription could be made more complete, it is an improvement on the methods commonly used by others when inferring the properties of the nuclear viscosity from fusion-fission data [5].

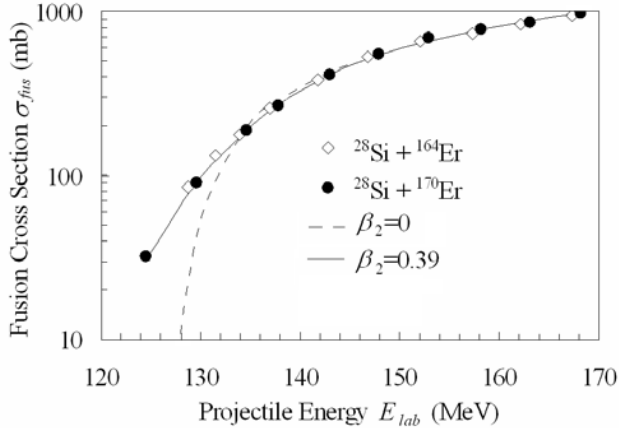

Fig. 28. Measured fusion cross sections (symbols) for the reactions $^{28}$Si + $^{164,170}$Er [68]. The dashed and solid curves show model calculations assuming spherical nuclei and an effective static deformation for Er nuclei of $\beta_2$=0.39, respectively.

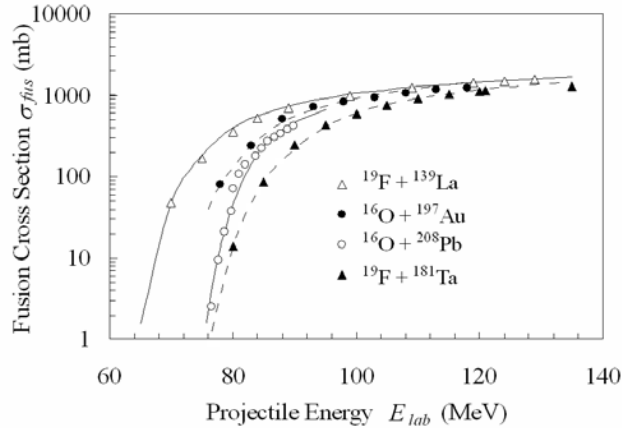

Fig. 29. Measured fusion cross sections for some reactions involving $^{16}$O and $^{19}$F projectiles on various non-actinide target nuclei [10,58,69,70]. The curves show model calculations where the radius scaling parameter $r_{fus}$ and shape parameter $\beta_2$ are adjusted to fit the data (see Table III).

Fig. 28 shows measured fusion cross sections for the reactions $^{28}$Si + $^{164,170}$Er [68]. The dashed and solid curves show model calculations assuming spherical nuclei and an effective static deformation for Er nuclei of $\beta_2$=0.39, respectively. The known deformations for $^{164}$Er and $^{170}$Er nuclei are $\beta_2$=0.333 and 0.336, respectively [64]. Fig. 29 shows measured fusion cross sections for some reactions involving $^{16}$O and $^{19}$F projectiles on various non-actinide target nuclei [10,58,69,70]. The curves show model calculations where the fusion radius scaling parameter $r_{fus}$, and shape parameter $\beta_2$ are adjusted to fit the fusion data. Table III contains the parameters $r_{fus}$ and $\beta_2$ that reproduce fusion cross section data for a range of reactions. The $\beta_2$ values listed in Table III are displayed by the solid circles in Fig. 30. The effective $\beta_2$ obtained from fitting the fusion cross sections are either close to or larger than the known static deformation [64] shown by the open circles in Fig. 30. This is expected as per the above discussion on vibrational and transfer degrees of freedom.

Table III. Model parameters $r_{fus}$ and $\beta_2$ that reproduce fusion cross section data for a range of reactions.

| Reaction | $Z_t$ | $r_{fus}$ | $\beta_2$ |
|---|---|---|---|
| $^{19}$F+$^{139}$La [58] | 57 | 0.99 | 0.41 |
| $^{18}$O+$^{150}$Sm [58] | 62 | 0.98 | 0.50 |
| $^{19}$F+$^{159}$Tb [58] | 65 | 0.98 | 0.32 |
| $^{28}$Si+$^{170}$Er [68] | 68 | 0.99 | 0.39 |
| $^{28}$Si+$^{164}$Er [68] | 68 | 0.99 | 0.41 |
| $^{19}$F+$^{169}$Tm [58] | 69 | 0.98 | 0.50 |
| $^{19}$F+$^{181}$Ta [70] | 73 | 0.99 | 0.45 |
| $^{18}$O+$^{192}$Os [58] | 76 | 0.98 | 0.50 |
| $^{16}$O+$^{197}$Au [10] | 79 | 1.01 | 0.38 |
| $^{16}$O+$^{208}$Pb [69] | 82 | 1.01 | 0.20 |
| $^{12}$C,$^{16}$O+$^{232}$Th [57,61] | 90 | 1.01 | 0.30 |
| $^{12}$C+$^{238}$U [62] | 92 | 1.01 | 0.29 |

In section III, experimental data for many reactions are analyzed using the statistical-model code JOANNE4. Emphasis is placed on several reactions for which both the fission and evaporation residue cross sections (and thus fusion cross sections) and prescission neutron multiplicities have been measured. The spin distributions for these reactions are calculated as a function of beam energy using the parameters $r_{fus}$ and $\beta_2$ given in Table III. The thick solid curves in Fig. 31 show calculated fusion spin distributions for the reaction $^{19}$F + $^{181}$Ta with $^{19}$F beam energies of 90 MeV and 120 MeV, using the corresponding parameters in Table III. The corresponding fusion cross sections are 200 mb and 1170 mb, respectively. The dashed curves show calculations assuming spherical nuclei. The thin solid curves show spin distributions corresponding to the parameterization given by Eq. (81) with $\delta_l$ = 4.7 [58]. Fortunately, when the fission cross section is larger than ~200 mb, the calculated fission cross sections and prescission emission properties are relatively insensitive to the assumed spin distribution. This is partial justification for why, in many papers involving a statistical-model analysis of heavy-ion fusion-fission data, the details of the assumed fusion spin distributions are either only briefly described or not mentioned at all. However, when the fission cross sections are small, the calculations become very sensitive to the assumed high-spin tail of the fusion spin distribution and model calculations cannot extrapolate to lower fission cross sections unless a reasonable estimate of the beam energy dependence of the spin distribution is used. Some additional analysis is performed in section III using measured fission cross sections for which there are no corresponding fusion cross sections. For reactions involving targets not listed in Table III the fusion cross sections and the corresponding spin distributions are calculated as a function of beam energy assuming $r_{fus} = 1.00$, and $\beta_2$ obtained from the fusion data with neighboring targets (see the crosses in Fig. 30).

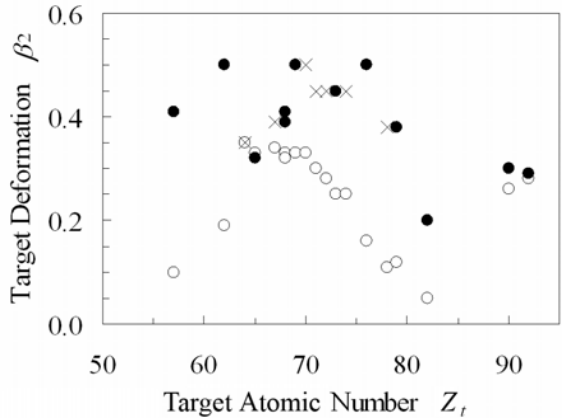

Fig. 30. The effective fusion $\beta_2$ values tabulated in Table III versus the atomic number of the target nucleus $Z_t$ (solid circles). The known static deformations [64] are shown by the open circles. Inferred effective fusion $\beta_2$ values are displayed by the crosses (see text).

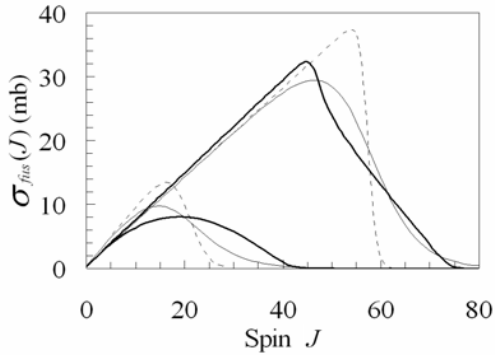

Fig. 31. The thick solid curves show calculated fusion spin distributions for the reaction $^{19}$F + $^{181}$Ta with $^{19}$F beam energies of 90 MeV and 120 MeV, using the corresponding parameters in Table III. The dashed curves show calculations assuming spherical nuclei. The thin solid curves show spin distributions corresponding to the parameterization given by Eq. (81) with $\delta_J = 4.7$ [58].

## I. PARTICLE EVAPORATION

Modeling the evaporation of small particles from hot compound systems is much simpler than the modeling of fission as described above. This is because, in the case of small particle evaporation, the transition states can be viewed as a small perturbation of the parent configuration. The transition states consist of the evaporated particle plus a daughter compound nucleus. The daughter can be assumed to be very similar to the parent, except for the energy, nucleons, and angular momentum removed by the evaporated particle. The decay width for particle evaporation can be estimated using the Bohr-Wheeler expression, Eq. (3). For evaporation from an equilibrated system, the deformations of the parent and daughter are generally not large (like fission saddle points) and not very different from each other. The level density associated with collective motion and the orientation degree of freedom can be neglected because their effect on the transition state density of the daughter is cancelled by their corresponding effect on the total level density of the parent. No Kramer's reduction factor is needed for the emission of small particles because, when small particles reach their emission barriers, the motion of the system is well approximated by two-body motion with the small particle moving in a conservative potential. This is not the case in fission, where the shape, motion, and internal energy of the nascent fragments is not locked in at the fission transition point.

The statistical-model code JOANNE4 uses a method to model the evaporation of particles from hot compound nuclei that is similar to those commonly used by other codes. Assuming the total initial spin of the system $J_i$ is much larger than the intrinsic spin of the evaporated particle $s$ and that the emission is from a nearly-spherical system, JOANNE4 assumes that the decay width for the emission of a particle with a center-of-mass kinetic energy range from $\varepsilon_p - 1/2$ MeV to $\varepsilon_p + 1/2$ MeV, with orbital angular momentum $L$, from a parent system with excitation energy $E_i$, leaving a daughter system with final spin $J_f$, can be approximated by

$$\Gamma_x(E_i, J_i, \varepsilon_p, L, J_f)$$

$$\sim \frac{2s+1}{2\pi} \frac{\int_{\varepsilon_p - 1/2}^{\varepsilon_p + 1/2} \rho(E - B_x - \varepsilon - E_{rot}^D(J_f)) T_L(\varepsilon_p) d\varepsilon}{\rho(E - E_{rot}^P(J_i))},$$

(91)

The particle binding energies $B_x$ are determined using the experimental mass of the evaporated particle, and the liquid-drop model (LDM) masses [47,48] of the parent and daughter systems. This is done because JOANNE4 contains no shell corrections and thus the excitation energies of the hot parent and daughter systems are relative to their LDM ground states. The rotational energies of the parent and daughter systems $E_{rot}(J)$ are determined using the FRLDM ground state energies [30] obtained via the subroutine BARFIT written by Sierk as done in other codes, we use neutron and proton transmission coefficients $T_L(\varepsilon_p)$, calculated using the optical-model potentials of Perey and Perey [71], and $\alpha$-particle transmission coefficients determined using the potential of Huizenga and Igo [72]. The level density as a function of thermal excitation energy is assumed to be as given in Eq. (55) with $n$=2. The total decay width for the evaporation of a given particle type is determined within JOANNE4 using

$$\Gamma_x(E_i, J_i)$$
$$\sim \sum_{i=0}^{\infty} \sum_{L=0}^{\infty} \sum_{J_f=|J_i-L|}^{J_i+L} \Gamma_x(E_i, J_i, \varepsilon_p = i + 1/2 \text{ MeV}, L, J_f) . \quad (92)$$

Hot compound nuclei are not spherical, but experience an ensemble of shapes about their ground-state positions. Fortunately, the dominant cooling process in heavy-ion fusion-fission reactions is the evaporation of neutrons whose emission properties are relatively insensitive to the nuclear shape. Due to Coulomb forces, the properties of the charged-particle emission are sensitive to the assumed nuclear shape. However, charged-particle emission is, in general, more than two orders of magnitude weaker than the neutron emission for all but very neutron-deficient systems, and inadequacies in the charged-particle emission do not significantly affect calculated fission cross sections and prescission neutron multiplicities. In the analysis presented in the present paper, only fission and evaporation-residue cross sections and prescission neutron-multiplicity data are used. An analysis of the available prescission charged-particle data from heavy-ion fusion-fission reactions [21,73,74] would require a more detailed model incorporating the effects of nuclear shape on the charged-particle emission process.

The solid curves in Fig. 32, Fig. 33, and Fig. 34 show JOANNE4 model calculations for the lifetime of neutron, proton, and $\alpha$-particle evaporation, and fission of $^{210}$Po compound systems for various combinations of spin and excitation energy. The Fermi-gas level-density parameter for nearly-spherical systems is assumed to be $a$=$A$/8.6 MeV$^{-1}$. The fission lifetime calculations assume $\beta$=3$\times$10$^{21}$ s$^{-1}$ and $\alpha$=0.016 MeV$^{-2}$ (see section II.F). Fig. 32 shows the spin dependence of the lifetime of the dominant decay processes for $^{210}$Po systems with a fixed total initial excitation energy of 80 MeV. The particle-evaporation lifetimes increase with increasing spin because more of the total excitation energy is tied up in collective rotation with increasing spin. Despite the decrease in the thermal excitation energy with increasing spin, the fission lifetime decreases because the fission barriers decrease with increasing spin. The dashed curve in Fig. 32 shows the results of a standard Bohr-Wheeler model estimate of the fission lifetime with no Kramer's reduction factor and $a_{sp}/a_{gs}$=1.04. As discussed in earlier sections, this model is inadequate. The use of this inadequate model causes fission to dominate at high spin, and causes the calculated prescission emission to be artificially suppressed at high beam energies. Some authors have compensated for this artificial decrease in the prescission emission at high beam energies by arbitrary modifications to the model of fission.

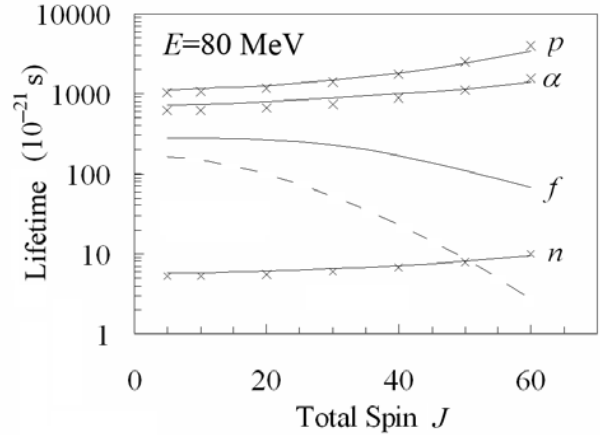

Fig. 32. The spin dependence of the lifetime for neutron, proton, and $\alpha$-particle evaporation from, and fission of, $^{210}$Po compound systems with a fixed total initial excitation energy of $E$=80 MeV. The solid lines show results obtained with the code JOANNE4. Model parameters are $a$=$A$/8.6 MeV$^{-1}$, $\beta$=3$\times$10$^{21}$ s$^{-1}$, and $\alpha$=0.016 MeV$^{-2}$. The dashed curve shows the standard Bohr-Wheeler model of the fission lifetime with no Kramers' reduction factor and $a_{sp}/a_{eq}$=1.04. The crosses show particle evaporation lifetimes estimated using Eq. (95) (see text).

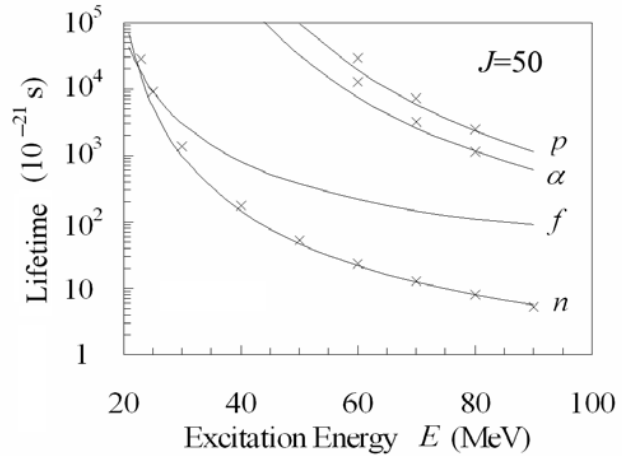

Fig. 33. The excitation-energy dependence of the JOANNE4 lifetimes (solid curves) of the dominant decay processes for $^{210}$Po systems with a fixed total spin of $J$=50. The model parameters are as for Fig. 32. The crosses show estimates obtained using Eq. (95) (see text).

Fig. 33 shows the excitation-energy dependence of the lifetimes of the dominant decay processes for $^{210}$Po systems with a fixed total spin of $J$=50. Notice that even at this high spin, the time scale for neutron emission at high excitation energy is shorter than the corresponding fission time scale. The ratio of the fission to neutron-emission lifetimes decreases with decreasing excitation energy, with fission becoming faster than neutron emission at low excitation energy. This behavior means that highly excited high-spin systems will fission, but not before emitting a number of prescission neutrons. Fig. 34 shows the excitation-energy dependence of the lifetimes of the dominant decay processes for fissioning $^{210}$Po systems formed in the reaction $^{18}$O + $^{192}$Os. The relationship between excitation energy and spin is assumed to be as given in Table I. The dashed curve in Fig. 34 shows the results of a standard Bohr-Wheeler model estimate of the fission lifetime with no Kramer's reduction factor and $a_{sp}/a_{gs}$=1.04 (as shown in Fig. 23).

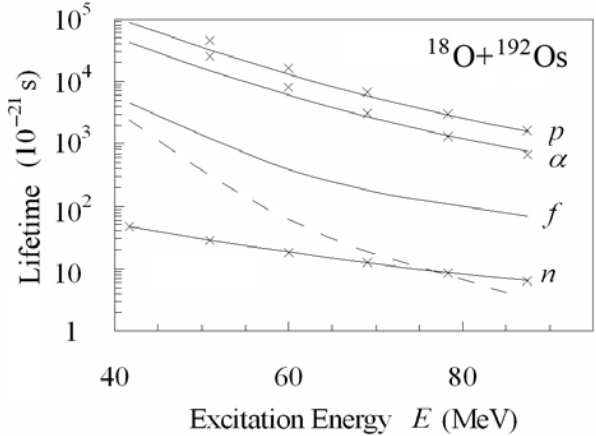

Fig. 34. The excitation-energy dependence of the JOANNE4 lifetimes (solid curves) of the dominant decay processes for fissioning $^{210}$Po systems formed in the reaction $^{18}$O + $^{192}$Os. The model parameters are as for Fig. 32. The dashed curve shows the results of a standard Bohr-Wheeler model of the fission lifetime with no Kramer's reduction factor and $a_{sp}/a_{eq}$=1.04. The crosses show particle evaporation lifetimes estimated using Eq. (95) (see text).

To obtain a better intuitive feel for particle evaporation from hot equilibrated systems, is it useful to make some semi-classical approximations so that emission lifetimes can be estimated via a simple analytical expression instead of numerically as Eqs. (91) and (92). In the classical limit, the particle transmission coefficients are 1 and 0 for particle orbital angular momentum below and above

$$Lh = r_B \sqrt{2\varepsilon\mu} \ , \qquad (93)$$

where $r_B$ is the radius of the emission barrier, $\varepsilon$ is the kinetic energy of the particle-daughter system in the corresponding center of mass as the emission barrier is crossed, and $\mu$ is the reduced mass of the particle-daughter system. If, in addition to this assumption, the mass and orbital spin of the evaporated particle are assumed to be negligible, then Eqs. (91) and (92) can be written as

$$\Gamma_x \sim \frac{2s+1}{\pi\hbar^2}\mu r_B^2 T_D^2 \exp\left(\frac{-2(B_x + B_E)}{T_P + T_D}\right) \ , \qquad (94)$$

where $T_P$ is the temperature of the parent, $T_D$ are the temperature of the daughter systems assuming the kinetic energy in the exit channel is equal to the corresponding emission barrier height $B_E$, and $B_x$ is the particle binding energy. For neutron emission, the height of the emission barrier is zero. The corresponding mean lifetime for particle evaporation can be written as

$$t_x \sim \frac{87 \ \text{fm}^2 \ \text{MeV}^2 \times 10^{-21} s}{(2s+1)A_x \, r_B^2 \, T_D^2} \exp\left(\frac{2(B_x + B_E)}{T_P + T_D}\right) \ , \qquad (95)$$

where $A_x$ is the mass number of the evaporated particle. For neutron emission, we assume that the emission barrier is at the corresponding real-nuclear-potential radius parameter [71] plus three times the corresponding diffuseness parameter. For systems with A~200, the corresponding neutron-emission barrier radius is $r_n$~1.27×$A^{1/3}$+1.98 fm. The crosses displayed in Fig. 32, Fig. 33, and Fig. 34 show $^{210}$Po particle evaporation lifetimes estimated using Eq. (95). The $^{210}$Po neutron, proton, and $\alpha$-particle binding energies are 6.62 MeV, 5.56 MeV, −4.23 MeV, respectively. Using the optical model potentials of refs [71,72], the $^{210}$Po proton and $\alpha$-particle barrier radii are 10.2 fm and 10.8 fm, respectively. The corresponding emission-barrier heights are 10.9 MeV, and 20.6 MeV, respectively. If a proton barrier height of 10.9 MeV is used, then Eq. (95) overestimates the proton emission lifetimes by a factor of ~2. This overestimation is largely due to the assumption that protons interact with the barrier in a classical fashion. Protons are light enough that quantum-mechanical barrier penetration needs to be included. This leads to significant proton emission at sub-barrier energies. This can be mimicked classically by dropping the proton emission barrier height. The proton calculations obtained via Eq. (95) shown by the crosses in Fig. 32, Fig. 33, and Fig. 34 use a proton emission barrier 5% lower than the barrier height obtained using the proton-nucleus potential of ref [71]. The $\alpha$-particle is heavy enough that, in obtaining an emission lifetime, the interaction with the barrier can be assumed to be classical. However, Eq. (95) significantly overestimates the mean lifetime for $\alpha$-particle emission from high-spin systems because the mass of the $\alpha$-particle is large enough to carry enough angular momentum and mass from the parent to invalidate the assumptions used to obtain Eq. (95). If the finite mass of the evaporated particle is accounted for, the emission decay width can be expressed as

$$\Gamma_x \sim \frac{2s+1}{2\pi} \exp\left(\frac{-(B_x + B_E)}{T}\right)\exp\left(\frac{-J^2\hbar^2}{2T}\left(\frac{1}{I_D} - \frac{1}{I_P}\right)\right)$$
$$\times \int_{\varepsilon=0}^{\infty} \exp\left(\frac{-\varepsilon}{T}\right)\int_{L=0}^{L_{max}} \int_{\Delta J=-L}^{L} \exp\left(\frac{-\Delta J \, J \hbar^2}{I_D \, T}\right)d\Delta J \, dL \, d\varepsilon \ , \qquad (96)$$

where $I_P$ and $I_D$ are the moments of inertia of the parent and daughter systems. After some algebraic manipulation of Eq. (96), one can show that the effect of the finite mass of the evaporated particle can be approximated using Eq.

(95) if the particle binding energy is replaced with a spin-dependent effective binding energy

$$B_x^{eff}(J) \sim B_x - 40 \frac{\text{MeV}}{\text{fm}^2} \frac{A_x r_B^2}{A_D^{10/3}} J^2 + 60 \text{MeV} \frac{A_x}{A_D^{8/3}} J^2 \quad . \quad (97)$$

The first $J^2$ term in Eq. (97) corrects for the angular momentum removed from the parent, and the second $J^2$ term corrects for the removed mass. The typical angular momentum removed from the parent by the evaporated particle is

$$\Delta J \sim \frac{0.6}{\text{fm}^2} \frac{A_x r_B^2 J}{A_D^{5/3}} \quad . \quad (98)$$

The $\alpha$-particle calculations obtained via Eq. (95) shown by the crosses in Fig. 32, Fig. 33, and Fig. 34 use a $^{210}$Po spin-dependent effective $\alpha$-particle binding energy of

$$B_\alpha^{eff}(J) = -4.23 \text{ MeV} - 1.9 \times 10^{-4} \text{ MeV} \times J^2 \quad . \quad (99)$$

It must be stressed that JOANNE4 and other commonly used statistical-model codes do not use analytical expressions, like Eqs. (94)-(99), to estimate the particle evaporation rates, but calculate the emission rates as a function of kinetic energy, orbital spin, and final compound nuclei spin using Eq. (91) or slight variations thereof, with particle transmission coefficients obtained by numerical means. The approximations summarized by Eqs. (93)-(99) are only introduced to give the reader a better intuitive feel for the particle-evaporation process.

### J. GAMMA-RAY EMISSION

If the thermal excitation energy of a compound system falls below the neutron binding energy, and if the fission barrier is lower than the neutron binding energy, then the fission probability at this low excitation is governed by the competition between $\gamma$-ray emission and fission. For heavy-ion fusion-fission reactions involving compound systems with $A<220$, most fissions occur at excitation energies well in excess of the neutron binding energy, and thus model calculations of fission and evaporation residue cross sections and prescission neutron emission are very insensitive to the assumed properties for the $\gamma$-ray emission. Despite this insensitivity, it is prudent to include a simple estimate of the $\gamma$-ray emission. By including a simple estimate of the $\gamma$-ray emission, one can test that model results of interest are not sensitive to one's assumed properties for the $\gamma$-ray emission. Of course, if the $\gamma$-ray emission is, itself, a topic of interest, then a more complete model would be required.

The $\gamma$-ray decay width is [13]

$$\Gamma_\gamma(E_i, J_i) = \frac{1}{2\pi \rho_i(E_i, J_i)}$$

$$\times \int_{\varepsilon_\gamma=0}^{\infty} \sum_L f_L(\varepsilon_\gamma) \, \varepsilon_\gamma^{2L+1} \sum_{J_f=|J_i-L|}^{J_i+L} \rho_f(E_f, J_f) \, d\varepsilon_\gamma \quad . \quad (100)$$

The statistical-model code JOANNE4 was written to calculate heavy-ion fusion-fission cross sections and to calculate the corresponding properties of the prescission particle emission. JOANNE4 is not intended for detailed modeling of high-energy $\gamma$-rays from heavy-ion reactions. For simplicity, JOANNE4 assumes only $L=1$ photons and that $f_L(\varepsilon_\gamma)$ is independent of the photon energy and proportional to $A^{2/3}$ [75], and estimates the $\gamma$-ray decay widths using

$$\Gamma_\gamma(E_i, J_i) = \frac{1}{2\pi \, \rho(E_i - E_{rot}(J_i))} \sum_{i=0}^{\infty} 3 C_\gamma A^{2/3} (i+1/2 \text{ MeV})^3$$

$$\times \int_{\varepsilon=i-1/2}^{i+1/2} \rho(E_i - E_{rot}(J_i) - \varepsilon) d\varepsilon \quad . \quad (101)$$

A value of $C_\gamma$=6.4×10$^{-9}$ MeV$^{-3}$ gives the best fit to measured decay widths just above the neutron binding energy of 40 nuclei, spanning the compound nuclei mass range from $A\sim150$ to $250$ [75]. Fig. 35 shows a comparison between modeled $\gamma$-ray decay widths with $C_\gamma$=6.4×10$^{-9}$ MeV$^{-3}$, and the corresponding decay widths just above the neutron binding energies. Typical differences between the modeled and experimental decay widths are less than a factor of 2. The simplicity of the $\gamma$-ray emission model contained within JOANNE4 is justified because an increase or decrease of $C_\gamma$ by a factor of 10 does not significantly change the JOANNE4 model calculations presented in section III.

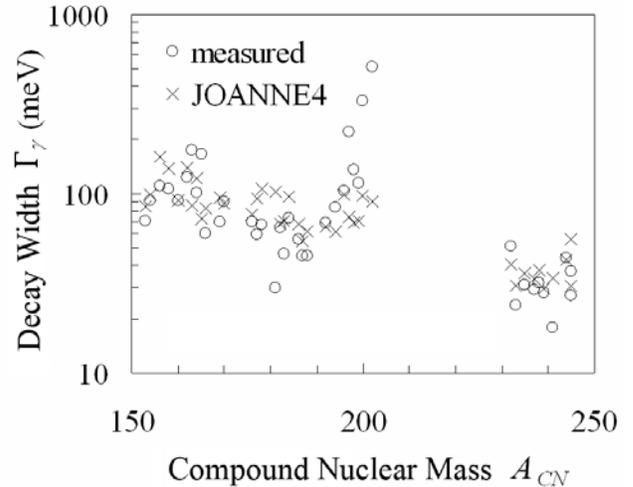

Fig. 35. Calculated $\gamma$-ray decay widths with $C_\gamma$=6.4×10$^{-9}$ MeV$^{-3}$, and the corresponding measured decay widths just above the neutron binding energies [75] as a function of compound nucleus mass $A_{CN}$.

Further manipulation of Eq. (101) leads to an approximate expression for the lifetime for $\gamma$-ray decay from systems with $A\sim200$,

$$t_\gamma \sim \frac{\text{MeV}^4}{T^4} 10^{-15} \text{s} \quad . \quad (102)$$

An estimate for the neutron lifetime of systems with $A\sim200$ and neutron binding energies of $\sim7$ MeV is (see Eq. (95))

$$t_n \sim \frac{\text{MeV}^2}{T_D^2} \exp\left(\frac{14\,\text{MeV}}{T_D + T_P}\right) 10^{-21}\text{s} \quad . \quad (103)$$

Fig. 36 compares the $\gamma$–ray and neutron lifetimes given by Eqs. (102) and (103). The time scales for these two emission processes are comparable only at excitation energies just above the neutron binding energy. As the excitation energy is increased, the neutron lifetimes decrease rapidly relative to the $\gamma$–ray emission time scale. However, Eqs. (101) and (102) should not be used at excitation energies well in excess of the neutron binding energy. This is because the $\gamma$–ray emission strength $f_L(\varepsilon_\gamma)$ is not a constant, and increases as the photon energy approaches the energy of the giant dipole resonance [12,13]. However, including this energy dependence of $f_L(\varepsilon_\gamma)$ can only reduce the $\gamma$–ray emission lifetime at high excitation energies by no more than 2 orders of magnitude. Therefore, neutron evaporation remains the dominant cooling mechanism at high excitation energy, and the details of the $\gamma$–ray emission at high excitation energy have no effect on calculated fission and evaporation residue cross sections and particle emission properties.

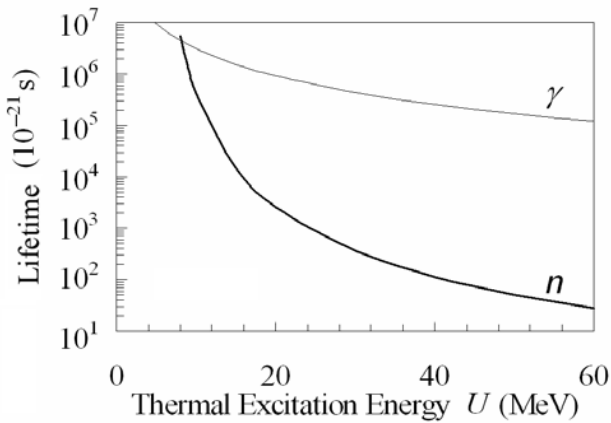

Fig. 36. $\gamma$–ray and neutron lifetimes for $A\sim200$ systems versus thermal excitation energy, estimated using Eqs. (102) and (103).

## III. MODELING FUSION-FISSION REACTIONS WITH JOANNE4

The statistical-model code JOANNE4 [25] was written to model fission and residue cross sections and prescission particle emission from heavy-ion fusion-fission reactions. The methods used to calculate the fusion spin distribution and the widths of the decay processes, are described in section II. The code inputs are: the number of cascades in the simulation; the atomic and mass numbers of the projectile and target; the laboratory beam energy of the projectile; the inverse level-density parameter for spherical systems $k=A/a$; the scaling parameter $r_{fus}$ and the shape of the target $\beta_2$ used to calculate the fusion cross section and the fusion-spin distribution; the parameters $\alpha$ and $r_S$, which control the temperature and deformation dependence of the effective potential energy of the compound nuclei; and a logical switch which controls the assumed fission decay width for systems with no fission barrier (discussed later in this section). The parameter $r_S$ is a scaling of the MLDM default radii used to calculate the surface and Coulomb energies, and will be described in greater detail later in this section.

JOANNE4 is a Monte-Carlo code. The initial total excitation energy is defined by the kinetic energy in the center of mass and the fusion Q-value. For each cascade, an initial compound nucleus spin is randomly sampled from the fusion spin distribution and the fission decay width and the partial decay widths for all the possible ways neutrons, protons, $\alpha$-particles, and $\gamma$-rays can be emitted are calculated. The first-chance fission probability is the ratio of the fission decay width to the total decay width. The energy, angular momentum, and nucleons associated with a randomly chosen emission mode are then removed from the compound nucleus. All decay modes are then recalculated for the new daughter compound nucleus, and the fission probability and tallies associated with prescission emission are updated. The cascade is allowed to continue until the fission decay width drops below $10^{-6}$ of the total decay width and the system is then assumed to form an evaporation residue. By simulating a large number of randomly chosen cascades, the fission and residue cross sections and the properties of the emission preceding fission are determined.

In heavy-ion fusion-fission reactions involving fissile nuclei with masses $A_{CN} >220$, the residue probability becomes very small, difficult to measure, and influenced by decay processes at low excitation energy at the end of emission cascades where shell corrections and the $\gamma$–ray emission strengths are of importance. To avoid complexities associated with sensitivities to assumed shell corrections and the $\gamma$–ray emission strength, we here restrict the use of JOANNE4 to compound nuclei with $A_{CN}<220$, where the decision to fission is made predominately at high excitation energies. For light compound systems ($A_{CN}<175$), fission is increasingly restricted to high spins in the tail of the fusion spin distribution. This makes calculated fission cross sections very sensitive to the assumed spin distributions, and we therefore restrict the analysis presented here to $A_{CN}>175$.

## A. ANALYSIS OF CROSS SECTION AND NEUTRON EMISSION DATA

When reliable measured fusion cross sections exist, the JOANNE4 inputs $r_{fus}$ and $\beta_2$, are adjusted to reproduce the fusion excitation function as described in section II.H. This procedure assumes complete fusion. We are, therefore, restricted to projectile energies less than $\sim8$ MeV per nucleon. JOANNE4 assumes fully equilibrated systems and should only be used to model prescission emission data from reactions where emission is predominantly from systems with a fission barrier. Projectiles with masses larger than $A_p \sim 26$ bring in enough angular momentum that the contribution from fast-fission reactions becomes significant before the excitation energy can get high. We therefore restrict ourselves to projectile masses $A_p \leq 26$. Given these restrictions, we focus on an impressive data set measured by the Australian National University (ANU) nuclear reactions group in the 1980's where fission/residue/fusion cross sections and prescission neutron emission data were obtained as a function of oxygen and fluorine projectile energy for a wide range of compound nuclear masses. The $A_{CN}$ =175-220 data from this systematic experimental investigation [10,58,76,77] are displayed in Fig. 37, along with some additional data for the same reactions obtained by others [78,79].

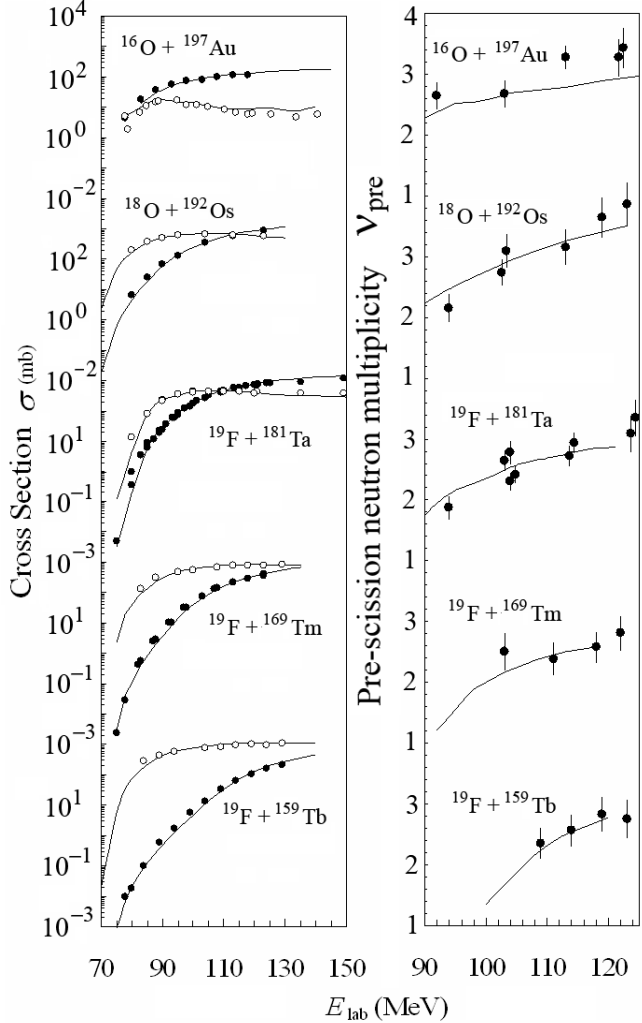

Fig. 37. JOANNE4 model predictions for the projectile energy dependence of cross sections and prescission neutron multiplicities for five reactions (solid curves). The experimental data are from refs [10,58,76,77,78,79]. The fission and residue cross sections are shown by solid and open symbols, respectively.

With earlier statistical-model codes, many authors have used a scaling of the FRLDM barrier heights $f_B$ and the ratio of the level density for fission and neutron emission, $a_f/a_n$, as adjustable parameters [77]. The adjustment of these parameters generally leads to a reasonable reproduction of fission and residue cross sections. Fission probabilities define a range of correlated values for the parameters $f_B$ and $a_f/a_n$. Given a reasonable model for fission decay widths and a choice for $f_B$ (~1.0), one can generally find a value of $a_f/a_n$ to reproduce cross section data. If $f_B$ is increased, then fission slows and the fission cross sections decrease. This can be compensated for by increasing $a_f/a_n$, which speeds fission up. In this way, a variety of models with different dissipation strengths can be made to reproduce cross section data. If only cross section data are available for a given reaction, then the properties of the nuclear viscosity can only be obtained if the $T$=0 potential-energy surfaces and the deformation dependence of the level-density parameters are known to good accuracy. This is not the case, and thus it is difficult to test a specific model type with only cross section data.

To test a given class of fission model, it is important to measure emission processes in coincidence with fission. This is because emission probabilities are sensitive to the excitation-energy dependence of the fission width controlled by $a_f/a_n$. If $a_f/a_n$ is increased, then $f_B$ can be increased to keep cross sections the same. Even though such an interplay between $a_f/a_n$ and $f_B$ keeps the fission probability the same, the excitation-energy dependence of the fission decay width is altered. If $a_f/a_n$ and $f_B$ are both increased in a fashion where the fission probability remains fixed, fission becomes more likely at higher excitation energy and less likely at lower excitation energy. This increases the probability of 1st and 2nd chance fission and causes the amount of emission in coincidence with fission to decrease. Therefore, if cross section and emission data are available then, for a given specific model of fission decay widths, the parameters $f_B$ and $a_f/a_n$ can be constrained and the corresponding beam energy dependence of the data is a test of the model. This has been known since the 1980's [77] and is why experimental studies in the 1980's and 1990's focused on emission in coincidence with fission for reactions where the cross sections were known. Based on this type of analysis, it has been determined that in heavy-ion reactions, the standard Bohr-Wheeler model of fission is inadequate.

We have shown in section II that the standard methods used to implement the Bohr-Wheeler statistical model are inadequate for reasons other then a lack of understanding of the nuclear dissipation processes. Fission in heavy-ion reactions can not be accurately modeled as a function of the excitation energy, using the $J$ dependence of the $T$=0 fission barriers, and a fixed value of $a_f/a_n$. Detailed modeling requires knowledge of the shape of the potential-energy surface about the ground states and the fission saddle points, the heights of the fission barriers, and the shape dependence of the level-density parameter. The influence of a shape dependence of the level density can be modeled via a $(1-\alpha T^2)$ dependence of the surface energy. The parameter $\alpha$ in JOANNE4, therefore, performs a role similar to $a_f/a_n$ in earlier models. However, using an effective potential with a $(1-\alpha T^2)$ dependence of the surface energy is a more complete approach. Within JOANNE4, for each $Z$, $A$, $J$, $K$, and $T$, the effective fission saddle point (transition point) is found by looking for the unstable equilibrium point in the effective potential energy. This means that, for a given system, the location of the fission transition point is being determined as a function of $J$, $K$, and $T$, and in the language of earlier statistical-model codes, the deformation dependence and thus spin dependence of $a_f/a_n$ is being taken into account.

In other statistical-model codes, the heights of fission barriers are often uniformly scaled by a parameter $f_B$. In JOANNE4, we instead scale the MLDM radii from the default values used to calculate the surface and Coulomb energies with the parameter $r_S$. The surface energy is proportional to the square of $r_S$, while the Coulomb energy is inversely proportional to $r_S$. A value of $r_S$=1 is the standard MLDM [46] with fission-barrier heights in agreement with the FRLDM [30]. Raising $r_S$ above one increases the surface energy and decreases the Coulomb energy. This stabilizes the systems and causes the fission barriers to increase. Fig. 38 shows $^{210}$Po, $T$=0 and $K$=0 MLDM barrier heights as a function of total spin $J$, with values of $r_S$ =0.995, 1.000, and 1.005. Notice that the barrier heights are not changed by a constant scaling factor. The advantage of using $r_S$ instead of a simple constant barrier height scaling is that the barrier locations and heights, and the angular frequencies at the ground states and the fission transition points are all being determined in a self-consistent manner as a function of $J$, $K$, and $T$.

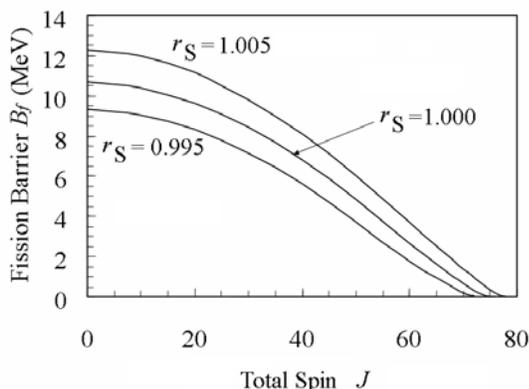

Fig. 38. $^{210}$Po, $T$=0 and $K$=0 MLDM barrier heights as a function of total spin $J$, with values of the MLDM radius scaling $r_S$ =0.995, 1.000, and 1.005.

All JOANNE4 calculations presented here assume $k$=$A/a$=8.6 MeV [31] and $\beta$=3×10$^{21}$ s$^{-1}$ [28,29] as discussed in section II. The only parameters available to fit fission and residue cross sections and neutron emission data are $\alpha$ and $r_S$. For each reaction with data displayed in Fig. 37, the parameters $\alpha$ and $r_S$ are adjusted to reproduce a single fission cross section and a single prescission neutron multiplicity at the same projectile kinetic energy, corresponding to the second lowest prescission neutron multiplicity measurement. Fig. 39 shows how the $E_{lab}$~103 MeV $^{18}$O + $^{192}$Os fission cross section [58] and the prescission neutron multiplicity [10] constrain the adjustable parameters to $\alpha$=0.017±0.006 MeV$^{-2}$ and $r_S$=1.002±0.002. The fission cross section at $E_{lab}$~103 MeV constrains $\alpha$ and $r_S$ to lie in the region between the solid curves shown in Fig. 39. As $r_S$ is increased the fission barriers increase and thus the fission cross sections decrease. This can be compensated for by increasing $\alpha$, which decreases the barriers at high excitation energy. The prescission neutron multiplicity depends more strongly on $\alpha$ than $r_S$. As $\alpha$ is increased, the effective fission barriers decrease more rapidly with increasing excitation energy. This enhances the earlier fission at the higher excitation energies and thus suppresses the emission in coincidence with fission. The $^{18}$O + $^{192}$Os prescission neutron multiplicity at $E_{lab}$~103 MeV constrains $\alpha$ and $r_S$ to lie in the region between the dashed curves shown in Fig. 39.

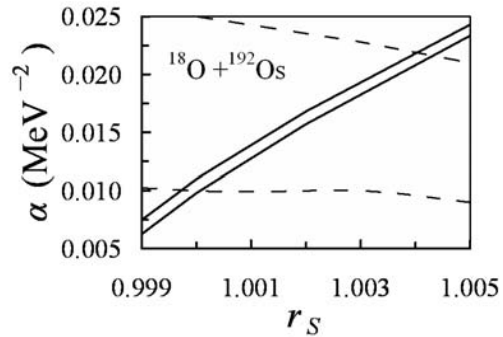

Fig. 39. The $E_{lab}$~103 MeV $^{18}$O + $^{192}$Os fission cross section [58] and neutron multiplicity [10] constrain the parameters $\alpha$ and $r_S$ to the regions between the solid and dashed curves, respectively.

Fig. 40 shows how the neutron multiplicity at the second lowest beam energy and the corresponding fission cross sections constrain the parameters $\alpha$ and $r_S$ for each of the other four reactions displayed in Fig. 37. No single combination of $\alpha$ and $r_S$ will reproduce the data for all five reactions. The parameters $\alpha$ and $r_S$ are displayed as a function of initial compound nucleus mass in Fig. 41. The inferred values of $\alpha$ are in the range of theoretical estimates [31-37] but appear to have a parabolic dependence on $A_{CN}$. The $r_S$ values scatter about 1.000, which suggests the $T$=0 potential energy surfaces are close to those predicted by the FRLDM [30]. The solid curves in Fig. 37 show the JOANNE4 model predictions for the projectile energy dependence of fission and residue cross sections and prescission neutron multiplicities, using the $\alpha$ and $r_S$ values represented by the symbols in Fig. 41. These predictions are consistent with the data. It is important to remember that $\alpha$ and $r_S$ were adjusted to reproduce data at a single beam energy for each reaction and no adjustment has been made to fit the beam energy dependences of the data shown in Fig. 37. To reproduce the data set displayed in Fig. 37, the model calculations of others would require either large fission dynamical delays [10] or a strong temperature dependence of the nuclear viscosity as shown in Fig. 26. It must be emphasized that the statistical-model results presented here should not be used to support the assumed value of $\beta$=3×10$^{21}$ s$^{-1}$ at fission transition points. Equally good reproductions of the data can be obtained by changing $\alpha$ by ~0.0025 MeV$^{-2}$ for each change in $\beta$ of 10$^{21}$ s$^{-1}$. For example, if $\beta$ is reduced to 10$^{21}$ s$^{-1}$ then the required $\alpha$ scatter about ~0.011 MeV$^{-2}$ instead of the value of ~0.016 MeV$^{-2}$ as shown in Fig. 41. The required $r_S$ are very insensitive to changes in the assumed value of $\beta$. The main purpose of the present work is not to justify a specific choice in $\beta$ but to show that the data set considered here is consistent with a temperature-independent dissipation coefficient.

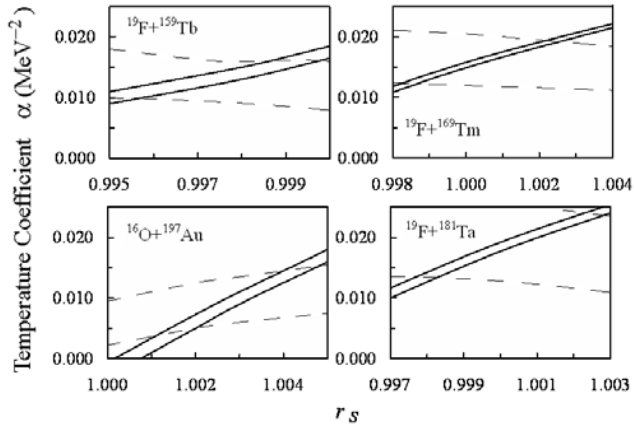

Fig. 40. The neutron multiplicities at the second lowest measured beam energy and the corresponding fission cross sections [10,58] constrain the parameters $\alpha$ and $r_S$ to the regions between the dashed and solid curves, respectively.

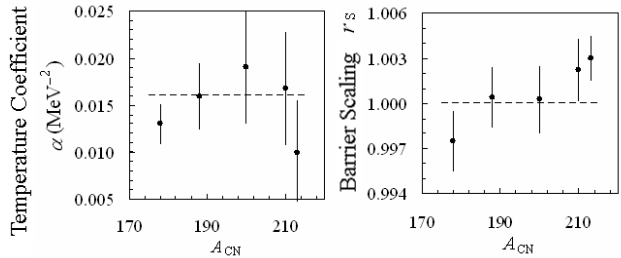

Fig. 41. Fit parameters $\alpha$ and $r_S$ for the five reactions displayed in Fig. 37. The dashed lines show the values corresponding to the model calculations of refs [30,31].

In the present study, JOANNE4 is used in a mode where no dynamical effects associated with transient delays or the saddle-to-scission transit times are included. We are thus assuming that most of the fission is proceeding through systems with a finite barrier that is high enough that the transient delay and the saddle-to-scission descent can be ignored. This assumption will break down at high beam energies where the combined effect of high angular momentum and high temperature will lead to systems that are unstable with respect to fission, i.e. systems where no fission barrier exists. To determine when this transition to fast fission occurs, JOANNE4 allows systems with no fission barriers to be treated in two very different ways. In one of these methods Eqs. (73), (74), and (51) are used even when the $K$=0 barrier vanishes. For $K$ values for which no barrier exists, the barrier heights are set to zero, and the angular frequencies at the equilibrium positions are set to $\omega_{gs}$=$\omega_{bf}$=10$^{21}$ s$^{-1}$. The probability of being in the low $K$ states with no fission barrier is estimated by extrapolating from the higher $K$ states for which barriers exist. In the other approach, when the $K$=0 barrier vanishes, it is assumed that fission is instantaneous and no prescission emission is allowed. JOANNE4 model calculations are assumed valid if calculations using these two very different and artificial estimates for the time scale for fast fission yield results within a few percent of each other.

Fission and residue cross sections are insensitive to the transition to fast fission because, for those partial waves where the barrier vanishes, the fission probability is very high and thus unaffected by the time scale assigned to the fast-fission reactions. However, the emission in coincidence with fission at high beam energies is affected by the fast-fission time scale. For the reactions shown in Fig. 37, the calculated neutron emissions determined using the two different fast fission approaches discussed above, start to deviate significantly above beam energies from ~120-125 MeV. The neutron multiplicity calculations shown in Fig. 37 are terminated when the effect of fast fission becomes significant. The calculation of the prescission neutron emission above these beam energies would require a model that couples statistical emission with a dynamical treatment of the nuclear fluid motion from fusion through to scission. This is beyond the scope of the present study.

## B. ANALYSIS OF FISSION CROSS SECTION DATA

The measurement of fission cross sections is a relatively easy task compared to the measurement of evaporation-residue cross sections and prescission emission data. Therefore, fission cross-section data exist for dozens of reactions for which there are presently no residue cross-section or prescission emission data. The statistical-model analysis of only fission cross section data from a single reaction should carry less weight than the analysis of a fission/residue/fusion cross-section and prescission emission data set from a similar reaction, because when using only fission cross section data, additional assumptions are required to estimate the fusion spin distributions and to constrain the model parameters $\alpha$ and $r_S$. Despite the added uncertainty associated with using reactions with no fusion cross section or prescission emission data, the large volume of fission data warrants a statistical-model analysis. For reactions involving targets not listed in Table III, we estimate the fusion cross sections and spin distributions assuming $r_{fus}=1.00$ and use a $\beta_2$ for the target nucleus obtained from fusion data with a neighboring target (see Fig. 30). Given the uncertainties associated with this procedure, we restrict the analysis of fission cross-section data to projectile energies above the Coulomb barrier. In this section, we assume the MLDM radius scaling $r_S$ is exactly one and adjust $\alpha$ to obtain a match to the fission excitation function below projectile energies of 8 MeV per nucleon.

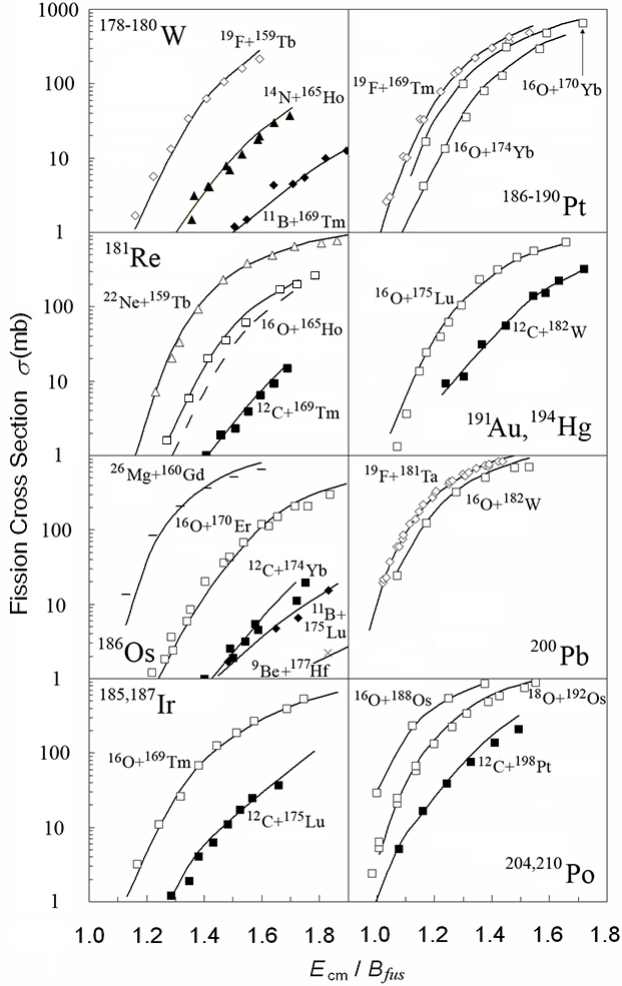

Fig. 42. The symbols show measured fission cross sections for 23 fusion-fission reactions [58,70,79-84] with compound nuclear atomic numbers spanning the range from $Z_{CN}$=74 to 84. The curves are JOANNE4 model calculations described in the text.

The symbols in Fig. 42 show measured fission cross sections for 23 fusion-fission reactions [58,70,79-84]

with compound nuclear atomic numbers spanning the range from $Z_{CN}$=74 to 84. Plotting the fission cross sections versus the kinetic energy in the center-of-mass relative to the corresponding Coulomb barrier (see Eq. (86)), allows reactions with different projectiles to be displayed together without overlapping data sets. The measured fission excitation functions are reproduced by the JOANNE4 model calculations shown by the solid curves. The corresponding values for $\alpha$ are displayed in Fig. 43. The inferred surface-energy temperature coefficients $\alpha$ scatter about a value of $\sim$0.011 MeV$^{-2}$. There appears to be a maximum of $\alpha$ $\sim$0.017 MeV$^{-2}$ at $Z_{CN}$=82 and a minimum of $\alpha$ $\sim$0.006 MeV$^{-2}$ at $Z_{CN}$=75. The possibility that the peak at $Z_{CN}$=82 is associated with the corresponding proton shell should be investigated further. However, it is possible that the dependence of $\alpha$ on $Z_{CN}$ displayed in Fig. 43 could disappear if accurate fusion cross sections were available for all the reactions displayed in Fig. 42, and if a more detailed fusion model were used. For example, three of the highest $\alpha$ values displayed in Fig. 43 are for reactions involving $^{19}$F projectiles, which contain a weakly-bound proton. This suggests that it is possible that the procedure used here to estimate fusion spin distributions is failing in $^{19}$F-induced reactions in a way that is being artificially compensated for by higher values of $\alpha$. The reader should also remember, as discussed in III.A, that the inferred $\alpha$ are sensitive to the assumed value of the dissipation coefficient, $\beta$.

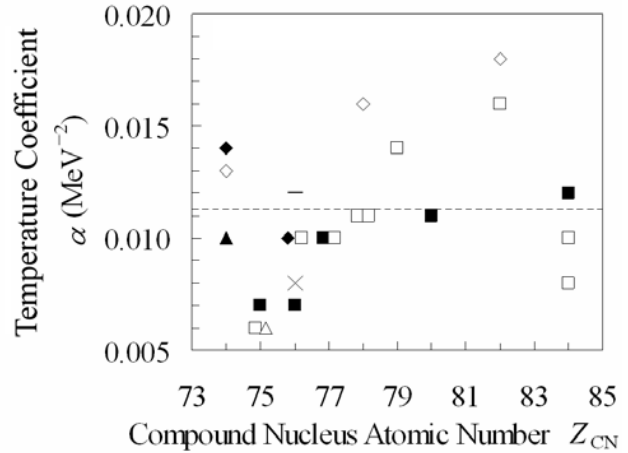

Fig. 43. The values for $\alpha$ corresponding to the JOANNE4 model calculations displayed by the solid curves in Fig. 42. The symbols are the same as for the corresponding reactions shown Fig. 42. Each projectile atomic number is represented by a different symbol: crosses (Be), solid diamonds (B), solid squares (C), solid triangles (N), open squares (O), open diamonds (F), open triangle (Ne), and sidewards bar (Mg).

The dashed curve in Fig. 42 shows a JOANNE4 model calculation for the $^{16}$O + $^{165}$Ho reactions with an unchanged value of $\alpha$ =0.006 MeV$^{-2}$, and $r_{fus}$ changed from 1.00 to 0.98 and $\beta_2$ from 0.45 to 0.39. Agreement with the data can be reestablished by changing $\alpha$ to 0.011 MeV$^{-2}$. This highlights the sensitivity to the assumed fusion spin distributions. Future work is needed to accurately determine fusion spin distributions in heavy-ion reactions and how these distributions vary based on the properties of the projectile and target nuclei. Despite uncertainties associated with the fusion spin distributions, we conclude that fusion-fission excitation functions for a large number of reactions spanning the compound nucleus mass range from 175–215 amu, are consistent with a Kramers-modified statistical model. If the nuclear dissipation is assumed to be $\beta=3\times10^{21}$ s$^{-1}$ [28,29] (independent of temperature) and the $T=0$ potential energy surfaces are estimated using the MLDM [46] then the temperature dependence of the effective potential required to reproduce fission excitation functions is in the range of theoretical estimates [31-37].

## IV. SUMMARY AND CONCLUSIONS

The main purpose of the present study is to illustrate that the standard method for implementing the Bohr-Wheeler statistical model of fission lifetimes is inadequate for heavy-ion reactions, for reasons other than a lack of understanding of the nature of nuclear dissipation. Three pieces of physics are commonly not included in Bohr-Wheeler model calculations. These are the determination of the total level density of the compound system, taking into account the collective motion of the system about the ground-state position; the calculation of the location and height of fission saddle-points as a function of excitation energy using the derivative of the free energy; and the incorporation of the orientation ($K$-state) degree of freedom. Each of these three pieces of physics slows calculated fusion-fission lifetimes at high excitation energy, relative to methods commonly used by others. The inadequacies in commonly used fission models can be compensated for by using an artificial rapid onset of the nuclear dissipation above an excitation energy of ~40 MeV. The strong increase in the nuclear viscosity above a temperature of ~1 MeV deduced by others [12,13] is an artifact generated by an inadequate model of the fission process.

Other authors have assumed that their ability to model nuclear fission is complete enough that the properties of the temperature-dependent nuclear viscosity can be extracted from fission cross section and precission emission data. Calculated fission lifetimes are very sensitive to the assumed deformation dependence of the potential energy and the Fermi-gas level-density parameter. We believe that this strong sensitivity makes it difficult to extract the properties of the nuclear viscosity from fission cross section and precission emission data, even when an adequate model of fission is used. Instead of trying to extract the nuclear viscosity from fission cross section and precission emission data, we instead assume that the nuclear dissipation near fission transition points has been previously constrained to be $\beta\sim3\times10^{21}$ s$^{-1}$ by the surface-plus-window dissipation model [28,29] using the mean kinetic energy of fission fragments, and the width of giant isoscalar resonances. The MLDM potential energy surfaces and the deformation dependence of the level-density parameter are adjusted to reproduce fission cross sections and precission neutron-emission data. The effects associated with a deformation dependence of the level-density parameters are modeled by using a $(1-\alpha T^2)$ dependence of the surface energy. A satisfactory reproduction of fusion-fission cross-section and precission neutron-emission data is obtained over a wide range of excitation energies and compound-nucleus masses. These data suggest that $T=0$ potential-energy surfaces are close to those obtained by the FRLDM [30] and that the surface-energy temperature coefficient is $\alpha\sim0.016$ MeV$^{-2}$, close to the theoretical estimate of Töke and Swiatecki [31]. Our estimate of $\alpha\sim0.016$ MeV$^{-2}$ may be biased on the high side for several reasons, including the small number of reactions involved in the analysis and/or uncertainties associated with fusion-spin distributions for reactions involving $^{19}$F projectiles. The inferred $\alpha$ is mainly constrained by the precission neutron-emission data because of its sensitivity to the excitation-energy dependence of the fission decay widths. This may be altered if a temperature dependence of the level-density parameter is added to the model [33,79]. The analysis of a large volume of fission cross-section data, for a wide range of projectiles (assuming $r_S$=1.000) suggests a lower value of $\alpha\sim0.011$ MeV$^{-2}$, close to the theoretical estimate of Ignatyuk [35] and Reisdorf [36]. We find that the data provides no evidence to indicate a need for a temperature dependence of the nuclear dissipation.

## ACKNOWLEDGMENTS

We wish to thank A. J. Sierk for the many lengthy discussions on nuclear fission, and his assistance in preparing this manuscript.

## Appendix A.

When the MLDM was originally published [46], the modified surface energy S'(q), Coulomb energy C(q), and inertias perpendicular and parallel to the elongation axis, $I_\perp(q)$, and $I_\parallel(q)$, were only tabulated in steps of $q/R_o$=0.05. However, the nuclear potential energy is a delicate balance between surface and Coulomb energies and poor results can be obtained by a simple interpolation of the S'(q), C(q), $I_\perp(q)$, and $I_\parallel(q)$ values published in ref [46]. To obtain an accurate potential-energy surface, one must use a spacing in $q/R_o$ of, or smaller than, ~0.01. The recommended values of S'(q), C(q), $I_\perp(q)$, and $I_\parallel(q)$ are presented in Table A.1 in steps of $q/R_o$=0.01. With these values, the nuclear potential energy can be easily estimated using Eq. (40) as a function of deformation q, Z, A, the total spin J, and the spin about the elongation axis K.

Table A.1. Modified Liquid-Drop Model (MLDM) values for S'(q), C(q), $I_\perp(q)$, and $I_\parallel(q)$ in steps of $q/R_o$=0.01.

| $q/R_o$ | S'(q) | C(q) | $I_\perp(q)$ | $I_\parallel(q)$ | $q/R_o$ | S'(q) | C(q) | $I_\perp(q)$ | $I_\parallel(q)$ |
|---|---|---|---|---|---|---|---|---|---|
| 0.50 | 1.073798 | 0.968713 | 0.972005 | 1.498125 | 1.55 | 1.150466 | 0.906826 | 2.233349 | 0.529803 |
| 0.51 | 1.066468 | 0.971642 | 0.966207 | 1.468274 | 1.56 | 1.152630 | 0.905335 | 2.255177 | 0.527542 |
| 0.52 | 1.059680 | 0.974383 | 0.961195 | 1.439665 | 1.57 | 1.154786 | 0.903846 | 2.277133 | 0.525332 |
| 0.53 | 1.053439 | 0.976931 | 0.956974 | 1.412339 | 1.58 | 1.156931 | 0.902360 | 2.299214 | 0.523172 |
| 0.54 | 1.047645 | 0.979313 | 0.953428 | 1.386131 | 1.59 | 1.159066 | 0.900876 | 2.321422 | 0.521061 |
| 0.55 | 1.042272 | 0.981537 | 0.950520 | 1.360859 | 1.60 | 1.161192 | 0.899396 | 2.343756 | 0.519000 |
| 0.56 | 1.037319 | 0.983604 | 0.948253 | 1.336523 | 1.61 | 1.163307 | 0.897919 | 2.366367 | 0.517036 |
| 0.57 | 1.032787 | 0.985514 | 0.946625 | 1.313122 | 1.62 | 1.165413 | 0.896446 | 2.389021 | 0.515055 |
| 0.58 | 1.028660 | 0.987275 | 0.945582 | 1.290517 | 1.63 | 1.167510 | 0.894977 | 2.411800 | 0.513111 |
| 0.59 | 1.024871 | 0.988894 | 0.945105 | 1.268539 | 1.64 | 1.169594 | 0.893512 | 2.434704 | 0.511206 |
| 0.60 | 1.021402 | 0.990386 | 0.945150 | 1.247268 | 1.65 | 1.171687 | 0.892052 | 2.457732 | 0.509338 |
| 0.61 | 1.018253 | 0.991751 | 0.945718 | 1.226704 | 1.66 | 1.173766 | 0.890596 | 2.480885 | 0.507509 |
| 0.62 | 1.015425 | 0.992989 | 0.946807 | 1.206847 | 1.67 | 1.175833 | 0.889145 | 2.504162 | 0.505718 |
| 0.63 | 1.012918 | 0.994100 | 0.948436 | 1.187753 | 1.68 | 1.177886 | 0.887699 | 2.527563 | 0.503965 |
| 0.64 | 1.010655 | 0.995109 | 0.950484 | 1.169298 | 1.69 | 1.179927 | 0.886259 | 2.551089 | 0.502251 |
| 0.65 | 1.008630 | 0.996016 | 0.952963 | 1.151405 | 1.70 | 1.181954 | 0.884824 | 2.574739 | 0.500574 |
| 0.66 | 1.006841 | 0.996823 | 0.955874 | 1.134074 | 1.71 | 1.183969 | 0.883395 | 2.598514 | 0.498935 |
| 0.67 | 1.005291 | 0.997530 | 0.959217 | 1.117306 | 1.72 | 1.185970 | 0.881972 | 2.622413 | 0.497335 |
| 0.68 | 1.003977 | 0.998136 | 0.962991 | 1.101102 | 1.73 | 1.187958 | 0.880556 | 2.646436 | 0.495773 |
| 0.69 | 1.002904 | 0.998641 | 0.967197 | 1.085321 | 1.74 | 1.189933 | 0.879146 | 2.670584 | 0.494249 |
| 0.70 | 1.002012 | 0.999066 | 0.971762 | 1.070047 | 1.75 | 1.191879 | 0.877743 | 2.694856 | 0.492762 |
| 0.71 | 1.001284 | 0.999411 | 0.976687 | 1.055195 | 1.76 | 1.193803 | 0.876280 | 2.719194 | 0.491265 |
| 0.72 | 1.000720 | 0.999678 | 0.981974 | 1.040765 | 1.77 | 1.195712 | 0.874882 | 2.743795 | 0.489813 |
| 0.73 | 1.000319 | 0.999865 | 0.987622 | 1.026757 | 1.78 | 1.197605 | 0.873489 | 2.768497 | 0.488393 |
| 0.74 | 1.000082 | 0.999974 | 0.993630 | 1.013170 | 1.79 | 1.199483 | 0.872101 | 2.793298 | 0.487006 |
| 0.75 | 0.999999 | 1.000004 | 0.999985 | 1.000025 | 1.80 | 1.201345 | 0.870719 | 2.818200 | 0.485652 |
| 0.76 | 1.000094 | 0.999963 | 1.006665 | 0.987213 | 1.81 | 1.203192 | 0.869343 | 2.843202 | 0.484330 |
| 0.77 | 1.000291 | 0.999858 | 1.013652 | 0.974722 | 1.82 | 1.205024 | 0.867972 | 2.868305 | 0.483041 |
| 0.78 | 1.000592 | 0.999692 | 1.020945 | 0.962552 | 1.83 | 1.206840 | 0.866608 | 2.893507 | 0.481784 |
| 0.79 | 1.000995 | 0.999467 | 1.028544 | 0.950703 | 1.84 | 1.208640 | 0.865249 | 2.918811 | 0.480560 |
| 0.80 | 1.001501 | 0.999184 | 1.036450 | 0.939176 | 1.85 | 1.210425 | 0.863897 | 2.944214 | 0.479368 |
| 0.81 | 1.002110 | 0.998847 | 1.044662 | 0.927970 | 1.86 | 1.212195 | 0.862550 | 2.969718 | 0.478209 |
| 0.82 | 1.002822 | 0.998455 | 1.053181 | 0.917086 | 1.87 | 1.213949 | 0.861210 | 2.995322 | 0.477082 |
| 0.83 | 1.003636 | 0.998012 | 1.061999 | 0.906574 | 1.88 | 1.215697 | 0.859876 | 3.021026 | 0.475988 |
| 0.84 | 1.004553 | 0.997520 | 1.071073 | 0.896209 | 1.89 | 1.217473 | 0.858548 | 3.046830 | 0.474927 |
| 0.85 | 1.005573 | 0.996979 | 1.080412 | 0.886087 | 1.90 | 1.219227 | 0.857226 | 3.072735 | 0.473898 |
| 0.86 | 1.006695 | 0.996392 | 1.090016 | 0.876208 | 1.91 | 1.220959 | 0.855911 | 3.098740 | 0.472902 |
| 0.87 | 1.007920 | 0.995760 | 1.099885 | 0.866572 | 1.92 | 1.222668 | 0.854603 | 3.124735 | 0.471987 |
| 0.88 | 1.009256 | 0.995085 | 1.110019 | 0.857179 | 1.93 | 1.224355 | 0.853301 | 3.150740 | 0.471074 |
| 0.89 | 1.010566 | 0.994368 | 1.120418 | 0.848029 | 1.94 | 1.226020 | 0.852006 | 3.176831 | 0.470190 |
| 0.90 | 1.011940 | 0.993612 | 1.131082 | 0.839121 | 1.95 | 1.227662 | 0.850718 | 3.203009 | 0.469335 |
| 0.91 | 1.013377 | 0.992817 | 1.141946 | 0.830265 | 1.96 | 1.229282 | 0.849436 | 3.229274 | 0.468510 |
| 0.92 | 1.014879 | 0.991986 | 1.153105 | 0.821735 | 1.97 | 1.230880 | 0.848162 | 3.255625 | 0.467713 |
| 0.93 | 1.016444 | 0.991119 | 1.164491 | 0.813391 | 1.98 | 1.232456 | 0.846894 | 3.282063 | 0.466946 |
| 0.94 | 1.018073 | 0.990218 | 1.176102 | 0.805234 | 1.99 | 1.234009 | 0.845634 | 3.308589 | 0.466209 |
| 0.95 | 1.019769 | 0.989285 | 1.187938 | 0.797263 | 2.00 | 1.235540 | 0.844381 | 3.335200 | 0.465500 |
| 0.96 | 1.021484 | 0.988320 | 1.199999 | 0.789478 | 2.01 | 1.237018 | 0.843135 | 3.361898 | 0.464821 |
| 0.97 | 1.023238 | 0.987326 | 1.212286 | 0.781879 | 2.02 | 1.238450 | 0.841896 | 3.388684 | 0.464170 |
| 0.98 | 1.025032 | 0.986303 | 1.224799 | 0.774466 | 2.03 | 1.239852 | 0.840665 | 3.415555 | 0.463549 |
| 0.99 | 1.026866 | 0.985253 | 1.237537 | 0.767240 | 2.04 | 1.241224 | 0.839441 | 3.442514 | 0.462958 |
| 1.00 | 1.028740 | 0.984176 | 1.250600 | 0.760200 | 2.05 | 1.242566 | 0.838225 | 3.469559 | 0.462395 |
| 1.01 | 1.030647 | 0.983075 | 1.263682 | 0.753302 | 2.06 | 1.243878 | 0.837016 | 3.496691 | 0.461862 |
| 1.02 | 1.032580 | 0.981950 | 1.276966 | 0.746545 | 2.07 | 1.245160 | 0.835816 | 3.523910 | 0.461358 |
| 1.03 | 1.034538 | 0.980802 | 1.290453 | 0.739927 | 2.08 | 1.246412 | 0.834622 | 3.551212 | 0.460518 |

| | | | | | | | | |
|---|---|---|---|---|---|---|---|---|
| 1.04 | 1.036523 | 0.979632 | 1.304141 | 0.733450 | 2.09 | 1.247634 | 0.833437 | 3.578649 | 0.459995 |
| 1.05 | 1.038534 | 0.978442 | 1.318032 | 0.727113 | 2.10 | 1.248826 | 0.832260 | 3.606178 | 0.459506 |
| 1.06 | 1.040570 | 0.977232 | 1.332125 | 0.720917 | 2.11 | 1.249987 | 0.831091 | 3.633798 | 0.459050 |
| 1.07 | 1.042633 | 0.976004 | 1.346420 | 0.714861 | 2.12 | 1.251119 | 0.829929 | 3.661510 | 0.458627 |
| 1.08 | 1.044708 | 0.974757 | 1.360917 | 0.708945 | 2.13 | 1.252221 | 0.828776 | 3.689313 | 0.458238 |
| 1.09 | 1.046822 | 0.973494 | 1.375617 | 0.703169 | 2.14 | 1.253292 | 0.827632 | 3.717208 | 0.457881 |
| 1.10 | 1.048950 | 0.972215 | 1.390606 | 0.697387 | 2.15 | 1.254334 | 0.826495 | 3.745193 | 0.457559 |
| 1.11 | 1.051095 | 0.970921 | 1.405741 | 0.691797 | 2.16 | 1.255345 | 0.825367 | 3.773271 | 0.457269 |
| 1.12 | 1.053254 | 0.969612 | 1.421065 | 0.686312 | 2.17 | 1.256327 | 0.824247 | 3.801439 | 0.457013 |
| 1.13 | 1.055429 | 0.968289 | 1.436576 | 0.680930 | 2.18 | 1.257278 | 0.823136 | 3.829699 | 0.456790 |
| 1.14 | 1.057620 | 0.966954 | 1.452276 | 0.675652 | 2.19 | 1.258200 | 0.822034 | 3.858050 | 0.456600 |
| 1.15 | 1.059826 | 0.965606 | 1.468163 | 0.670478 | 2.20 | 1.259091 | 0.820940 | 3.886492 | 0.456444 |
| 1.16 | 1.062048 | 0.964246 | 1.484239 | 0.665408 | 2.21 | 1.259953 | 0.819855 | 3.915026 | 0.456321 |
| 1.17 | 1.064285 | 0.962875 | 1.500503 | 0.660442 | 2.22 | 1.260784 | 0.818779 | 3.943651 | 0.456231 |
| 1.18 | 1.066552 | 0.961494 | 1.516955 | 0.655580 | 2.23 | 1.261585 | 0.817658 | 3.972608 | 0.456292 |
| 1.19 | 1.068807 | 0.960103 | 1.533596 | 0.650822 | 2.24 | 1.262356 | 0.816596 | 4.001499 | 0.456329 |
| 1.20 | 1.071069 | 0.958702 | 1.550424 | 0.646168 | 2.25 | 1.263098 | 0.815543 | 4.030469 | 0.456406 |
| 1.21 | 1.073337 | 0.957292 | 1.567316 | 0.641824 | 2.26 | 1.263809 | 0.814499 | 4.059518 | 0.456524 |
| 1.22 | 1.075610 | 0.955873 | 1.584524 | 0.637338 | 2.27 | 1.264490 | 0.813464 | 4.088645 | 0.456682 |
| 1.23 | 1.077889 | 0.954446 | 1.601891 | 0.632933 | 2.28 | 1.265141 | 0.812437 | 4.117853 | 0.456881 |
| 1.24 | 1.080175 | 0.953010 | 1.619419 | 0.628610 | 2.29 | 1.265762 | 0.811419 | 4.147138 | 0.457119 |
| 1.25 | 1.082466 | 0.951568 | 1.637106 | 0.624369 | 2.30 | 1.266353 | 0.810409 | 4.176503 | 0.457399 |
| 1.26 | 1.084760 | 0.950117 | 1.654954 | 0.620209 | 2.31 | 1.266914 | 0.809409 | 4.205946 | 0.457719 |
| 1.27 | 1.087056 | 0.948660 | 1.672961 | 0.616131 | 2.32 | 1.267445 | 0.808417 | 4.235469 | 0.458079 |
| 1.28 | 1.089352 | 0.947195 | 1.691128 | 0.612135 | 2.33 | 1.267946 | 0.807433 | 4.265071 | 0.458480 |
| 1.29 | 1.091647 | 0.945724 | 1.709456 | 0.608220 | 2.34 | 1.268417 | 0.806459 | 4.294751 | 0.458921 |
| 1.30 | 1.093941 | 0.944246 | 1.727943 | 0.604387 | 2.35 | 1.268858 | 0.805493 | 4.324511 | 0.459402 |
| 1.31 | 1.096235 | 0.942761 | 1.746590 | 0.600636 | 2.36 | 1.269269 | 0.804536 | 4.354349 | 0.459924 |
| 1.32 | 1.098528 | 0.941270 | 1.765397 | 0.596966 | 2.37 | 1.269650 | 0.803588 | 4.384266 | 0.460486 |
| 1.33 | 1.100820 | 0.939777 | 1.784294 | 0.593295 | 2.38 | 1.270000 | 0.802648 | 4.414176 | 0.461089 |
| 1.34 | 1.103112 | 0.938292 | 1.803358 | 0.589766 | 2.39 | 1.270321 | 0.801717 | 4.444151 | 0.461089 |
| 1.35 | 1.105404 | 0.936804 | 1.822555 | 0.586302 | 2.40 | 1.270612 | 0.800795 | 4.474192 | 0.461089 |
| 1.36 | 1.107695 | 0.935314 | 1.841884 | 0.582901 | 2.41 | 1.270872 | 0.799881 | 4.504300 | 0.461089 |
| 1.37 | 1.109985 | 0.933821 | 1.861347 | 0.579566 | 2.42 | 1.271103 | 0.798977 | 4.534475 | 0.461089 |
| 1.38 | 1.112275 | 0.932327 | 1.880942 | 0.576294 | 2.43 | 1.271303 | 0.798081 | 4.564716 | 0.461089 |
| 1.39 | 1.114564 | 0.930830 | 1.900671 | 0.573087 | 2.44 | 1.271474 | 0.797193 | 4.595024 | 0.461089 |
| 1.40 | 1.116853 | 0.929332 | 1.920532 | 0.569944 | 2.45 | 1.271614 | 0.796315 | 4.625399 | 0.461089 |
| 1.41 | 1.119135 | 0.927832 | 1.940526 | 0.566865 | 2.46 | 1.271725 | 0.795445 | 4.655841 | 0.461089 |
| 1.42 | 1.121422 | 0.926332 | 1.960653 | 0.563851 | 2.47 | 1.271805 | 0.794584 | 4.686349 | 0.461089 |
| 1.43 | 1.123702 | 0.924830 | 1.980913 | 0.560901 | 2.48 | 1.271856 | 0.793731 | 4.716924 | 0.461089 |
| 1.44 | 1.125977 | 0.923328 | 2.001305 | 0.558015 | 2.49 | 1.271876 | 0.792888 | 4.747566 | 0.461089 |
| 1.45 | 1.128245 | 0.921825 | 2.021831 | 0.555194 | 2.50 | 1.271876 | 0.792053 | 4.778275 | 0.461089 |
| 1.46 | 1.130507 | 0.920322 | 2.042575 | 0.552385 | 2.51 | 1.271876 | 0.791226 | 4.809050 | 0.461089 |
| 1.47 | 1.132764 | 0.918819 | 2.063267 | 0.549676 | 2.52 | 1.271876 | 0.790409 | 4.839892 | 0.461089 |
| 1.48 | 1.135013 | 0.917316 | 2.084085 | 0.547018 | 2.53 | 1.271876 | 0.789600 | 4.870600 | 0.461089 |
| 1.49 | 1.137257 | 0.915814 | 2.105029 | 0.544409 | 2.54 | 1.271856 | 0.788800 | 4.901300 | 0.461089 |
| 1.50 | 1.139495 | 0.914313 | 2.126100 | 0.541850 | 2.55 | 1.271876 | 0.788009 | 4.932000 | 0.461089 |
| 1.51 | 1.141709 | 0.912813 | 2.147297 | 0.539341 | 2.56 | 1.271876 | 0.787226 | 4.962700 | 0.461089 |
| 1.52 | 1.143913 | 0.911313 | 2.168621 | 0.536882 | 2.57 | 1.271876 | 0.786452 | 4.993400 | 0.461089 |
| 1.53 | 1.146107 | 0.909816 | 2.190070 | 0.534472 | 2.58 | 1.271876 | 0.785687 | 5.024100 | 0.461089 |
| 1.54 | 1.148291 | 0.908320 | 2.211647 | 0.532112 | 2.59 | 1.271876 | 0.784930 | 5.054900 | 0.461089 |
| | | | | | 2.60 | 1.271856 | 0.784182 | 5.085600 | 0.461089 |

[1] V. V. Sargsyan, Yu. V. Palchikov, Z. Kanokov, G. G. Adamian, and N. V. Antonenko, Phys. Rev. C **76**, 064604 (2007).

[2] P. N. Nadtochy, A. Kelić, K.-H. Schmidt, Phys. Rev. C **75**, 064614 (2007).

[3] V. A. Rubchenya, Phys. Rev. C **75**, 054601 (2007).

[4] Ying Jia and Jing-Dong Bao, Phys. Rev. C **75**, 034601 (2007).

[5] P. D. Shidling *et al.*, Phys. Rev. C **74**, 064603 (2006).

[6] E. Holub *et al.*, Phys. Rev. C **28**, 252 (1983).

[7] W. P. Zank *et al.*, Phys. Rev. C **33**, 519 (1986).

[8] A. Gavron *et al.*, Phys. Lett. B **176**, 312 (1986).

[9] A. Gavron *et al.*, Phys. Rev. C **35**, 579 (1987).

[10] D. J. Hinde *et al.*, Nucl. Phys. **A452**, 550 (1986).

[11] D. Hilscher and H. Rossner, Ann. Phys. (Paris) **17**, 471 (1992).

[12] P. Paul and M. Thoennessen, Annu. Rev. Nucl. Part. Sci. **44**, 65 (1994).

[13] I. Dioszegi, N. P. Shaw, I. Mazumdar, A. Hatzikoutelis, and P. Paul, Phys. Rev. C **61**, 024613 (2000).

[14] N. P. Shaw *et al.*, Phys. Rev. C **61**, 044612 (2000).

[15] I. Dioszegi *et al.*, Phys. Rev. C **63**, 014611 (2000).

[16] F. Pulnhofer, Nucl. Phys. **A280**, 267 (1977).

[17] M. Blann and T. A. Komoto, Lawrence Livermore National Laboratory Report No. UCID 19390 (1982).

[18] M. Blann and J. Bisplinghoff, Lawrence Livermore National Laboratory Report No. UCID 19614 (1982).

[19] A. Gavron, Phys. Rev. C 21, **230** (1980).

[20] H. Rossner *et al.*, Phys. Rev. C **40**, 2629 (1989).

[21] J. P. Lestone *et al.*, Nucl. Phys. **A559**, 277 (1993).

[22] V. M. Strutinsky, Phys. Lett. **47B**, 121 (1973).

[23] I. I. Gontchar, P. Fröbrich, and N. I. Pischasov, Phys. Rev. C **47**, 2228 (1993).

[24] R. J. Charity, Phys. Rev. C **53**, 512 (1996).

[25] J. P. Lestone, Phys. Rev. C **59**, 1540 (1999).

[26] A. Bohr and B. R. Mottelson, Nuclear Structure Vol. II (W. A. Benjamin, 1975).

[27] S. G. McCalla and J. P. Lestone, Phys. Rev. Lett. **101**, 032702 (2008).

[28] J. R. Nix and A. J. Sierk, J. Madras University Section B, **50** ,38 (1987) [Los Alamos National Laboratory Report, LA-UR-87-133 (1987)].

[29] J. R. Nix and A. J. Sierk, International School-Seminar on Heavy Ion Physics, Dubna, USSR, Septmeber 23-30, 1986.

[30] A. J. Sierk, Phys. Rev. C **33**, 2039 (1986).

[31] J. Töke and W. J. Swiatecki, Nucl. Phys. **A372**, 141 (1981).

[32] M. Prakash, J. Wambach, and Z. Y. Ma, Phys. Lett. **128B**, 141 (1983).

[33] J. P. Lestone, Phys. Rev. C **52**, 1118 (1995).

[34] S. Shlomo, Nucl. Phys. **A539**, 17 (1992).

[35] A.V. Ignatyuk, M. G. Itkis, V. N. Smirenkin, and A. S. Tiskin, Yad. Fiz. **21**, 1185 (1975) [Sov. J. Nucl. Phys. **21**, 612 (1975)].

[36] W. Reisdorf, Z. Phys. A **300**, 227 (1981).

[37] X. Campi and S. Stringari, Z. Phys. A 309, 239 (1983)

[38] N. Bohr and J. A. Wheeler, Phys. Rev. **56**, 426 (1939).

[39] J. P. Lestone, Mod. Phys. Lett. A, **23**, 1067 (2008).

[40] A. Bohr and B. R. Mottelson, Nuclear Structure Vol I (W. A. Benjamin, 1969).

[41] D. Boilley, E. Suraud, Y. Abe, and S. Ayik, Nucl. Phys. **A556**, 67 (1993).

[42] H. A. Kramers, Physika **7**, 284 (1940).

[43] P. Grangé, Li Jun-Qing, and H. A. Weidenmüller, Phys. Rev. C **27**, 2063 (1983).

[44] K. H. Bhatt, P. Grangé, and B. Hiller, Phys. Rev. C **33**, 954 (1986).

[45] J. R. Nix *et al.*, Nucl. Phys. **A424**, 239 (1984).

[46] J. P. Lestone, Phys. Rev. C **51**, 580 (1995).

[47] W. D. Myers and W. J. Swiatecki, Nucl. Phys. **81**, 1 (1966).

[48] W. D. Myers and W. J. Swiatecki, Ark. Fys. **36**, 343 (1967).
[49] P. Möller and J. R. Nix, Phys. Rev. Lett. **37**, 1461 (1976).
[50] J. J. Griffin and M. Dworzecka, Nucl. Phys. **A455**, 61 (1986).
[51] C. Yannouleas, Nucl. Phys. **A439**, 336 (1985).
[52] A. J. Sierk, private communication.
[53] P. N. Nadtochy and G. D. Adeev, Phys. Rev. C **72**, 054608 (2005).
[54] R. Vandenbosch and J. R. Huizenga, Nuclear Fission (Academic Press, New York, 1973).
[55] T. Døssing and J. Randrup, Nucl. Phys. **A433**, 215 (1985).
[56] J. Randrup, Nucl. Phys. **A383**, 468 (1982).
[57] J. P. Lestone, A. A. Sonzogni, M. P. Kelly, and R. Vandenbosch, J. Phys. G: Nucl. Part. Phys. **23**, 1349 (1997).
[58] R. J. Charity *et al.*, Nucl. Phys. **A457**, 441 (1986).
[59] H. Esbensen, Nucl. Phys. **A352**, 147 (1981).
[60] M. Lozano and G. Madurga, Nucl. Phys. **A334**, 349 (1980).
[61] J. P. Lestone *et al.*, Phys. Rev. C **55**, R16 (1997).
[62] J. P. Lestone *et al.*, Phys. Rev. C **56**, R2907 (1997).
[63] K. T. R. Davies and J. R. Nix, Phys. Rev. C **14**, 1977 (1976).
[64] S. Raman *et al.*, Atom. Data and Nucl. Data Tables, **36**, 1 (1987).
[65] D. J. Hinde *et al.*, Phys. Rev. C **53**, 1290 (1996).
[66] J. R. Leigh *et al.*, Phys. Rev. C **52**, 3151 (1995).
[67] A. A. Sonzogni *et al.*, Phys. Rev. C **57**, 722 (1998).
[68] D. J. Hinde *et al.*, Nucl. Phys. **A398**, 308 (1983).
[69] C. Morton *et al.*, Phys. Rev. C **52**, 243 (1995).
[70] D. J. Hinde *et al.*, Nucl. Phys. **A385**, 109 (1982).
[71] M. Perey and F. G. Perey, At. Nucl. Data Tables **17**, 1 (1976).
[72] J. R. Huizenga and G. Igo, Nucl. Phys. **29**, 462 (1962).
[73] H. Ikezoe *et al.*, Phys. Rev. C **46**, 1922 (1992).
[74] H. Ikezoe *et al.*, Phys. Rev. C **49**, 968 (1994).
[75] J. E. Lynn, The Theory of Neutron Resonance Reactions (Clarendon Press, Oxford, 1968)
[76] D. Ward, R. J. Charity, D. J. Hinde, J. R. Leigh, and J. O. Newton, Nucl. Phys. **A403**, 189 (1983).
[77] J. O. Newton *et al.*, Nucl. Phys. **A483**, 126 (1988).
[78] K. –T. Brinkmann *et al.*, Phys. Rev. C **50**, 309 (1994).
[79] A. L. Caraley, B. P. Henry, J. P. Lestone, and R. Vandenbosch, Phys. Rev. C **62**, 054612 (2000).
[80] T. Sikkeland, Phys. Rev. **135**, 669 (1964).
[81] T. Sikkeland *et al.*, Phys. Rev. C **3**, 329 (1971).
[82] J. van der Plicht *et al.*, Phys. Rev. C **28**, 2022 (1983).
[83] B. G. Glagola, B. B. Back, and R. R. Betts, Phys. Rev. C **29**, 486 (1984).
[84] Forster *et al.*, Nucl. Phys. **A464**, 497 (1987).